%% file: main.tex
\documentclass[a4paper,11pt]{article}
\usepackage{jheppub} 
\usepackage{lineno}
\usepackage[T1]{fontenc} 
\usepackage{cancel}
\usepackage{xcolor}
\usepackage{enumerate}
\usepackage{bm}
\usepackage{amssymb}
\usepackage[numbers]{natbib}
\usepackage{subcaption} 

\usepackage{physics}
\usepackage{graphicx}
\usepackage{amsmath}
\usepackage{bbold}
\usepackage{yfonts}
\usepackage{placeins}
\usepackage{bm} 
\usepackage{nicefrac}
\usepackage{slashed}
\usepackage{feynmp}
\usepackage{tikz-feynman}
\tikzfeynmanset{compat=1.1.0}
\usepackage{accents}

\newcommand{\nn}{\nonumber\\}

\title{Diffusion coefficient matrix for multiple conserved charges: a Kubo approach}

\author[1]{Sourav Dey,}
\author[1,2]{Amaresh Jaiswal}
\author[1]{and Hiranmaya Mishra}
\affiliation[1]{School of Physical Sciences, National Institute of Science Education and Research, An OCC of Homi Bhabha National Institute, Jatni-752050, India}
\affiliation[2]{Institute of Theoretical Physics, Jagiellonian University, ul. St. \L ojasiewicza 11, 30-348 Krakow, Poland}

\emailAdd{sourav.dey@niser.ac.in}
\emailAdd{a.jaiswal@niser.ac.in}
\emailAdd{hiranmaya@niser.ac.in}

\abstract{
The strongly interacting matter created in relativistic heavy-ion collisions possesses several conserved quantum numbers, such as baryon number, strangeness, and electric charge. The diffusion process of these charges can be characterized by a diffusion matrix that describes the mutual influence of the diffusion of various charges. We derive the Kubo relations for evaluating diffusion coefficients as elements of a diffusion matrix. We further demonstrate that in the weak coupling limit, the diffusion matrix elements obtained through Kubo relations reduce to those obtained from kinetic theory with an appropriate identification of the relaxation times. We illustrate this evaluation in a toy model of two interacting scalar fields with two conserved charges.
}

\input{shortcut}

\begin{document}
\today
\maketitle
\flushbottom


\section{Introduction}
\label{sec:intro}
The computation of transport coefficients in relativistic systems holds significant importance across various fields, such as relativistic astrophysics, cosmology, and high-energy heavy-ion physics. In the context of heavy ion collisions, the strongly interacting medium that is produced in such a process shows collective motion and its space-time evolution has been described successfully using relativistic dissipative hydrodynamics along with modelling for the early stage and the freeze out for the hadrons \cite{Jeon:2015dfa, Gale:2013da, Florkowski:2010zz}. The dissipative effects like the shear and bulk viscosity play a significant role in the evolution of QGP and influence various observables like flow coefficients \cite{Heinz:2013th, Kovtun:2004de} and the hadron transverse momentum spectra, the hadron transverse momentum spectrum \cite{Dusling:2011fd, Bozek:2009dw}. Besides the viscosity coefficients, the other transport coefficient that has become relevant in the context of heavy ion collisions is the electrical conductivity of the strongly interacting matter. This has been discussed in the context of charge fluctuations \cite{Ling:2013ksb}, the evolution of electromagnetic fields \cite{Tuchin:2014hza, Gursoy:2014aka}. Further, it has been suggested that it can be extracted from the flow parameters in heavy ion collisions \cite{Hirono:2012rt}. There has been a recent interest in the diffusion of conserved charges due to temperature and density gradients, particularly for low energy collisions\cite{Greif:2017byw, Fotakis:2019nbq, Fotakis:2021diq, Das:2021bkz, Fotakis:2022usk}. As a dissipative phenomenon, diffusion emerges whenever variations in a conserved quantity occur. 

Let us note that the fluctuations of conserved charges play an important role to find the critical point of QCD \cite{Asakawa:2015ybt, Jeon:2000wg, Shuryak:2000pd, Pratt:2019pnd}. In this context, the process of diffusion plays an important role as the time evolution of conserved charges can be caused by such processes. When dealing with a system with multiple conserved charges, such as electric charge, baryon number, and strangeness, the diffusion processes associated with these charges are not independent \cite{Onsager:1931kxm}. Instead, they must be described through a set of coupled diffusion equations. To account for this phenomenon, the conventional diffusion coefficients corresponding to each conserved charge are replaced by a diffusion coefficient matrix \cite{Greif:2017byw, Fotakis:2019nbq, Das:2021bkz, Fotakis:2021diq, Fotakis:2022usk}. This matrix quantifies the coupling between the various conserved quantum numbers. The diagonal elements of the matrix represent the familiar charge diffusion coefficients, while the off-diagonal elements characterize the diffusive coupling between different charge currents.

To estimate the transport coefficients, it must be kept in mind that these coefficients capture the dynamics of the underlying microscopic theory. Thus, any first principle calculation must include the challenges of strong coupling.  In the context of diffusion matrix, the initial estimation of the matrix elements of this matrix have been derived within the ambit of kinetic theory, both with Chapman Enskog (CE) expansion\cite{Greif:2017byw, Fotakis:2019nbq, Fotakis:2022usk} and a relaxation time approximation (RTA) \cite{Fotakis:2019nbq, Das:2021bkz}. Although the relaxation time approximation allows one to use a much simpler collision kernel for the Boltzmann kinetic equation, it is not possible to have control on the systematic degree of accuracy of the method. Moreover, RTA, with momentum-dependent relaxation time, is in contradiction with the macroscopic conservation laws although novel approaches have been proposed to overcome them \cite{Bhatnagar:1954zz, Rocha:2021zcw}. Due to its simplicity of estimating the transport coefficients, the RTA has been used widely both for hadronic and partonic matter \cite{Muller:1967zza, Chapman:1970, Israel:1979wp, York:2008rr, Betz:2008me, Denicol:2010xn, Denicol:2012cn, Jaiswal:2014isa, jaiswal:2015mxa, Gabbana:2017uvc}.  On the other hand, CE approach is a variational approach that allows one to obtain solutions with arbitrary accuracy depending on the order of approximation used. The CE method is also consistent with macroscopic conservation laws.

In contrast to kinetic theory approach, the other often used method to estimate the transport coefficients is the Green-Kubo correlator approach. Using the linear response theory,  the Green-Kubo formula relates  the transport coefficients  to the spectral functions of  the relevant current-current correlators. For example, the shear and bulk viscosities are related to energy-momentum correlators, the electrical conductivity corresponds to the vector currents correlators of the light quarks and the heavy quark diffusion to the color electric field correlators. If the temperature is sufficiently high or the theory is weakly coupled, transport coefficients can be computed in a perturbative expansion using the Kubo relations. On the other hand, such relation is also valid, in general, for strong couplings. This, therefore, opens up the possibility of applying lattice QCD at finite temperature as a nonperturbative tool to compute transport coefficients provided the analytic continuation from euclidean correlators to spectral functions can be done reliably. Indeed, shear and bulk viscosity coefficients have been obtained within  the ambit of lattice QCD and Green-kubo relations \cite{Meyer:2011gj}. The electrical conductivity has also been estimated for a 2+1 flavor system in Ref.~\cite{Aarts:2014nba} using maximum entropy method. Keeping in mind the importance of cross-diffusion coefficients and the nonperturbative features of Green-Kubo formulation, we attempt here to derive the Green-Kubo relations for the diffusion matrix elements.

In Kubo approach, the equilibrium correlation functions can be utilized to compute the response, offering the advantage of evaluating transport coefficients using equilibrium Green's functions formulated in imaginary time. In the present investigation, we employ the correlation-function method developed by Zubarev to derive transport coefficients of a relativistic fluid \cite{Zubarev:1974dn}. In particular, we obtain Kubo formulas for diffusion matrix for a system with multiple conserved charges, such as electric charge, baryon number, and strangeness. The Zubarev method is founded on the concept of a non-equilibrium statistical operator (NESO), which extends the equilibrium Gibbs statistical operator to non-equilibrium states. By expanding the operator using small gradients of the thermodynamic parameters, this approach allows us to derive kinetic transport coefficients from correlation functions. Typically, the state of the system can be described by fields of temperature, chemical potential, and momentum, which may vary in space and time. The specific parameter set required for a complete system description depends on the problem at hand. The transport coefficients are obtained by perturbing the non-equilibrium statistical operator linearly around its equilibrium value. This allows to relate the transport coefficients to retarded equilibrium correlation functions of the microscopic theory. Such an approach has been successfully applied to compute viscosity coefficients of strongly interacting matter \cite{Harutyunyan:2017ttz} also in the presence of magnetic field \cite{Huang:2011dc, Harutyunyan:2016rxm}. We follow a similar approach to estimate diffusion matrix elements for a system of two complex interacting scalar fields.

This paper is organized as follows: In Sec.~\ref{sec:neso}, we review the basics of Zubarev's method of non-equilibrium statistical operators to derive transport coefficients and derive Kubo formulas for transport coefficients, including the diffusion matrix. In Sec.~\ref{sec:transport}, we employ Kubo relations to derive the diffusion coefficient matrix for two interacting charged scalar fields coupled to each other and with self-interaction. We summarize our results in Sec.~\ref{sec:summary}.

We use natural units ($c=\hbar=k_{B}=1$) and mostly negative metric sign convention $\eta^{\mu\nu}=\;$diag$\,(+1,-1,-1,-1)$. The bold font denotes three vectors. The scalar products for three- and four-vectors are denoted by a dot, i.e., $a \cdot b = a^0 b^0 - {\bm a} \cdot {\bm b}$.


\section{Non-equilibrium statistical operators and Kubo relations} \label{sec:neso}

Hydrodynamics is an effective field theory to describe the collective behaviour of a system, and its universal features are based on conservation laws. Zubarev's method of non-equilibrium statistical operator (NESO) provides a suitable framework for describing systems in the hydrodynamic regime while incorporating thermodynamic parameters such as temperature and chemical potential, which can be defined locally. In this method, the relevant NESO, which characterizes the nonequilibrium state of the system, can be expressed as \cite{Zubarev:1974dn}
\begin{align}
\hat{\varrho}(t) =\frac{1}{\mathcal{Q}} \exp \left[-\int d^{3} \textbf{x} \, \mathcal{\hat{Z}}(\textbf{x}, t)\right], \label{NESO}
\end{align}
where $\mathcal{Q}=\text{Tr}\exp \left[-\int d^{3} \textbf{x} \,\mathcal{\hat{Z}}(\textbf{x}, t)\right]$ is the normalization factor. The operator $\mathcal{\hat{Z}}(\textbf{x}, t)$ in Eq.~\eqref{NESO} has the form \cite{Zubarev:1974dn}
\begin{align}
    \mathcal{\hat{Z}}(\textbf{x}, t) = \varepsilon \int_{-\infty}^{t} d t_{1} \, e^{\varepsilon\left(t_{1}-t\right)}\Big[&\beta^{\nu}\left(\textbf{x}, t_{1}\right) \hat{T}_{0 \nu}\left(\textbf{x}, t_{1}\right) - \sum_{a}\alpha_{a}\left(\textbf{x}, t_{1}\right) \hat{J}^{0}_{a}\left(\textbf{x}, t_{1}\right)\Big]~,
    \label{eq:Z_function_def}
\end{align}
where $\varepsilon$ tends to zero after taking the thermodynamic limit. 

The operators $\hat{T}^{\mu\nu}$ and $\hat{J}_{a}^{\mu}$ correspond to the energy-momentum tensor and the charge currents, respectively. They satisfy the conservation equations given by
\begin{align}
    \partial_{\mu}\hat{T}^{\mu\nu}=0\, ~,~~ \partial_{\mu}\hat{J}_{a}^{\mu}=0.
    \label{eq:Con_Eq_in_Operator_form}
\end{align}
The quantities $\beta^{\mu}(\textbf{x},t)$ and $\alpha_{a}(\textbf{x},t)$ in Eq.~\eqref{eq:Z_function_def} are related to temperature, chemical potential, and fluid velocity as
\begin{align}
\beta^{\mu}(\textbf{x},t) &= \beta(\textbf{x},t) \, u^{\mu}(\textbf{x},t), \\
\alpha_{a}(\textbf{x},t) &= \sum_{A}q_{aA} \, \alpha_{A}(\textbf{x},t)=\beta(\textbf{x},t)\sum_{A}q_{aA} \, \mu_{A}(\textbf{x},t),
\end{align}
where $\beta(\textbf{x},t)$ represents the inverse temperature, $u^{\mu}(\textbf{x},t)$ is the fluid velocity, and $\mu_{a}(\textbf{x},t)=\sum_{A}q_{aA} \, \mu_{A}(\textbf{x},t)$ is the chemical potential associated with the species labeled by $a$. In general, each species can have multiple conserved charges $q_{a A}$ with the associated chemical potential $\mu_{A}(\textbf{x},t)$. For instance, in the context of heavy ion collision, $a\in (u,d,s)$ while $A\in (B,Q,S)$. Here $(u,d,s)$ represents up, down and strange quarks, respectively, and $(B,Q,S)$ corresponds to baryon number, electric charge and strangeness. 
Performing integration by parts in Eq.~\eqref{eq:Z_function_def}, we obtain,
\begin{align}
    \mathcal{\hat{Z}}(\textbf{x}, t) =&~ \beta^{\nu}\left(\textbf{x}, t\right) \hat{T}_{0 \nu}\left(\textbf{x}, t\right) - \sum_{a}\alpha_{a}\left(\textbf{x},t \right) \hat{J}_{a}^{0}\left(\textbf{x}, t\right)
    \nn
    & - \varepsilon \int_{-\infty}^{t} d t_{1} e^{\varepsilon\left(t_{1}-t\right)}\frac{d}{dt_{1}}\Big[\beta^{\nu}\left(\textbf{x}, t_{1}\right) \hat{T}_{0 \nu}\left(\textbf{x}, t_{1}\right)-\sum_{a}\alpha_{a}\left(\textbf{x}, t_{1}\right) \hat{J}_{a}^{0}\left(\textbf{x}, t_{1}\right)\Big]~.
    \label{eq:Z_function_byparts}
\end{align}
Using the conservation equations given in Eq.~\eqref{eq:Con_Eq_in_Operator_form} in the above equation, one can write the NESO of Eq.~\eqref{NESO} as
\begin{align}
    \hat{\varrho}(t)=\frac{1}{\mathcal{Q}}\text{exp}\Big[\hat{\mathcal{A}}-\hat{\mathcal{B}}\Big].\label{NESO1}
\end{align}
In the above equation, we have
\begin{align}
    \mathcal{\hat{A}} &=\int d^{3} \textbf{x}\left[\beta^{\nu}(\textbf{x}, t) \hat{T}_{0 \nu}(\textbf{x}, t)-\sum_{a}\alpha_{a}(\textbf{x}, t) \hat{J}_{a}^{0}(\textbf{x}, t)\right]~,\label{eq:equi_part_Z} \\
    \mathcal{\hat{B}} &=\int d^{3} \textbf{x} \int_{-\infty}^{t} d t_{1} e^{\varepsilon\left(t_{1}-t\right)} \, \mathcal{\hat{C}}(\textbf{x}, t_{1})~, \label{eq:non_equi_part_Z}
\end{align}
and
\begin{align}
    \mathcal{\hat{C}}(\textbf{x},t_{1})=\hat{T}^{\mu \nu}\left(\textbf{x}, t_{1}\right) \partial_{\mu} \beta_{\nu}\left(\textbf{x}, t_{1}\right)-\sum_{a}\hat{J}_{a}^{\mu}\left(\textbf{x}, t_{1}\right) \partial_{\mu} \alpha_{a}\left(\textbf{x}, t_{1}\right)~, \label{eq:opera_thermo_force}
\end{align}
where $\frac{\partial}{\partial t_1} \to \partial_\mu$ has been performed by adding a surface term.

The expression $\hat{\mathcal{A}}$ in Eq.~\eqref{eq:equi_part_Z} having terms without any gradient of temperature and chemical potentials corresponds to the equilibrium part of the NESO whereas the operator $\hat{\mathcal{B}} $ in Eq.~\eqref{eq:non_equi_part_Z} corresponds to the non-equilibrium part, with the expression of $\mathcal{\hat{C}}$ in Eq.~\eqref{eq:opera_thermo_force} containing gradient of temperature and chemical potentials representing thermodynamic \textit{forces}. The statistical operator in Eq.~\eqref{NESO1} can then be used to derive transport equations. To this end, we will treat the non-equilibrium part $\hat{\mathcal{B}}$ as a perturbation. We will keep up to linear order in $\hat{\mathcal{B}}$ so that the relations between the thermodynamic processes and dissipative currents are linear. Expanding up to linear order in $\mathcal{\hat{B}}$, i.e., first-order gradient of thermodynamic variables, the NESO can be written as
\begin{align}
    \hat{\varrho}=\left[1+\int_{0}^{1} d \lambda\left(\mathcal{\hat{B}}_{\lambda}-\langle \mathcal{\hat{B}}_{\lambda} \rangle_{l}\right)\right] \hat{\varrho}_{l}~, \label{eq:rho}
\end{align}
where we have used the shorthand notation for any operator $\mathcal{\hat{X}}$ as
\begin{align}
    \mathcal{\hat{X}}_{\lambda} =  e^{-\lambda \mathcal{\hat{A}} } \mathcal{\hat{X}} e^{\lambda \mathcal{\hat{A}} }~.\label{TTofX}
\end{align}
Here, $\hat{\varrho}_{l}$ is the equilibrium part of NESO given as 
\begin{align}
    \hat{\varrho}_{l}(t)=\mathcal{Q}_{l}^{-1}\text{exp}\Big[-\hat{\mathcal{A}}\Big],\label{Leqd}
\end{align}
with $\mathcal{Q}=\text{Tr}\,\exp\Big[-\hat{\mathcal{A}}\Big]$.
Further, in Eq.~\eqref{eq:rho}, $\langle\hat{O}\rangle_{l}$ is the average over the local equilibrium that is $\langle\hat{O}\rangle_{l}=\text{Tr}\Big(\hat{\varrho_{l}}(t) \, \hat{O}\Big)$.

Using Eq.~\eqref{eq:rho} we can write down the deviation of the statistical average of any arbitrary tensor operator from its equilibrium value as 
\begin{align}
    \delta\left\langle\hat{\mathcal{O}}^{\mu_1 \mu_2...\mu_n}(\mathbf{x}, t)\right\rangle &= \operatorname{Tr}\left[ \hat{\varrho}_{l} \int_{0}^{1} d \lambda \, \mathcal{\hat{B}}_{\lambda} \, \hat{\mathcal{O}}^{\mu_1 \mu_2...\mu_n}(\mathbf{x}, t)\right]  - \operatorname{Tr}\left[\hat{\varrho}_{l} \, \langle \mathcal{\hat{B}}_{\lambda} \rangle_{l} \, \hat{\mathcal{O}}^{\mu_1 \mu_2...\mu_n}(\mathbf{x}, t)\right]~. \label{eq:opera_devia_O}
\end{align}
Substituting the expression for the non-equilibrium part of  $\hat{\mathcal{B}}$ from Eq.~\eqref{eq:non_equi_part_Z} in  Eq.~\eqref{eq:opera_devia_O}, we can write the operator deviation as
\begin{align}
    \delta\left\langle\hat{\mathcal{O}}^{\mu_1 \mu_2...\mu_n}(\mathbf{x}, t)\right\rangle=& \int d^{3} \mathbf{x}_{1} \int_{-\infty}^{t} d t_{1} e^{\varepsilon\left(t_{1}-t\right)} 
    \nn
    & \times\int_{0}^{1} d\lambda  \left\langle \hat{\mathcal{O}}^{\mu_1 \mu_2...\mu_n}(\mathbf{x}, t) \left[\mathcal{\hat{C}}_{\lambda}\left(\mathbf{x}_{1}, t_{1}\right) - \left\langle \mathcal{\hat{C}}_{\lambda} \left(\mathbf{x}_{1}, t_{1}\right)\right\rangle_{l}\right]\right\rangle_{l}~.
    \label{eq:oper_devi_Kubo_nota}
\end{align}
Compactly, the off-equilibrium contribution of a hydro variable $\hat{\mathcal{O}}$ is connected with the gradient forces through the Kubo-Mori-Bogoliubov (KMB) inner product between the operator $\hat{\mathcal{O}}$ and $\hat{\mathcal{C}}$,
\begin{align}
    \delta\left\langle\hat{\mathcal{O}}^{\mu_1 \mu_2...\mu_n}(\mathbf{x}, t)\right\rangle =\int d^{3} \mathbf{x}_{1} \int_{-\infty}^{t} d t_{1} \, e^{\varepsilon\left(t_{1}-t\right)}\left(\hat{\mathcal{O}}^{\mu_1 \mu_2...\mu_n}(\mathbf{x}, t), \mathcal{\hat{C}}\left(\mathbf{x}_{1}, t_{1}\right)\right)~,
    \label{eq:EMT_in_Kubo_form}
\end{align}
where the Kubo-Mori-Bogoliubov (KMB) inner product is given by,
\begin{equation}
    \left(\hat{\mathcal{O}}_{a}(\mathbf{x}, t), \hat{\mathcal{O}}_{b}(\mathbf{x}_{1}, t_{1})\right)\equiv \int_{0}^{1} d \lambda\left\langle\hat{\mathcal{O}}_{a}(\mathbf{x}, t)\left[\hat{\mathcal{O}}_{2\lambda}\left(\mathbf{x}_{1}, t_{1}\right) - \left\langle\hat{\mathcal{O}}_{2\lambda}\left(\mathbf{x}_{1}, t_{1}\right)\right\rangle_{l}\right]\right\rangle_{l}~.
    \label{eq:Kubo_corr_nota}
\end{equation}
We shall use Eq.~\eqref{eq:EMT_in_Kubo_form} to define the transport coefficients in the calculation of dissipative contributions. To do so, one needs to distinguish between the equilibrium and dissipative contribution.

To separate the dissipative contributions in conserved hydrodynamic currents, we decompose the energy-momentum tensor and the charge currents. In general, we construct the operator, $\hat{T}^{\mu\nu}$, $
\hat{J}_{a}^{\mu}$ in the form
\begin{align}\label{decomposition}
    \hat{T}^{\mu\nu}&=\hat{\epsilon}u^{\mu}u^{\nu}-\hat{P}\Delta^{\mu\nu}+\heat^{\mu}u^{\nu}+\heat^{\nu}u^{\mu}+\mathcal{\hat{T}}^{\mu\nu}, 
    \\
    \hat{J}_{a}^{\mu}&=\hat{n}_{a}u^{\mu}+\hat{\mathcal{N}}_{a}^{\mu},
\end{align}
with $u_{\mu}$ being the fluid velocity and the  dissipative operators $\hat{\mathcal{T}}^{\mu\nu},\hat{\mathcal{Q}}^{\mu}$ and $\hat{\mathcal{N}}^{\mu}_{a}$ are orthogonal to $u_{\mu}$, and $\hat{\mathcal{T}}^{\mu\nu}$ is traceless~\cite{Harutyunyan:2021rmb, Huang:2011dc, Harutyunyan:2017lrm}. For further computation it is convenient to decompose the operator $\hat{\mathcal{C}}$ containing the thermodynamic forces into different dissipative processes as in Eq.~\eqref{decomposition},
\begin{align}
    \hat{\mathcal{C}}=\hat{\epsilon}D\beta-\hat{P}\beta\theta-\sum_{a}\hat{n}_{a}D\alpha_{a}+\heat_{\mu}\Big(\beta Du^{\mu}+\del^{\mu}\beta\Big)-\sum_{a}\mathcal{\hat{N}}_{a}^{\mu}\nabla_{\mu}\alpha_{a}+\beta\mathcal{\hat{T}}^{\mu\nu}\nabla_{\mu}u_{\nu},\label{hatC}
\end{align}
where we used the notation $D=u^{\mu}\del_{\mu},\, \theta=\del_{\mu}u^{\mu},\,\nabla_{\alpha}=\Delta_{\alpha\beta}\del^{\beta}$ with $\Delta_{\alpha\beta}=\eta_{\alpha\beta}-u_{\alpha}u_{\beta}$ being the projector orthogonal to $u_{\alpha}$. One can eliminate the derivatives $D\beta$, $D\alpha_{a}$ and $Du^{\mu}$ using the equations of ideal hydrodynamics. From the conservation of energy-momentum tensor and currents of ideal hydrodynamics $\partial_{\mu}\hat{T}_{(0)}^{\mu\nu}=0\,,\,\partial_{\mu}\hat{J}_{a\,0}^{\mu}=0$, we have
\begin{align}
D\epsilon &= -h\theta ,\label{Depsilon}\\
\quad Dn_{a} &= -n_{a}\theta ,\label{Dn}\\
Du_{\alpha}&=\frac{1}{h}\nabla_{\alpha}P,\label{acc}
\end{align}
where, $\theta=\partial_{\alpha}u^{\alpha}$ is the expansion scalar and $h=\epsilon+P$ is the enthalpy. One can choose the energy density $\epsilon$ and number density $n_{a}$ as independent thermodynamic variables to write 
\begin{align}
D\beta &= \left(\frac{\partial\beta}{\partial\epsilon}\right)_{n_{a}}D\epsilon + \sum_{a}\left(\frac{\partial\beta}{\partial n_{a}}\right)_{\epsilon,n_{b}\neq n_{a}}Dn_{a} \\
&= -\theta\Bigg[\left(\frac{\partial\beta}{\partial\epsilon}\right)_{n_{a}} h+\sum_{a}\left(\frac{\partial\beta}{\partial n_{a}}\right)_{\epsilon,n_{b}\neq n_{a}} n_{a}\Bigg],\label{Dbeta}
\\
D\alpha_{c} &= \left(\frac{\partial \alpha_{c}}{\partial\epsilon}\right)_{n_{a}}D\epsilon + \sum_{a}\left(\frac{\partial\alpha_{c}}{\partial n_{a}}\right)_{\epsilon,n_{b}\neq n_{a}}Dn_{a} \\
&= -\theta\Bigg[\left(\frac{\partial \alpha_{c}}{\partial\epsilon}\right)_{n_{a}} h+\sum_{a}\left(\frac{\partial\alpha_{c}}{\partial n_{a}}\right)_{\epsilon,n_{b}\neq n_{a}} n_{a}\Bigg].\label{Dalpha}
\end{align}

Next, we use the Gibbs-Duhem relations in thermodynamics, $\beta dP=-h d\beta+\sum_{a}n_{a} d\alpha_{a}$, which leads to
\begin{align}
    h=-\beta\left(\frac{\partial P}{\partial\beta}\right)_{\alpha_a}\, ,\,n_{a}=\beta\left(\frac{\partial P}{\partial\alpha_{a}}\right)_{\beta,n_{b}\neq n_{a}}.
\end{align}
Substituting the above equation into Eq.~\eqref{Dbeta} and \eqref{Dalpha}, one can write
\begin{align}
    D\beta &=\beta\,\theta\,\gamma = \beta\,\theta\left(\frac{\partial P}{\partial \epsilon}\right)_{n_{a}}, \label{Dbetam}
    \\
    D\alpha_{c} &=-\delta_{c}\,\beta\,\theta = -\beta\,\theta\left(\frac{\partial P}{\partial \alpha_{c}}\right)_{\epsilon,n_{b}\neq n_{c}}. \label{Dalpham}
\end{align}
Similarly, using the Gibbs-Duhem relation in Eq.~\eqref{acc}, we have 
\begin{align}
    Du_{\mu}=-T\,\nabla_{\mu}\beta + T\sum_{a}\frac{n_{a}}{\epsilon+P} \, \nabla_{\mu}\alpha_{a}.\label{accb}
\end{align}
Substituting the expression for the differentials in Eqs.~\eqref{Dbetam}, \eqref{Dalpham} and \eqref{acc} into Eq.~\eqref{hatC}, one obtains the operator $\hat{\mathcal{C}}$ at first order in gradients in a compact form as
\begin{align}
\hat{\mathcal{C}}&=-\beta\,\theta\,\hat{P}^{*}+\beta\,\mathcal{\hat{T}}^{\mu\nu}\sigma_{\mu\nu}-\sum_{a}\mathcal{\hat{J}}_{a}^{\mu}\,\nabla_{\mu}\alpha_{a}.\label{frame}
\end{align}
Here, $\sigma_{\mu\nu}\equiv\frac{1}{2}\left[ \Delta^\alpha_\mu \Delta^\beta_\nu + \Delta^\alpha_\nu \Delta^\beta_\mu -(2/3)\Delta^{\alpha\beta} \Delta_{\mu\nu} \right]\nabla_\alpha u_\beta$ and, the operators $\hat{P}^{*}$ and $\hat{\mathcal{J}}^{\mu}_{a}$ are defined as\,\footnote{ By matching condition, one can locally demand the densities of thermodynamic parameters to match its expectation value with respect to the local relevant statistical operator. But the relevant statistical operator as a function of non-uniform thermodynamic parameters gives an additional shift; that term has been absorbed into $\hat{P}^{\star}$~\cite{Harutyunyan:2017lrm}. Moreover, one can define a combination of energy flux and the particle flux, $\mathcal{\hat{J}}_{a}^{\mu}$, which remains invariant under any first-order transformation in choosing fluid frame $u^{\mu}$.\cite{Kovtun:2012rj} }
\begin{align}
   \hat{P}^{*}&=\hat{P}-\gamma\,\hat{\epsilon}-\sum_{a}\delta_{a}\,\hat{n}_{a}, \\
   \hat{\mathcal{J}}^{\mu}_{a}&=\,
   \hat{\mathcal{N}}^{\mu}_{a}-\frac{n_{a}}{h}\,\heat^{\mu}. \label{dmj}
\end{align}
In the present work, we perform our calculation with the above quantities which are valid in generic fluid frames. Since, hydrodynamic modes need to be subtracted as given in second expression with a proper definition of number density and enthalpy in equilibrium.\cite{Kovtun:2019hdm,Hosoya:1983id,Harutyunyan:2021rmb,Fukushima:2017lvb,Zubarev:1979afm}

Let us note that $\hat{J}^{\mu}_{a}$ and the corresponding density $\hat{n}_{a}$ is defined for the particle species basis. To calculate the cross-conductivity matrix elements (or the diffusion matrix in charge basis), we need to go to the conserved charge basis~\cite{Gavassino:2023qnw}. The chemical potentials in the basis of species are $\mu_{a}=\sum_{A}q_{aA}\,\mu_{A}$ where $q_{aA}$ is the corresponding matrix of charges of the species `$a$'. Here, we refer to the capital alphabet `$A$' for conserved charge quantum numbers. It is easy to check using the fact that  $\sum_{a}\mu_{a}\hat{n}_{a}=\sum_{A}\mu_{A}\hat{n}_{A}$, the number densities in the species basis are related to those in conserved charge basis as $n_{a}=\sum_{A}q^{-1}_{a A}n_{A}$~\cite{Gavassino:2023qnw}. Accordingly, the diffusion current $\hat{\mathcal{J}}^
{\mu}_{A}$ in the conserved charge basis can be written as 
\begin{align}
    \hat{\mathcal{J}}^{\mu}_{A}=\sum_{a}q_{a A}\,\hat{\mathcal{J}}^{\mu}_{a} = \hat{\mathcal{N}}^{\mu}_{A}-\frac{n_{A}}{h}\, \hat{Q}^{\mu}.\label{Atoa}
\end{align}
In the present work, we are interested in evaluating the diffusion matrix in the conserved charge basis, so for that, one needs to employ the transformation matrix $q_{aA}$, which connects the particle basis to the charge basis.

Now, using Eqs.~\eqref{eq:EMT_in_Kubo_form} and \eqref{hatC}, one can extract the off-equilibrium correction to each conserved current. Therefore, we notice that a similar correction to the particle diffusion currents in particle basis, i.e., the linear response of the particle diffusion current can be defined as
\begin{align}
    \mathcal{J}^{\mu}_{a}(\mathbf{x}, t)=\left\langle\mathcal{\hat{J}}^{\mu}_{a}(\mathbf{x}, t)\right\rangle=-\int d^{3} \mathbf{x}_{1} \int_{-\infty}^{t} d t_{1} e^{\varepsilon\left(t_{1}-t\right)}\left(\hat{\mathcal{J}}_{a}^{\mu}(\mathbf{x}, t), \mathcal{\hat{C}}\left(\mathbf{x}_{1}, t_{1}\right)\right)~.
\end{align}
Keeping terms till linear order in gradients, we obtain
\begin{align}
     \mathcal{J}^{\mu}_{a}(\mathbf{x}, t) =\left\langle\mathcal{\hat{J}}^{\mu}_{a}(\mathbf{x}, t)\right\rangle_{(1)}=-\sum_{b}\int d^{3} \mathbf{x}_{1} & \int_{-\infty}^{t} d t_{1}\, e^{\varepsilon\left(t_{1}-t\right)}\, \nabla_{\nu}\alpha_{b} \left(\mathbf{x},t\right) \nonumber\\
     &\times\left(\mathcal{\hat{J}}_{a}^{\mu}(\mathbf{x}, t), \hat{\mathcal{J}}_{b}^{\nu}\left(\mathbf{x}_{1}, t_{1}\right)\right)+\mathcal{O}(\nabla^{2}). \label{Charge kubo1}
\end{align}
The above equation can be expresses in the form 
\begin{align}
    \mathcal{J}^{\mu}_{a}(\mathbf{x},t)=\sum_{b}\mathcal{K}^{\mu\nu}_{ab}\, \nabla_{\nu}\alpha_{b}(\mathbf{x},t)\,, \label{eq:jmua}
\end{align}
where the diffusion tensor is defined as
\begin{align}
    \mathcal{K}^{\mu\nu}_{ab}=-\int d^{3} \mathbf{x}_{1} \int_{-\infty}^{t} d t_{1} e^{\varepsilon\left(t_{1}-t\right)}\left(\mathcal{\hat{J}}_{a}^{\mu}(\mathbf{x}, t), \hat{\mathcal{J}}_{b}^{\nu}\left(\mathbf{x}_{1}, t_{1}\right)\right)\,.
\end{align}

In order to construct the form of $\mathcal{K}^{\mu\nu}_{ab}$ tensor, we note that heat flow $\hat{\mathcal{Q}}^{\mu}$ and particle current $\mathcal{\hat{N}}_{a}^{\mu}$ satisfy the orthogonality with the time-like vector $\beta^{\mu}$, i.e., $\beta^{\mu}\hat{\mathcal{Q}}_{\mu}=\beta^{\mu}\hat{\mathcal{N}}_{a\,\mu}=0$. This is satisfied by demanding
\begin{equation}
    u_{\mu}\,\mathcal{K}^{\mu\nu}_{ab}=0\,.\label{orthogonality condition}
\end{equation}
Furthermore, the KMB inner product must satisfy the Onsager reciprocal relations which is expressed as\,\footnote{The proof of this relation is discussed in details in Refs. \cite{Huang:2011dc, Kovtun:2012rj}}
\begin{align}
    \mathcal{K}^{\mu\nu}_{ab}=\mathcal{K}^{\nu\mu}_{ba}\,.\label{onsager condition}
\end{align}
Using the properties of diffusion tensor in Eqs.~\eqref{orthogonality condition} and \eqref{onsager condition}, we can express it as
\begin{equation}
    \mathcal{K}^{\mu\nu}_{a\,b}=\Delta^{\mu\nu}\kappa_{ab}, \label{kmunuab}
\end{equation}
which leads to the expression of diffusion matrix as
\begin{align}
    \kappa_{ab}&=\frac{1}{3}\Delta_{\mu\nu}\mathcal{K}^{\mu\nu}_{ab}=-\frac{1}{3} \int d^{3} \mathbf{x}_{1} \int_{-\infty}^{t} d t_{1} e^{\varepsilon\left(t_{1}-t\right)}\left(\hat{\mathcal{J}}_{a}^{\mu}(\mathbf{x}, t), \hat{\mathcal{J}}_{b\,\mu}\left(\mathbf{x}_{1}, t_{1}\right)\right)\,.\label{Dab}
\end{align}
The above expression of diffusion matrix can be converted in the conserved charges basis by employing the conversion matrix as
\begin{align}
    \kappa_{AB}=-\frac{1}{3} \int d^{3} \mathbf{x}_{1} \int_{-\infty}^{t} d t_{1}\, e^{\varepsilon\left(t_{1}-t\right)}\left(\hat{\mathcal{J}}_{A}^{\mu}(\mathbf{x}, t), \hat{\mathcal{J}}_{B\,\mu}\left(\mathbf{x}_{1}, t_{1}\right)\right)=\sum_{a,b}q_{aA}\,q_{bB}\,\kappa_{ab}\,.\label{DAB}
\end{align}
This matrix is symmetric subject to the symmetry conditions of the diffusion matrix elements in species basis $\kappa_{ab}$. In the present work, however, we shall confine our discussion to the diffusion coefficient $\kappa_{ab}$ in particle basis respectively. In the next section, we evaluate these diffusion matrix elements in species basis for a system of two complex interacting scalar field theory. We then employ Eq.~\eqref{DAB} to translate this diffusion matrix into charge basis.

The computation of the dissipative diffusion coefficients boils down to the evaluation of the KMB product of the dissipative diffusive currents. We give a detailed derivation of this evaluation in Appendix-\ref{A2}. This leads to the diffusion coefficients getting related to the corresponding retarded Green's function $G^{R}_{\mathcal{\hat{J}}^{\mu}_{a},\mathcal{\hat{J}}_{\mu\,b}}$, resulting in
\begin{align}
    \kappa_{ab}&=\left.\frac{T}{3}\frac{\partial}{\partial \omega}\text{Im}\,\Big(G^{R}_{\mathcal{\hat{J}}^{\mu}_{a},\mathcal{\hat{J}}_{\mu\,b}}\Big)(\mathbf{0}, \omega)\right|_{\omega \rightarrow 0}~.\label{dimjj}
\end{align}
Using Eq.~\eqref{dGtorho} one can further relate $\kappa_{ab}$ to the spectral function as,
\begin{align}
    \kappa_{ab} &= \left.\frac{T}{3}\frac{\rho_{\hat{\mathcal{J}}_{a}\hat{\mathcal{J}}_{b}}(\omega,\mathbf{0})}{\omega}\right|_{\omega \rightarrow 0}~\, , \label{dspmjj}
\end{align}
Thus the diffusion coefficients are determined by the small frequency limit of the zero momentum spectral functions of corresponding composite operators. These two point correlators are to be evaluated at constant values of thermodynamic parameters, i.e. as if the system is in global thermal equilibrium. In the next section, a similar condition helps to set the thermal field theory with uniform temperature $T$ and chemical potentials $\mu_{a}$ for particle and then extract the transport coefficients.\footnote{ Here, one may think of uniform values of thermodynamic parameters as an average over a local patch of the fluid cells. The deviation due to the non-uniform profile has been absorbed in the higher order gradient correction than the transport coefficients at a given gradient order and discarded by changing the local equilibrium distribution to a global one. This non-locality in the thermodynamical forces is crucial for causal transport equations of hydrodynamics\cite{Harutyunyan:2021rmb,Koide:2006ef,Koide:2008nw,Hosoya:1983id}.}

For the sake of completeness, we also extract the off-equilibrium contribution to the trace-less and trace part of the energy-momentum tensor. A similar linear response correction to the energy-momentum tensor can be written as
\begin{equation}
\pi^{\mu\nu}\left(\mathbf{x},t\right) =\left\langle\hat{\mathcal{T}}^{\mu\nu}(\mathbf{x}, t)\right\rangle_{(1)} \, , \quad 
\Pi\left(\mathbf{x},t\right)=\left\langle P^{*}\left(\mathbf{x},t\right)\right\rangle_{(1)}\,.
\end{equation}
Further, up to linear order in gradients, the above relations can be expressed as
\begin{align}
\pi^{\mu\nu}\left(\mathbf{x},t\right)&=\beta\left(\mathbf{x}, t\right)\sigma_{\rho\sigma}\left(\mathbf{x}, t\right)\int d^{3} \mathbf{x}_{1} \int_{-\infty}^{t} d t_{1} e^{\varepsilon\left(t_{1}-t\right)}\left(\hat{\mathcal{T}}^{\mu\nu}(\mathbf{x}, t), \mathcal{\hat{T}}^{\rho\sigma}\left(\mathbf{x}_{1}, t_{1}\right)\right)~+\mathcal{O}(\nabla^{2}),\label{otherkubo1}\\
\Pi\left(\mathbf{x},t\right)&=-\beta\left(\mathbf{x},t\right)\theta\left(\mathbf{x},t\right)\int d^{3} \mathbf{x}_{1} \int_{-\infty}^{t} d t_{1} e^{\varepsilon\left(t_{1}-t\right)}\left(\hat{P}^{*}(\mathbf{x}, t), \hat{P}^{*}\left(\mathbf{x}_{1}, t_{1}\right)\right)~+\mathcal{O}(\nabla^{2}).\label{otherkubo2}
\end{align}
From this equation, one can extract the coefficient of shear and bulk viscosity by comparing with the Navier-Stokes equation, which is
\begin{align}
\pi^{\mu\nu}\left(\mathbf{x},t\right)&=2\eta\,\sigma^{\mu\nu}\left(\mathbf{x},t\right)\\
    \Pi\left(\mathbf{x},t\right)&=-\zeta\,\theta\left(\mathbf{x},t\right)
\end{align}
In the isotropic medium, one can expect to calculate the shear and bulk viscosity coefficient from KMB correlation function Eq.~\eqref{eq:Kubo_corr_nota},
\begin{align}
    \eta&=\frac{\beta}{10}\int d^{3} \mathbf{x}_{1} \int_{-\infty}^{t} d t_{1} e^{\varepsilon\left(t_{1}-t\right)}\left(\hat{\mathcal{T}}^{\mu\nu}(\mathbf{x}, t), \mathcal{\hat{T}}_{\mu\nu}\left(\mathbf{x}_{1}, t_{1}\right)\right)\,,\label{kuboeta}\\
    \zeta&=\beta\int d^{3} \mathbf{x}_{1} \int_{-\infty}^{t} d t_{1} e^{\varepsilon\left(t_{1}-t\right)}\left(\hat{P}^{*}(\mathbf{x}, t), \hat{P}^{*}\left(\mathbf{x}_{1}, t_{1}\right)\right)\,.\label{kubozeta}
\end{align}
Employing Eqs.~\eqref{GRAB} and \eqref{ImGs}, one can represent these two KMB correlators in terms of the corresponding spectral functions (See Appendix-[\ref{A2}]) which can be evaluated in global thermal equilibrium. 


\section{Cross-diffusion in a toy model: scalar field theory}\label{sec:transport}

Having defined the matrix of diffusion coefficient $\kappa_{ab}$ in Eq.~\eqref{dspmjj} we shall next estimate this perturbatively from field theory using the spectral function $\rho_{\hat{\mathcal{J}}_{a}\hat{\mathcal{J}}_{b}}$ as given in Eq.~\eqref{defSAJJ}. We shall explicitly outline this derivation for scalar field theory with quartic interaction for simplicity. We note here that the transport coefficients, such as the viscosity coefficients, have been studied earlier for single component scalar field~\cite{Czajka:2017bod}. On the other hand, in the present work, we discuss the cross-diffusion matrix with multiple conserved charges for multicomponent systems. The dissipative coefficients for diffusion have not been investigated so far. We shall consider a system of two complex interacting scalar fields with a Lagrangian given as
\begin{align}
\mathcal{L}(x)
=\partial\phi^{\dagger}\partial\phi+\partial\xi^{\dagger}\partial\xi-V(\phi,\xi),\label{chargedlagrangiaan}
\end{align}
where, the potential part is given by
\begin{equation}
 V(\phi,\xi) = m^{2}_{\phi}\, \phi^{\dagger}\phi + m^{2}_{\xi}\,\xi^{\dagger}\xi+\frac{g}{2}\phi^{\dagger}\phi \, \xi^{\dagger}\xi+\frac{\lambda_{1}(\phi^{\dagger}\phi)^{2}}{4}+\frac{\lambda_{2}(\xi^{\dagger}\xi)^{2}}{4}. \label{intvertex}
\end{equation}
Here, $m_{\phi}$ and $m_{\xi}$ are the masses of the scalar fields $\phi$ and $\xi$, respectively, while $\lambda_{1}$ and $\lambda_{2}$ are the coupling for the self-interaction of these two fields, $g$ is the mutual interaction coupling. In the following, we shall evaluate the spectral function at finite temperature $T$ and chemical potentials $\mu_{a}$ using the imaginary time formalism as outlined in Refs.~\cite{Laine:2016hma, Jeon:1992kk}. Henceforth, we denote Minkowskian space-time coordinates by $\Ch=(t,\textbf{x})$ and momenta by $\Ka=(k^{0},\km)$, whereas their Euclidean counterparts are denoted by $X=(\tau,\textbf{x})$ and $K=(k_{n},\km)$. Scalar products are defined as $\Ka\cdot\Ch=k^{0}t-\km\cdot\xv$ and $K\cdot X=k_{n}\tau-\km\cdot\xv$ in the Minkowski and Euclidean manifold, respectively. Moreover the integral measures are $\int_{\Ch}=\int dt \int d^{3}\xv$ and $\int_{X}=\int_{0}^{\beta} d\tau\int d^{3}\xv$.

We shall assume the Lagrangian is invariant under two global U(1) symmetry having corresponding chemical potential $\mu_{a}$ with $(a=\phi,\xi)$. We also take the chemical potentials for conserved charges as $\mu_{A}$ with $(A=1,2)$. These two pairs of chemical potentials are related as $\mu_{a}= q_{aA}\, \mu_{A}$, i.e.,
\begin{align}
\begin{pmatrix}
\mu_{\phi}\\
\mu_{\xi}
\end{pmatrix}&=\begin{pmatrix}
q_{\phi 1}&q_{\phi 2} \\
q_{\xi 1}&q_{\xi 2}
\end{pmatrix}\begin{pmatrix}
\mu_{1} \\
\mu_{2}
\end{pmatrix}.\label{chargeMat}
\end{align}
Using the condition\,\footnote{Under such transformation, this condition ensures the invariance of energy due to the presence of finite chemical potentials in either particle or charge basis. One may infer that free energy for the Grand canonical partition function is invariant under such transformation, i.e., $\mathcal{Z}(T,\mu_{\phi},\mu_{\xi})= C\int \mathcal{D}\phi \mathcal{D}\xi \,e^{-S_{E}}=\mathcal{Z}(T,\mu_{1},\mu_{2})$. }
\begin{align}
    \mu_{\phi}n_{\phi}+\mu_{\xi}n_{\xi}=\mu_{1}n_{1}+\mu_{2}n_{2},
\end{align}
the number densities of the fields can be related to the charge densities as 
\begin{align}
    n_{a}=q^{-1}_{aA}n_{A}.
\end{align}
We shall calculate here the spectral function in the species basis and relate it to the diffusion matrix $\kappa_{AB}$ by employing the Eq.~\eqref{DAB}. Consistent with the Kubo-Martin-Schwinger (KMS) condition with finite chemical potential, the field operators can be Fourier expanded in the Euclidean space as
\begin{align}
    \phi_{a}(X)=\sumintb_{P}e^{i p_{n}\tau-i\pmo\cdot\xv}\Iphi_{a}(P),\label{FdPh}
\end{align}
where, we have used the notation 
\begin{equation}
    \sumintb_{P}=T\sum_{p_{n}}\int\frac{d^{3}\pmo}{(2\pi)^{3}}=T\sum_{p_{n}}\int d\pmo\, ,
    \label{eq:sumint}
\end{equation}
and $p_{n}$ are the Matsubara bosonic frequency $p_{n}=2n\pi T$. 

The two-point Euclidean correlation function can be written as
\begin{align}
    \mathcal{G}^{E}_{a,b}(X)&=\left\langle\phi_{a}(X)\, \phi_{b}^{\dagger}(0)\right\rangle_{E} = \sumtintb_{PQ}e^{i\tilde{p}_{n}\tau+ip_{i}\xv^{i}}\left\langle\Iphi_{a}(P)\, \Iphi_{b}^{\dagger}(Q)\right\rangle_{E},
\end{align} 
where, $\tilde{p}_{n}=p_{n}+i\mu_{a}$.\footnote{Additional term $e^{-\mu\tau}$ in the Fourier transformation with respect to $\tau$ helps to satisfy the KMS relation $\langle\phi_a(t-i\beta,\xv)\phi^{\dagger}(0,\textbf{0})\rangle=e^{-\mu\beta}\langle\phi_b^{\dagger}(0,\textbf{0})\phi_a(t,\xv)\rangle$ at $t=0$ for commuting bosonic fields.} The free propagator can be expressed as
\begin{equation}
 \langle\Iphi_{a}(P)\Iphi_{b}^{\dagger}(Q)\rangle^{\text{f}}=\delta_{ab}\frac{\deltacut(P-Q)}{(p_{n}+i\mu_{a})^{2}+\textbf{p}^{2}+m_{a}^{2}} \, ,
\end{equation}
where, `f' in the superscript denotes the free propagator and $\deltacut(P-Q)$ is defined as
\begin{align}
  \deltacut(P-Q)= \!\int_{X}\! e^{i(P-Q)\cdot X} =\!\int_{0}^{\beta}\! d\tau\,e^{i(p_{n}-q_{n})\tau} \!\int\! d^{3}\xv\, e^{-i(\pmo-\qm)\cdot\xv} &=\beta\, (2\pi)^{3}\delta^{3}(\pmo-\qm)\,\delta_{p_{n},q_{n}} \, .\label{delc} 
\end{align} 
The two-point Euclidean correlation function can be expressed as
\begin{align}
\mathcal{G}^{E\,\text{f}}_{a,b}(X)&=\delta_{ab}\sumintb_{P}\frac{e^{i\tilde{p}_{n}\tau+ip_{i}\xv^{i}}}{(p_{n}+i\mu_{a})^{2}+\vec{p}^{2}+m_{a}^{2}}\, ,\label{FGE}
\end{align}
and the two-point function in Euclidean momentum space can be written as,
\begin{align}
    \mathcal{G}^{E\,\text{f}}_{a,b}(K)=\int_{0}^{\beta}d\tau\int_{\xv} e^{(i k_{n}+\mu_{a})\tau-i\km\cdot\xv}\, \mathcal{G}^{E\,\text{f}}_{a,b}(X) . \label{FTGE}
\end{align}
The Euclidean correlator is time-ordered by definition $(0\leq\tau\leq\beta)$.

The two-point function $\mathcal{G}^{E}_{a,b}(K)$ can also be represented in terms of the single free particle spectral function\,\footnote{Fourier transformation of the commutator of $\phi(\Ch)$ and $\phi^{\dagger}(0)$ is defined as spectral function: $2\rho(\kappa)=\int_{\Ch}\,e^{i\kappa\cdot\Ch}\Big[\phi(\Ch),\phi^{\dagger}(0)\Big]$.}, i.e.,
\begin{align}
    \mathcal{G}^{E\,\text{f}}_{a,b}(K)=\delta_{ab}\int^{\infty}_{-\infty}\frac{dk^{0}}{\pi}\frac{\rho^{\text{f}}_{a}(k^{0},\km)}{k^{0}-i[k_{n}-i\mu_{a}]} , \label{GEtoSpdf}
\end{align}
where free particle spectral function has the form \cite{Laine:2016hma,Jeon:1994if,Kapusta:2023eix}
\begin{align}
    \rho_{a}^{\text{f}}(k_{0},\km)&=\frac{k_{0}}{|k_{0}|}\pi\delta(k_{0}^{2}-E_{a\,\km}^{2})=\frac{\pi}{2 E_{a\,\km}}\Big[\delta(k_{0}-E_{a\,\km})-\delta(k_{0}+E_{a\,\km})\Big],\label{deltaanti}
\end{align}
with $E_{a\,\km}=\sqrt{\km^{2}+m_{a}^{2}}$. The expression of the Euclidean free Green'sfunction in Eq.~\eqref{GEtoSpdf} can be generalised in the presence of the interaction with the single particle free spectral function (FSF) getting replaced by single particle resummed spectral function (RSF) given as,
\begin{align}
    \mathcal{G}^{E}_{a,b}(K)=\delta_{ab}\int^{\infty}_{-\infty}\frac{dk^{0}}{\pi}\frac{\rho_{a}(k^{0},\km)}{k^{0}-i[k_{n}-i\mu_{a}]}\label{GEtoSpd}.
\end{align}
Here, $\rho_{a}(k^{0},\km)$ is the single particle RSF which contains the information of interaction through the self-energy contribution. 
From this definition, one can invert the relation to obtain \cite{Laine:2016hma,Jeon:1992kk,Kapusta:2023eix}
\begin{align}
    \rho_{a}(\Ka)&=\frac{1}{2i}\text{Disc}\,\mathcal{G}^{E}_{a,a}\left[i(k_{n}-i\mu)\rightarrow k^{0}+i 0^{+},\km\right]\\
    &\equiv\frac{1}{2i}\Bigg[\mathcal{G}^{E}_{a,a}(i(k_{n}-i\mu)\rightarrow k^{0}+i 0^{+},\km)-\mathcal{G}^{E}_{a,a}(i(k_{n}-i\mu)\rightarrow k^{0}-i 0^{+},\km)\Bigg] , \label{SpdtoGE}
\end{align}
where, $\text{Disc}$ denoted discontinuity of the Green's function.

The relation between the spectral function and the two-point correlation function for single fields $\phi_{a}$ can be generalized for composite Hermitian bosonic fields. If $\mathcal{O}_{1}$,$\mathcal{O}_{2}$ are two composite bosonic fields which commute with the charge operator, the two-point correlation functions can be written as (See Eq.~\eqref{gfao2})
\begin{align}
    \mathcal{G}^{E}_{\hat{\Oc}_1\hat{\Oc}_2}(K)&=\int_{0}^{\beta}d\tau\int_{\xv} e^{i k_{n}\tau-i\km\cdot\xv}\mathcal{G}^{E}_{\hat{\Oc}_1\hat{\Oc}_2}(X) =\int^{\infty}_{-\infty}\frac{dk^{0}}{\pi}\frac{\rho_{\hat{\Oc}_1\hat{\Oc}_2}(k^{0},\km)}{k^{0}-i k_{n}} , \label{gfao}
\end{align}
where, $\mathcal{G}^{E}_{\hat{\Oc}_1\hat{\Oc}_2}(X)=\langle\Oc_{1}(X)\Oc^{\dagger}_{2}(0)\rangle_{E}$ is the Euclidean coordinate correlation function for the set of composite hermitian operators $\hat{\mathcal{O}}_{1}$, $\hat{\mathcal{O}}_{2}$. The difference between the Fourier representations in Eqs.~\eqref{GEtoSpd} and \eqref{gfao} may be noted. The reason for this difference lies in the fact that the single fields do not commute with the total charge operator. This is shown explicitly in Appendix-\ref{AppenA}. Further, similar to Eq.~\eqref{SpdtoGE}, the above equation can be inverted to obtain the relation for the composite operators as
\begin{align}
    \rho_{\hat{\Oc}_1\hat{\Oc}_2}(\Ka)&=\frac{1}{2i}\text{Disc}\,\mathcal{G}^{E}_{\hat{\Oc}_1\hat{\Oc}_2}(ik_{n}\rightarrow k^{0}+i 0^{+},\km).\label{spO}
\end{align}
We shall next use this expression to calculate the diffusion coefficient $\kappa_{ab}$ as given in Eq.~\eqref{dspmjj}. The diffusion current-current correlation can be computed by plugging this set of two operators $\hat{\mathcal{O}}_{1}=\hat{\mathcal{J}}_{a}^{\,\mu},\, \hat{\mathcal{O}}_{2}=\hat{\mathcal{J}}_{b\,\mu}$, into Eq.~\eqref{spO}.

Now, we can compute the thermodynamic dissipative coefficients with the aid of the Kubo formulas given in Sec-(\ref{sec:neso}). As a toy model, we can take a set of two fields, namely $\phi$ and $\xi$ with conserved currents under $U_{\phi}(1)\times U_{\xi}(1)$ as
\begin{align}
   \hat{J}^{\mu}_{a}&=-i\Big(\del^{\mu}\phd_{a}\phi_{a}-\phd_{a}\del^{\mu}\phi_{a}\Big)\,,
\end{align}
while the conserved charge currents are
\begin{align}
    \hat{J}^{\mu}_{A}=\sum_{a}q_{aA} \hat{J}^{\mu}_{a},
\end{align}
and the energy-momentum tensor is
\begin{align}
 \hat{T}^{\mu\nu}=\sum_{a}\del^{\mu}\phd_{a}\del^{\nu}\phi_{a}-g^{\mu\nu}\mathcal{L}(x).
\end{align}
Using Eq.~\eqref{decomposition}, one can write down the dissipative heat current and the dissipative number current operators $\hat{\mathcal{Q}}_{\mu}$ and $\hat{\mathcal{N}}_{\mu}$ as 
\begin{align}
    \hat{\mathcal{Q}}_{\mu}&=u_{\alpha}\Delta_{\mu\beta}\hat{T}^{\alpha\,\beta}\label{PQ} , \\
    \hat{\mathcal{N}}_{a\,\mu}&=\Delta_{\mu\alpha}\hat{J}_{a}^{\alpha}, \label{PN}
\end{align}
where the presence of projector $\Delta^{\mu\nu}=\eta^{\mu\nu}-u^{\mu}u^{\nu}$ makes these operators orthogonal to the fluid velocity $u^{\mu}$.

It is convenient to calculate the correlation function of interest, like Eq.~\eqref{dimjj} which can be evaluated using the thermal equilibrium Green's function within the imaginary-time Matsubara technique. In order to do so, one needs to convert corresponding operators into their Euclidean counterparts via Wick rotation. In the fluid rest frame $u_{\mu}=(1,0,0,0)$, $\Delta^{\mu\nu}=\text{diag}(0,-1,-1,-1)$, performing a Wick rotation $t\rightarrow-i\tau$, the current operators become
\begin{align}
\hat{\mathcal{Q}}^{E}_{i}=\hat{T}_{\tau i}&=i\sum_{a}\Big(\del_{\tau}\phd_{a}\del_{i}\phi_{a}+\del_{i}\phd_{a}\del_{\tau}\phi_{a}\Big), \label{Hph}\\
\hat{\mathcal{N}}_{a\,i}^{E}&=-i\Big(\del_{i}\phd_{a}\phi_{a}-\phd_{a}\del_{i}\phi_{a}\Big).\label{Nph}
\end{align}
Now, to calculate the required retarded Green's function in Eq.~\eqref{dimjj}, one can perform an analytic continuation process $i k_{n}\rightarrow k_{0}+i\epsilon$ to the thermal Green's function of desired operators \,$\hat{\mathcal{O}}_{1}=\hat{\mathcal{J}}_{a}^{i},\hat{\mathcal{O}}_{2}=\hat{\mathcal{J}}_{b}^{j}$ as given in Eq.~\eqref{gfao}. Therefore, one can write the Euclidean correlation function $\mathcal{G}^{E}_{\hat{\mathcal{J}}_{a}\hat{\mathcal{J}}_{b}}(X)$ for diffusion currents as
\begin{align}
    \mathcal{G}^{E}_{\hat{\mathcal{J}}_{a}\hat{\mathcal{J}}_{b}}(X)&=\Delta^{ij}\bigg\langle\hat{\mathcal{J}}_{a\,i}(X)\hat{\mathcal{J}}_{b\,j}(0)\bigg\rangle_{l}\nonumber\\
    &=-\Bigg[\bigg\langle\hat{\mathcal{N}}_{a\,i}(X)\hat{\mathcal{N}}_{b\,i}(0)\bigg\rangle_{l}-\frac{n_{b}}{h}\bigg\langle\hat{\mathcal{N}}_{a\,i}(X)\hat{\mathcal{Q}}_{i}(0)\bigg\rangle_{l}\nonumber\\
    &\quad-\frac{n_{a}}{h}\bigg\langle\hat{\mathcal{Q}}_{i}(X)\hat{\mathcal{N}}_{b\,j}(0)\bigg\rangle_{l}+\frac{n_{a}n_{b}}{h^{2}}\bigg\langle\hat{\mathcal{Q}}_{i}(X)\hat{\mathcal{Q}}_{j}(0)\bigg\rangle_{l}\Bigg],\label{Jabdef}
\end{align}
which will be used to calculate the different contributions to the spectral function given in Eq.~\eqref{gfao} and eventually for estimating the diffusion coefficient given in Eq.~\eqref{dspmjj}. 

A few comments here are in order. Firstly, it is important to note that while the expression for the dissipative coefficients is estimated using linear response, i.e., restricted to linear order in the gradient for the off-equilibrium correction, the Green's function that appears in Eq.~\eqref{Jabdef} is, in general, non-perturbative and not restricted to weak coupling only. We will try to estimate this correlator for the present Lagrangian. The detailed evaluation of these correlators using thermal field theory is given in Appendix-\ref{AppenB}. The Fourier-transformed correlation function $\mathcal{G}^{E}_{\hat{\mathcal{J}}_{a}\hat{\mathcal{J}}_{b}}(L)$ of Eq.~\eqref{Jabdef} is given by
\begin{align}
     \mathcal{G}^{E}_{\hat{\mathcal{J}}_{a}\hat{\mathcal{J}}_{b}}(L)=\delta_{ab}\mathcal{G}^{E}_{\hat{\mathcal{N}}_{a}\hat{\mathcal{N}}_{a}}(L)-\Big(\frac{n_{a}}{h}\mathcal{G}^{E}_{\hat{\mathcal{N}}_{b}\hat{\mathcal{Q}}}(L)+\frac{n_{b}}{h}\mathcal{G}^{E}_{\hat{\mathcal{N}}_{a}\hat{\mathcal{Q}}}(L)\Big)+\frac{n_{a}n_{b}}{h^{2}}\mathcal{G}^{E}_{\hat{\mathcal{Q}}\hat{\mathcal{Q}}}(L).\label{GJJEL}
\end{align}
The different components of the correlation function on the right hand side are given in Eqs.~\eqref{GENN}, \eqref{GENQ}, \eqref{GEQN} and \eqref{GEQQ} respectively in Appendix-[\ref{AppenB}].
Let us note that the retarded Green's function are related to the Euclidean Green's function by analytic continuation, as shown in Appendix-[\ref{AppenA}]. The former Green's functions are used to define the transport coefficients as in Eq.~\eqref{dimjj}, which are evaluated using the corresponding spectral function Eq.~\eqref{dspmjj}.

Further, one needs to take appropriate limits of vanishing of three momentum and of frequency for the spectral function $\rho_{\hat{\mathcal{J}}_{a}\hat{\mathcal{J}}_{b}}(\omega,\lm)$. The spectral function in the limit of momentum $\lm\rightarrow0$ in Eq.~\eqref{defSAJJ} leads to
\begin{align}
    \rho_{\hat{\mathcal{J}}_{a}\hat{\mathcal{J}}_{b}}(\omega,\textbf{0})&=\delta_{ab}\rho_{\hat{\mathcal{N}}_{a}\hat{\mathcal{N}}_{a}}(\omega,\textbf{0})-\frac{n_{b}}{h}\rho_{\hat{\mathcal{N}}_{a}\hat{\mathcal{Q}}}(\omega,\textbf{0})-\frac{n_{a}}{h}\rho_{\hat{\mathcal{N}}_{b}\hat{\mathcal{Q}}}(\omega,\textbf{0})+\frac{n_{a}n_{b}}{h^{2}}\rho_{\hat{\mathcal{Q}}\hat{\mathcal{Q}}}(\omega,\textbf{0})\,\nonumber\\
    &=\int\! d\pmo \!\int\!\frac{d\omega^{\prime}}{\pi}\sum_{c} \pmo^{2}\bigg[ 4\delta_{ab}\delta_{ca} \!-\frac{2}{h}\big(n_{a}\delta_{cb}\!+n_{b}\delta_{ca}\big)(2\omega^{\prime}+\omega)+\frac{n_{a}n_{b}}{h^{2}}(2\omega^{\prime}+\omega)^{2} \bigg] \nonumber\\
    &\quad\quad\quad\quad\quad\quad\quad\quad\quad \times \rho_{k}(\omega^{\prime},\pmo)\, \rho_{c}(\omega^{\prime}+\omega,\pmo)\Big[f_{c}(\omega^{\prime}+\omega)-f_{c}(\omega^{\prime})\Big] \,.\label{rhoJaJb}
\end{align}
Finally, taking the limit $\omega\rightarrow0$ in Eq.~\eqref{dspmjj} leads to the diffusion coefficients as\,\footnote{ To have a non-zero and finite result while taking $\lim_{\omega\rightarrow0}\Big(\frac{\rho_{\mathcal{J}_{a}\mathcal{J}_{b}}(\omega,\textbf{0})}{\omega}\Big)$, the numerator should have a term proportion to $\omega$. Such term in this case is $\lim_{\omega\rightarrow0}\Big(\frac{f_{k}(\omega^{\prime}+\omega)-f_{k}(\omega^{\prime})}{\omega}\Big)=-\beta f_{k}(\omega^{\prime})\big(1+f_{k}(\omega^{\prime})\big)$. } 
\begin{align}
    \kappa_{ab} = \frac{4}{3}\int\,d\pmo\int\frac{d\omega^{\prime}}{\pi}\sum_{c}\, \pmo^{2} \bigg[&\,\delta_{ab}\delta_{ca}-\frac{\omega^{\prime}}{h}\big(n_{a}\delta_{cb}+n_{b}\delta_{ca}\big)+n_{a}n_{b}\Big(\frac{\omega^{\prime}}{h}\Big)^{2} \bigg]\rho_{c}(\omega^{\prime},\pmo)^{2}\nonumber\\
    &\times f_{c}(\omega^{\prime})\Big[1+f_{c}(\omega^{\prime})\Big] \,.\label{kappaaab}
\end{align}
Hence, the evaluation of the diffusion coefficients reduces to the determination of the one-particle spectral function of the fields. Upto this point, the expressions are quite general without any approximatyions regarding coupling strength. Thus the  method is extendable beyond the weak coupling limit, with our results as above expressed in terms of spectral functions. In the non-perturbative regime, the non-triviality of evaluating spectral functions and transport coefficients can be addressed using e.g. large N techniques, which we plan to explore in future works.

However, this changes if the spectral function has a finite width.
Here, we shall consider the quasi-particle spectral function with a finite width to calculate these diffusion coefficients. The re-summed single-particle spectral function can be calculated from the analytically continued Euclidean propagator's discontinuity as described in Eq.~\eqref{SpdtoGE}. In an interacting theory, transport coefficients are finite due to the finite mean free path. This leads to a nonzero broadening of the spectral function around the peak of the quasi-particle excitation. The width of such broadening is inversely proportional to the lifetime of the excitation. In a weakly interacting thermal medium, the spectral function can be approximated by Lorentzian form given as \cite{Jeon:1994if}
\begin{equation}
    \rho_{a}(k_{0},\km)=\frac{\text{Im}(\Sigma_{a}(k_{0},\km))}{k_{0}^{2}-\km^{2}-M_{a}^{2}+|\text{Im}(\Sigma_{a}(k_{0},\km))|^{2}}\label{spdp}
\end{equation}
The origin of such a form of spectral function can be cast from the re-summed propagator. These resummed propagators carry information of the quasi-particle (of species a) interaction with the medium theough the self-energy function $\Sigma_{a}=\text{Re}(\Sigma_{a}(k_{0},\km))-i\text{Im}(\Sigma_{a}(k_{0},\km))$. One can show that the real part of the self-energy has the leading order contribution from the tadpole diagram which is momentum independent (See Appendix-[\ref{SETad}]). This spectral function in the weak coupling limit has a peak at $k_{0}^{2}(\km)=\km^{2}+m_{a}^{2}+\text{Re}(\Sigma_{a}(k_{0},\km))\equiv\km^{2}+M_{a}^{2}=E_{a\,\km}^{2}$. The width of the spectral function is connected to the imaginary part of the self energy by $\Gamma_{a\,\km}=\frac{\text{Im}(\Sigma_{a}(k_{0},\km))}{2 E_{a\,\km}}$. 

For sufficiently small coupling, the frequency dependence of the self-energy can be neglected over the width of the spectral function. Therefore, the single particle spectral function can be approximated with $|E_{a\,\km}|\sim\omega$, and we have
\begin{align}
    \rho_{a}(\omega,\pmo)&\approx\frac{2\omega\Gamma_{a\,p}}{[\omega^{2}-E_{p}^{2}]^{2}+4(\omega\Gamma_{a\,p})^{2}}.
\end{align}
Further in the small width approximation, $\rho_{a}(\omega,\pmo)$ can be written by a difference of two Lorentzians \cite{Czajka:2017bod, Jeon:1992kk},
\begin{align}
    \rho_{a}(\omega,\pmo)&\approx\frac{1}{2i}\Bigg[\frac{1}{\Big(\omega-i\Gamma_{\pmo a}\Big)^{2}-E_{\pmo a}^{2}}-\frac{1}{\Big(\omega+i\Gamma_{\pmo a}\Big)^{2}-E_{\pmo a}^{2}}\Bigg]\Bigg[1+\mathcal{O}\Bigg(\frac{\Gamma_{a\,p}}{E_{a\,p}}\Bigg)\Bigg]\nonumber\\
    &\equiv\frac{1}{2i}\Bigg[\Delta_{a\,1}(\omega,\pmo)-\Delta_{a\,2}(\omega,\pmo)\Bigg]. \label{rspdtoD}
\end{align}
Here, $\Gamma_{\pmo a}$ is the thermal width, which is the consequence of the scattering with constituents of the thermal bath. The validation of ignoring the $\mathcal{O}\Big(\frac{\Gamma_{a\,p}}{E_{a\,p}}\Big)$ can be justified at weak coupling limit as to the lowest order, the imaginary part of self-energy $\Sigma_{Ia}$ arises at the quadratic power of coupling constants. We explicitly compute the self-energy for the Lagrangian \eqref{chargedlagrangiaan} in the (See Appendix-[\ref{SETad}]).
In passing, we note that the propagators  $\Delta_{a\,1}(\omega,\pmo)$,$\Delta_{a\,2}(\omega,\pmo)$ correspond to the resummed retarded and advanced propagators of Ref~\cite{Czajka:2017bod}.

With the form of the single-particle spectral function Eq.~\eqref{rspdtoD}, one can extract the diffusion coefficients using Eq.~\eqref{kappaaab}. A few comments in this regard are in order. Let us note that in the expression for the diffusion coefficient, Eq.~\eqref{kappaaab}, the spectral function $\rho_{a}(k_{0},\km)$ occurs in quadratic order of the integrand. This will involve squared terms like $\Delta_{a\,1}^{2}(\omega,\pmo),~\Delta_{a\,2}^{2}(\omega,\pmo)$ and cross-term $\Delta_{a\,1}(\omega,\pmo)\Delta_{a\,2}(\omega,\pmo)$. In each of the squared terms, $\Delta_{a\,1}^{2}(\omega,\pmo),~\Delta_{a\,2}^{2}(\omega,\pmo)$, the integrand in Eq.~\eqref{kappaaab} will have second-order pole on the same side of the real axis, i.e., $\pm E_{a\pmo}+ i\Gamma_{a\pmo},~\pm E_{a\pmo}- i\Gamma_{a\pmo}$, respectively. Using the residue theorem for integration over frequency in Eq.~\eqref{kappaaab}, one can show that the squared terms lead to contributions which are proportional to a polynomial of order $\mathcal{O}\Big(\Gamma_{a\,p}/E_{a\,p}\Big)^{k}$ with $k\geq1$. On the other hand, the dominant contribution of the mixed term $\Delta_{a\,1}(\omega,\pmo)\Delta_{a\,2}(\omega,\pmo)$ lead to a contribution which is of the order $\mathcal{O}\Big(1/\Gamma_{a\,p}E_{a\,p}\Big)$ and other contribution is suppressed by $\mathcal{O}\Big(\Gamma_{a\,p}/E_{a\,p}\Big)^{k}$ with $k\geq0$. Therefore, the cross-terms give a dominant contribution in the small width approximation, which corresponds to the \emph{pinching pole} approximation of Ref.~\cite{Czajka:2017bod}. Hence, by plugging the spectral function of Eq.~\eqref{rspdtoD} into Eq.~\eqref{kappaaab} and performing the integration over $\omega^{\prime}$ by closing the contour in the upper half plane, we have
\begin{align}
    \kappa_{ab}&=\frac{1}{6}\int\,d\pmo\sum_{c}\, \frac{\pmo^{2}}{E_{c\,\pmo}^{2}\,\Gamma_{c\,\pmo}}\Bigg[\delta_{ab}\delta_{ca}-\theta_{c}\frac{E_{c\,\pmo}}{h}\big(n_{a}\delta_{cb}+n_{b}\delta_{ca}\big)+n_{a}n_{b}\left(\frac{\theta_{c}E_{c\,\pmo}}{h}\right)^{2} \Bigg]\nonumber\\
    &\quad\quad\quad\quad\quad\quad\quad\quad\quad\quad\quad
    \times\Big[f_{c}(E_{c\,\pmo}-\theta_{c}\mu_{c})\big(1+f_{c}(E_{c\,\pmo}-\theta_{c}\mu_{c})\big)\Big] \,.\label{Kappat}
\end{align}
Here $\theta_{k}=\pm1$ for particles and antiparticles, respectively, for each given species `$c$'. The method of this calculation within the pinching pole approximation is given in Appendix-[\ref{Pinchingpole}], explicitly using Eq.~\eqref{PinchingpoleD2}.

The different components of the diffusion matrix in the species basis are explicitly given by
\begin{align}
    \kappa_{\phi\phi}=\frac{1}{3}\int\,d\pmo\,\pmo^{2}\Bigg(&\frac{1}{2\Gamma_{\phi\,\pmo}E_{\phi\,\pmo}^{2}}\Bigg[\Big(1-\frac{n_{\phi}E_{\phi\,\pmo}}{h}\Big)^{2}f_{\phi}(E_{\phi\,\pmo})\Big(1+f_{\phi}(E_{\phi\,\pmo})\Big)+\Big(1+\frac{n_{\phi}E_{\phi\,\pmo}}{h}\Big)^{2}\nonumber\\
    &\times \Bar{f}_{\phi}(E_{\phi\,\pmo}) \Big(1+\Bar{f}_{\phi}(E_{\phi\,\pmo})\Big)\Bigg]+\frac{1}{2\Gamma_{\xi\,\pmo}E_{\xi\,\pmo}^{2}}\Big(\frac{n_{\xi}E_{\xi\,\pmo}}{h}\Big)^{2}\nonumber\\
    &\times\Bigg[f_{\xi}(E_{\xi\,\pmo})\Big(1+f_{\xi}(E_{\xi\,\pmo})\Big)+\Bar{f}_{\xi}(E_{\xi\,\pmo})\Big(1+\Bar{f}_{\xi}(E_{\xi\,\pmo})\Big)\Bigg]\Bigg) 
    \,,\label{Kphiphi}
\end{align}
\begin{align}
    \kappa_{\xi\xi}=\frac{1}{3}\int\,d\pmo\,\pmo^{2}\Bigg(&\frac{1}{2\Gamma_{\xi\,\pmo}E_{\xi\,\pmo}^{2}}\Bigg[\Big(1-\frac{n_{\xi}E_{\xi\,\pmo}}{h}\Big)^{2}f_{\xi}(E_{\xi\,\pmo})\Big(1+f_{\xi}(E_{\xi\,\pmo})\Big)+\Big(1+\frac{n_{\xi}E_{\xi\,\pmo}}{h}\Big)^{2} \nonumber\\
    &\times \Bar{f}_{\xi}(E_{\xi\,\pmo}) \Big(1+\Bar{f}_{\xi}(E_{\xi\,\pmo})\Big)\Bigg]+\frac{1}{2\Gamma_{\phi\,\pmo}E_{\phi\,\pmo}^{2}}\Big(\frac{n_{\phi}E_{\phi\,\pmo}}{h}\Big)^{2} \nonumber\\
    &\times\Bigg[f_{\phi}(E_{\phi\,\pmo}) \Big(1+f_{\phi}(E_{\phi\,\pmo})\Big)+\Bar{f}_{\phi}(E_{\phi\,\pmo})\Big(1+\Bar{f}_{\phi}(E_{\phi\,\pmo})\Big)\Bigg]\Bigg)\,.\label{Kphixi}
\end{align}
The symmetric off-diagonal elements are given by
\begin{align}
    \kappa_{\phi\xi}&=\kappa_{\xi\phi}=\frac{1}{3}\int\,d\pmo\,\pmo^{2}\Bigg(\frac{1}{2\Gamma_{\phi\,\pmo}E_{\phi\,\pmo}^{2}}\Bigg[\frac{n_{\xi}E_{\phi\,\pmo}}{h}\Big(\frac{n_{\phi}E_{\phi\,\pmo}}{h}-1\Big)f_{\phi}(E_{\phi\,\pmo})\Big(1+f_{\phi}(E_{\phi\,\pmo})\Big) \nonumber\\
    &+\frac{n_{\xi}E_{\phi\,\pmo}}{h}\Big(\frac{n_{\phi}E_{\phi\,\pmo}}{h}+1\Big)\Bar{f}_{\phi}(E_{\phi\,\pmo})\Big(1+\Bar{f}_{\phi}(E_{\phi\,\pmo})\Big)\Bigg]+\frac{1}{2\Gamma_{\xi\,\pmo}E_{\xi\,\pmo}^{2}}\Bigg[\frac{n_{\phi}E_{\xi\,\pmo}}{h}\Big(\frac{n_{\xi}E_{\xi\,\pmo}}{h}-1\Big)\nonumber\\
    &f_{\xi}(E_{\xi\,\pmo})\Big(1+f_{\xi}(E_{\xi\,\pmo})\Big)+\frac{n_{\phi}E_{\xi\,\pmo}}{h}\Big(\frac{n_{\phi}E_{\xi\,\pmo}}{h}+1\Big)\Bar{f}_{\xi}(E_{\xi\,\pmo})\Big(1+\Bar{f}_{\xi}(E_{\xi\,\pmo})\Big)\Bigg]\Bigg).\label{Kxixi}
\end{align}
In the above, $f_{a}(\omega)=(e^{\beta(\omega-\mu_{a})}-1)^{-1}$ and $\Bar{f}_{a}(\omega)=(e^{\beta(\omega+\mu_{a})}-1)^{-1}$ correspond to the particle and anti-particle distribution respectively. We might note here that we also have non-diagonal coefficients in the particle basis. Such a mixing term has its origin from the heat current $\mathcal{Q}^{i}$ contributing to the diffusion current as in Eq.~\eqref{dmj} and Eq.~\eqref{PQ}. Both species contribute to the heat flow within the single fluid frame. 

To translate the above results for elements of the diffusion matrix into the charge basis, we employ the transformation matrix as given in Eq.~\eqref{DAB} to obtain
\begin{align}
    \kappa_{AB}&=\frac{1}{3}\!\int\!d\pmo\,|\pmo|^{2}\sum_{a}\!\Bigg(\frac{1}{2\Gamma_{a\pmo}E^{2}_{a\pmo}}\Bigg[\Big(q_{aA} \!-\!\frac{n_{A}E_{a\pmo}}{h}\Big)\Big(q_{aB} \!-\! \frac{n_{B}E_{a\pmo}}{h}\Big)f_{a}(E_{a\pmo})\Big(1 \!+\! f_{a}(E_{a\pmo})\Big)\Bigg]\nonumber\\
    &\quad\quad+\frac{1}{2\Gamma_{a\pmo}E^{2}_{a\pmo}}\Bigg[\Big(q_{aA}+\frac{n_{A}E_{a\pmo}}{h}\Big)\Big(q_{aB}+\frac{n_{B}E_{a\pmo}}{h}\Big)\Bar{f}_{a}(E_{a\pmo})\Big(1+\Bar{f}_{a}(E_{a\pmo})\Big)\Bigg]\Bigg).\label{KAB}
\end{align}
It is interesting to see that in this weak coupling limit, the charge diffusion coefficients are identically the same as derived in Ref.~\cite{Das:2021bkz} using the kinetic theory approach with the identification of relaxation time $\tau_{k}=1/2\Gamma_{k}$ and including Bose enhancement factor. Such an expression for diffusion coefficients has the desirable features that $\kappa_{AB}$ is symmetric w.r.t. $A\leftrightarrow B$ consistent with the Onsangar reciprocal relation. Furthermore, $\kappa_{AA}$ is manifestly positive and definite. For the sake of completeness, we have also evaluated the thermal width $\Gamma_{k}$ to the lowest order, which is explicitly shown in Appendix-[\ref{SEsun}].

In a similar manner, the computation of the shear viscosity coefficient can be performed. Using the Eq.\eqref{kuboeta} one can show that
\begin{align}
      \eta&=\frac{\beta}{30}\!\int\!d\mathbf{p}\,|\mathbf{p}|^{4}\sum_{a}\!\Bigg(\frac{1}{2\Gamma_{a\mathbf{p}}E^{2}_{a\mathbf{p}}}\Bigg[f_{a}(E_{a\mathbf{p}})\Big(1 \!+\! f_{a}(E_{a\mathbf{p}})\Big)+\Bar{f}_{a}(E_{a\mathbf{p}})\Big(1+\Bar{f}_{a}(E_{a\mathbf{p}})\Big)\Bigg]\Bigg).\label{vis}
\end{align}
At vanishing chemical potential $ f_{a}(E_{a\textbf{p}})=\bar{f}_{a}(E_{a\textbf{p}})$, and the above result matches the same given in e.g. Ref.\cite{Czajka:2017bod}.

In addition, generalization of these transport coefficients given in Eq.\eqref{KAB} and Eq.\eqref{vis} for fermions can be done by inserting the current and energy-momentum operators for the fermionic fields in the KMB product as given in Eq.\eqref{DAB} and Eq.\eqref{kuboeta} respectively; which are
\begin{align}
    J^{\mu}_{a}&=\Bar{\psi}_{a}\gamma^{\mu}\psi_{a}\,,\\
    T_{\mu\nu}&=\frac{i}{4}\sum_{a}\Big(\Bar{\psi_{a}}\gamma_{\mu}\overleftrightarrow{\partial}_{\nu}\psi_{a}+\big(\mu\leftrightarrow\nu\big)\Big)-g_{\mu\nu}\big(\mathcal{L}_{0}+\mathcal{L}_{int}\big)\,.
\end{align}
Here, $\psi_{a}$ are the fermion field operators. In addition, $\mathcal{L}_{0}$ and $\mathcal{L}_{int}$ are correspond to the free part of Lagrangian and the interaction among fields, including self/ or mutual interactions, respectively. Here, we do not display explicit form of $\mathcal{L}_{int}$ but assume it has weak coupling so that a perturbative calculation for the self energies can be justified. We shall thus include the effect of the interactions through a thermal mass and a finite thermal width in the particle spectral functions i.e. with a quasi-particle approximations, so that the spectral function can be expressed as Ref.\cite{Pisarski:1993rf,Lang:2013lla}, 
\begin{align}
    \rho^{F}_{a} \left( p^0, \textbf{p}\right) &= \left(\slashed{\mathcal{P}} + M_{a}\right)\rho^0_{a} \left( p^0, \textbf{p}\right)\,,
\end{align}
where we note that $\rho^{F}_{a}\left( p^0, \textbf{p}\right)$ is the $4\times4$ matrix in the dirac space.

Here the form of the $\rho^{0}_{a}(\omega,\pmo)$ can be written down near the peaks $\omega\sim|E_{a\pmo}|$ in weak coupling limit,
\begin{align}
    \rho^{0}_{a}(\omega,\pmo)&\approx\frac{1}{2i}\Bigg[\frac{1}{\Big(\omega-i\Gamma_{\pmo a}\Big)^{2}-E_{\pmo a}^{2}}-\frac{1}{\Big(\omega+i\Gamma_{\pmo a}\Big)^{2}-E_{\pmo a}^{2}}\Bigg]\Bigg[1+\mathcal{O}\Bigg(\frac{\Gamma_{a\,p}}{E_{a\,p}}\Bigg)\Bigg]\nonumber\\
    &\equiv\frac{1}{2i}\Bigg[\Delta_{a\,1}(\omega,\pmo)-\Delta_{a\,2}(\omega,\pmo)\Bigg].
\end{align}
This spectral function in the weak-coupling limit has a peak at $\omega^{2}(\km)=\km^{2}+m_{a}^{2}+\Sigma^{a}_{R}\equiv\km^{2}+M_{a}^{2}=E_{a\,\km}^{2}$. The width of the spectral function is connected to the imaginary part of the self-energy by $\Gamma_{a\,\km}=\frac{\text{Im}(\Sigma_{a}(k_{0},\km))}{2 E_{a\,\km}}$. Explicit form of the self energy $\Sigma_{a}=\text{Re}(\Sigma_{a}(k_{0},\km))-i\text{Im}(\Sigma_{a}(k_{0},\km))$ depends upon the structure of $\mathcal{L}_{int}$.Similar to the bosonic case, the diffusion coefficient and shear viscosity coefficient for fermionic matter can be written as,
\begin{align}
    \kappa_{AB}&=\frac{2}{3}\!\int\!d\pmo\,|\pmo|^{2}\sum_{a}\!\Bigg(\frac{1}{2\Gamma_{a\pmo}E^{2}_{a\pmo}}\Bigg[\Big(q_{aA} \!-\!\frac{n_{A}E_{a\pmo}}{h}\Big)\Big(q_{aB} \!-\! \frac{n_{B}E_{a\pmo}}{h}\Big)f_{a}(E_{a\pmo})\Big(1 \!-\! f_{a}(E_{a\pmo})\Big)\Bigg]\nonumber\\
    &\quad\quad+\frac{1}{2\Gamma_{a\pmo}E^{2}_{a\pmo}}\Bigg[\Big(q_{aA}+\frac{n_{A}E_{a\pmo}}{h}\Big)\Big(q_{aB}+\frac{n_{B}E_{a\pmo}}{h}\Big)\Bar{f}_{a}(E_{a\pmo})\Big(1-\Bar{f}_{a}(E_{a\pmo})\Big)\Bigg]\Bigg)\,.\label{KABf}
\end{align}
and,
\begin{align}
      \eta&=\frac{\beta}{15}\!\int\!d\mathbf{p}\,|\mathbf{p}|^{4}\sum_{a}\!\Bigg(\frac{1}{2\Gamma_{a\mathbf{p}}E^{2}_{a\mathbf{p}}}\Bigg[f_{a}(E_{a\mathbf{p}})\Big(1 \!-\! f_{a}(E_{a\mathbf{p}})\Big)+\Bar{f}_{a}(E_{a\mathbf{p}})\Big(1-\Bar{f}_{a}(E_{a\mathbf{p}})\Big)\Bigg]\Bigg)\,\label{visf}
\end{align}
respectively. Here, one may notice that Pauli blocking factor can be seen through $f_{a}(1-f_{a})$ and $\bar{f}_{a}(1-\bar{f}_{a})$ factors, where $f_{a}(\omega)=\big(\text{exp}\beta(\omega-\mu_{a})+1\big)^{-1}$ and $\bar{f}_{a}(\omega)=\big(\text{exp}\beta(\omega+\mu_{a})+1\big)^{-1}$ are the fermionic distribution functions of `a'-type particles and anti-particles respectively.

\begin{figure}[t]  
    \centering
    \includegraphics[width=0.9\textwidth]{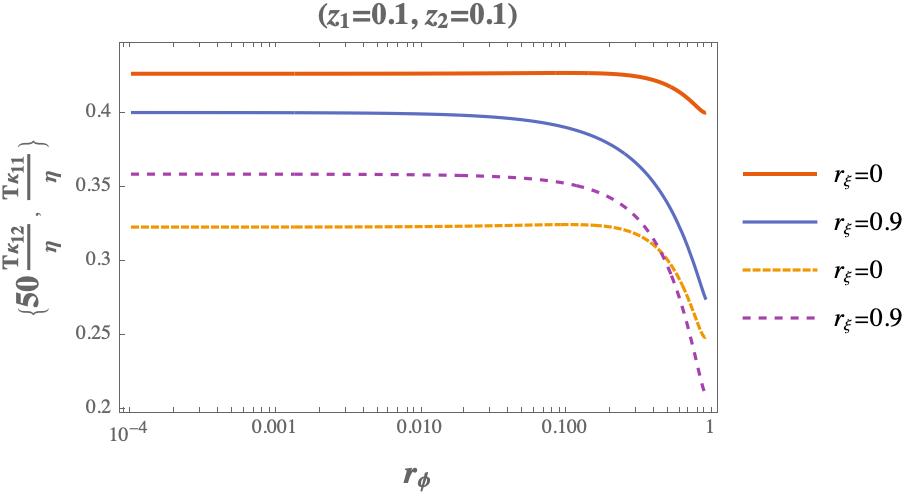}  
    \caption{Scaled diffusion coefficients $T\kappa_{11}/\eta$ (solid lines) and $T\kappa_{12}/\eta$ (dashed lines) as a function of $r_{\phi}=\frac{\mu_{\phi}}{m_{\phi}}$ for the bosonic case. We consider $z_1=z_2=0.1$ with $z\equiv m/T$ and draw curves for different $r_{\xi}$.}
    \label{fig:bosonic_cond}
\end{figure}

At this point, we can enumerate the transport coefficients as given in Eqs.\eqref{KAB}-\eqref{vis} and Eqs.\eqref{KABf}-\eqref{visf} for bosonic and fermionic systems resprctively, in terms of the computed integrals which are given in Appendix-\ref{Integrals}. Furthermore, in order to understand certain qualitative features of the transport coefficients, a simplified assumption can be made by considering the momentum independent single relaxation time scale irrespective of the species, as $\tau_{a}=\langle\frac{1}{2\Gamma_{a\mathbf{p}}}\rangle\approx\tau$. In such a case the diffusion and the shear viscosity coefficients are given as,
\\
\textbf{Boson:}
\begin{align}
    \kappa_{AB}&=\frac{\tau}{6\pi^2}\sum_{a}\Bigg(T^{3}\frac{q_{aA}q_{aB}}{z_{a}}\partial_{r_{a}}I^{-}_{4}(z_{a},r_{a})-\frac{T^{4}}{h}\Big(\frac{n_{A}q_{aB}}{z_{a}}+\frac{n_{B}q_{aA}}{z_{a}}\Big)\partial_{r_{a}}H^{-}_{5}(z_{a},r_{a})\nonumber\\
    &+\frac{T^{5}}{h^2}\frac{n_{A}n_{B}}{z_{a}}\partial_{r_{a}}G^{-}_{5}(z_{a},r_{a})\Bigg)\,,\label{KABbos}\\
    \eta&=\frac{\tau}{60\pi^2}\sum_{a}\frac{T^{4}}{z_{a}}\partial_{r_{a}}I^{-}_{6}(z_{a},r_{a})\,.\label{etaBos}
\end{align}
Where $n_{A}=\sum_{a}q_{aA}n_{a}(z_{a},r_{a})$ and $h(T,\{z_{a},r_{a}\})$ are the charge density in charge basis and enthalpy density for the bosonic system respectively, with $z_{a}=\beta m_{a}$ and $\mu_{a}=r_{a} z_{a}$. The inegrals $I^{-}_{n}(z_{a},r_{a}),G^{-}_{n}(z_{a},r_{a})$ and $H^{-}_{n}(z_{a},r_{a})$ are defined in Appendix-\ref{Integrals}. These thermodynamic densities  can be computed from these expression,
\begin{align}
    n_{A}(T,\{z_{k},\mu_{k}\})&=\frac{T^{3}}{2\pi^2}\Gamma(3)\sum_{a}q_{aA}G^{-}_{3}(z_{a},r_{a})\,,\\
    h(T,\{z_{k},\mu_{k}\})&=\frac{T^{4}}{2\pi^{2}}\sum_{a}\Big(\frac{4}{3}\Gamma(5)H^{-}_{5}(z_{a},r_{a})+z_{a}^{2}\Gamma(3)H^{-}_{3}(z_{a},r_{a})\Big)\,.
\end{align}
\textbf{Fermion:}
\begin{align}
    \kappa_{AB}&=\frac{\tau}{3\pi^2}\sum_{a}\Bigg(T^{3}\frac{q_{aA}q_{aB}}{z_{a}}\partial_{r_{a}}I^{+}_{4}(z_{a},r_{a})-\frac{T^{4}}{h}\Big(\frac{n_{A}q_{aB}}{z_{a}}+\frac{n_{B}q_{aA}}{z_{a}}\Big)\partial_{r_{a}}H^{+}_{5}(z_{a},r_{a})\nonumber\\
    &+\frac{T^{5}}{h^2}\frac{n_{A}n_{B}}{z_{a}}\partial_{r_{a}}G^{+}_{5}(z_{a},r_{a})\Bigg)\,,\label{KABfer}\\
    \eta&=\frac{\tau}{30\pi^2}\sum_{a}\frac{T^{4}}{z_{a}}\partial_{r_{a}}I^{+}_{6}(z_{a},r_{a})\,.\label{etafer}
\end{align}
Where $n_{A}(T,\{z_{k},\mu_{k}\})=q_{aA}n_{a}(z_{a},r_{a})$ and $h(T,\{z_{k},\mu_{k}\})$ are the charge density in charge basis and enthalpy density for the fermionic system respectively, where they can be expressed in terms of $I^{+}_{n}(z_{a},r_{a}),G^{+}_{n}(z_{a},r_{a})$ and $H^{+}_{n}(z_{a},r_{a})$ integrals, which are given by,
\begin{align}
    n_{A}(T,\{z_{k},\mu_{k}\})&=\frac{T^{3}}{2\pi^2}\Gamma(3)\sum_{a}q_{aA}G^{+}_{3}(z_{a},r_{a})\,,\\
    h(T,\{z_{k},\mu_{k}\})&=\frac{T^{4}}{2\pi^{2}}\sum_{a}\Big(\frac{4}{3}\Gamma(5)H^{+}_{5}(z_{a},r_{a})+z_{a}^{2}\Gamma(3)H^{+}_{3}(z_{a},r_{a})\Big)\,.
\end{align}
Next we show the results for the diffusion coefficients normalized by the shear viscosity of the matter. As may be observed in Eq.\eqref{KABbos} and Eq.\eqref{etaBos}, this ratio is independent of the relaxation time. In Fig.~\ref{fig:bosonic_cond}, we show the plots for the scaled diffusion coefficients $T\kappa_{11}/\eta$ and $T\kappa_{12}/\eta$ as a function of $r_{\phi}=\frac{\mu_{\phi}}{m_{\phi}}$ i.e. the chemical potential of $\phi$ fields normalized by its mass. Specifically, without loss of generality, we have taken the charge matrix of the Eq.\eqref{chargeMat} as,
\begin{align}
\begin{pmatrix}
\mu_{\phi}\\
\mu_{\xi}
\end{pmatrix}&=\begin{pmatrix}
\frac{1}{3}&\frac{2e}{3}\\
\frac{1}{3}&-\frac{e}{3}
\end{pmatrix}\begin{pmatrix}
\mu_{1} \\
\mu_{2}
\end{pmatrix}.
\end{align}
It may be noted that $r_{a}$ $(a=\phi,\xi)$ can vary from $0$-$1$ as beyond this range the expression for diffusion coefficients diverges due to the brunch cut in the bosonic integrals. As may be observed from Fig.~\ref{fig:bosonic_cond}, the ratio remains independent of $r_{\phi}$ till $r_{\phi}=0.1$ for all the values of $r_{\xi}$ considered here. This feature confirms the scaling behavior of the diffusion coefficients for small values of $r_{\phi}$. As $r_{\xi}$ increases the scaled diagonal diffusion coefficient decreases whereas the off-diagonal coefficient shows a non-monotonic behavior with smaller magnitudes.

It may be noted that different ratios of the transport coefficients in the strong coupling limit have been evaluated Ref.~\cite{Jain:2009bi, Jain:2010ip, Jain:2009pw}. It will be interesting to see how do the present results compare with that obtained in such a method particularly in the conformal limit. Since for the boson such a limit lead to divergence at finite chemical potentials we will consider only the massless limit for the fermions. In particular we shall consider the ratio of thermal conductivity to shear viscosity in the conformal limit. 
Another interesting quantity is the ratio of the heat conductivity to the shear viscosity. In a generic fluid frame, the dissipative correction in the charge diffusion current (as given in Eq.~\eqref{dmj}) and the heat diffusion current are related as \cite{Harutyunyan:2017lrm},
\begin{align}
    n_{A}\hat{H}^{\mu}=-\Bigg(\frac{h}{T}\Bigg)\hat{\mathcal{J}^{\mu}_{A}}\,.
\end{align}
Therefore, to find the thermal conductivity, one needs to insert the heat current operator into the KMB inner-product similar to the Eq.~\eqref{Dab} and Eq.~\eqref{DAB}, in such a way,
\begin{align}
    \kappa_{T}&=-\frac{\beta^{2}}{3}\int d^{3} \mathbf{x}_{1} \int_{-\infty}^{t} d t_{1} e^{\varepsilon\left(t_{1}-t\right)}\left(\hat{H}^{\mu}(\mathbf{x}, t), \hat{H}_{\mu}\left(\mathbf{x}_{1}, t_{1}\right)\right)\,.
\end{align}
 Which leads to the relation between the thermal conductivity $(\kappa_{T})$ to the diffusion matrix $(\kappa_{AB})$ as,
\begin{align}
   \kappa_{T}&=\Bigg(\frac{h}{T\sum_{A}\mu_{A}n_{A}}\Bigg)^2\sum_{AB}\mu_{A}\kappa_{AB}\mu_{B}\,.
\end{align}
In the conformal limit for fermions, the diffusion matrix $\kappa_{AB}$, the shear viscosity $\eta$ and the thermal conductivity $\kappa_{T}$ are given as
\begin{align}
    \kappa_{AB}&=\frac{\tau}{3\pi^2}\sum_{a}T^{3}\Bigg(q_{aA}q_{aB}\,\partial_{\lambda}L_{3}(\lambda)-\Big(\frac{T n_{A}q_{aB}}{h}+\frac{T n_{B}q_{aA}}{h}\Big)\partial_{\lambda}J_{4}(\lambda)\nonumber\\
    &\,\,+\frac{T^{2}}{h^2}n_{A}n_{B}\partial_{\lambda}L_{5}(\lambda)\Bigg)\Big|_{\lambda\rightarrow\alpha_{a}}\,,\\
    \eta&=\frac{\tau}{30\pi^2}\sum_{a}T^{4}\partial_{\lambda}L_{5}(\lambda)\Big|_{\lambda\rightarrow\alpha_{a}}\,,\\
    \kappa_{T}&=\Bigg(\frac{h}{T S}\Bigg)^2\sum_{a}T^{5}\Bigg[\alpha_{a}^{2}\,\partial_{\lambda}L_{3}(\lambda)-2\alpha_{a}\Big(\frac{ S}{h}\Big)\,\partial_{\lambda}J_{4}(\lambda)\nonumber\\
    &\,\,\quad+\Bigg(\frac{S}{h}\Bigg)^2\,\partial_{\lambda}L_{5}(\lambda)\Bigg]\Bigg|_{\lambda\rightarrow\alpha_{a}}\label{conKT}
\end{align}
Here, $S=\sum_{a}\mu_{a}n_{a}=\sum_{A}\mu_{A}n_{A}$, this sum is invariant under change of  label `a' or `A'. The functions $L_{n}(\alpha_{a})$ and $J_{n}(\alpha_{a})$ ($\alpha_{a}=\frac{\mu_{a}}{T}$) are related to the massless counterpart of the functions $G^{+}$,$H^{+}$ and $I^{+}$ of Eq.~\eqref{KABfer} and Eq.~\eqref{etafer} for fixed chemical potential $\mu_{a}$, and are defined in Appendix-\ref{Integrals}.

\begin{figure}[t]  
    \centering
    \includegraphics[width=0.9\textwidth]{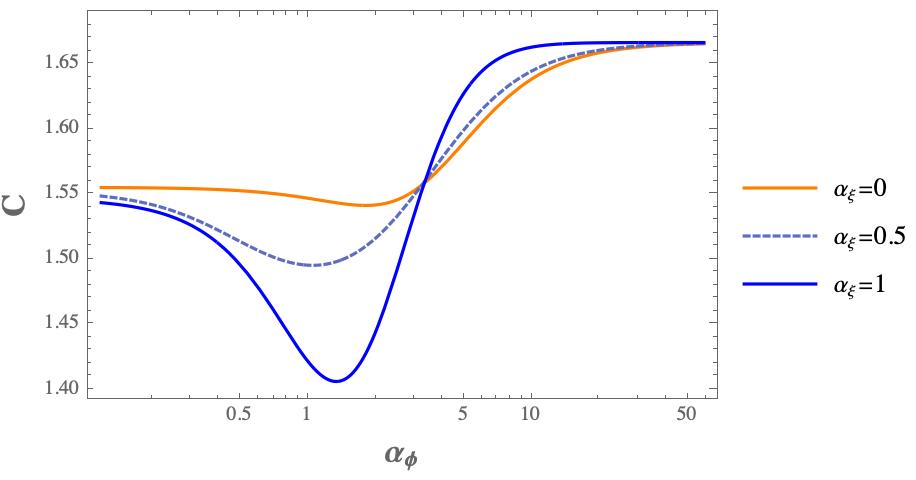}  
    \caption{Plot of the scaled thermal conductivity measure $C$, defined in Eq.~\eqref{C_def}, as a function of $\alpha_\phi\equiv\mu_\phi/T$ for the fermionic case in the conformal limit. The curves are drawn for different $\alpha_{\xi}$.}
    \label{fig:thermal cond}
\end{figure}

In the limit of both small and large values of $\alpha_{a}=\mu_{a}/T$, scaling properties can be shown from Eq.~\eqref{conKT} in two extreme limits.
\\
\textbf{Small values of chemical potentials ($\alpha_{a}\ll1$):}
\begin{align}\label{small_alpha}
    \frac{\kappa_{T}}{\eta}=\frac{7N\pi^{2}}{9}\frac{T}{\sum_{a}\mu_{a}^{2}}\Bigg(1-\frac{1}{\pi^2}\left(\frac{\sum_{a}\alpha_{a}^{4}}{\sum_{a}\alpha_{a}^{2}}-\frac{15}{7N}\sum_{a}\alpha^{2}_{a}\right)\Bigg)+\mathcal{O}\Bigg(\left(\sum_{a}\alpha_{a}^{4}/\sum_{a}\alpha_{a}^{2}\right)^2\Bigg)\,,
\end{align}
\textbf{Large values of chemical potentials ($\alpha_{a}\gg1$):}
\begin{align}\label{large_alpha}
    \frac{\kappa_{T}}{\eta}=\frac{5\pi^{2}T}{3}\frac{\sum_{a}\mu_{a}^{2}}{\sum_{a}\mu_{a}^{4}}\left(1+\frac{7}{5}\frac{N\pi^{2}}{\sum_{a}\alpha_{a}^2}-3\pi^2\frac{\sum_{a}\alpha^{2}_{a}}{\sum_{a}\alpha^{4}_{a}}\right)+\mathcal{O}\Bigg(\Big(\sum_{a}\alpha_{a}^{2}/\sum_{a}\alpha_{a}^{4}\Big)^2\Bigg)\,.
\end{align}
In each of the limiting cases, the ratio of heat conductivity to shear viscosity has the same scaling behavior with respect to $\mu_{a}\rightarrow\lambda\mu_{a}$, which is analogous to the Wiedemann-Franz law;
\begin{equation}\label{C_def}
   \frac{\kappa_{T}}{\eta}=C\pi^2\frac{T}{\sum_{a=1}^{N} \, \mu_{a}^{2}} \,,
\end{equation}
with different prefactors $\frac{7N\pi^2}{9}$ and $\frac{5\pi^2}{3}$, where $N$ is the total number of charges in our model.

\begin{figure}[t]  
    \centering
    \includegraphics[width=0.9\textwidth]{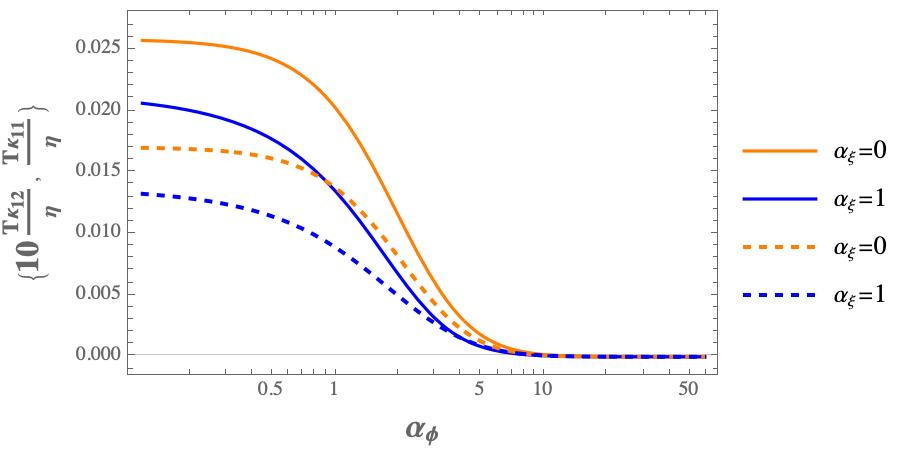} 
    \caption{Scaled diffusion coefficients $T\kappa_{11}/\eta$ (solid lines) and $T\kappa_{12}/\eta$ (dashed lines) as a function of $\alpha_\phi=\frac{\mu_{\phi}}{T}$ for the ferminoic case in the conformal limit. The curves are drawn for different $\alpha_{\xi}$.}
    \label{fig:fermionic_cond}
\end{figure}

In Fig.~\ref{fig:thermal cond}, we show the plot of the scaled thermal conductivity measure $C$, defined in Eq.~\eqref{C_def}, as a function of $\alpha_\phi\equiv\mu_\phi/T$ for the fermionic case in the conformal limit and for various values of $\alpha_\xi$. We observe a scaling behavior in the limits of both small and large $\alpha_\phi$, which is also demonstrated analytically in Eqs.~\eqref{small_alpha} and \eqref{large_alpha}. It is important to note that in the small limit $\alpha_\phi$, the value of $C$ depends on $N$, while this is not the case in the large $\alpha_\phi$ limit; see Eqs.~\eqref{small_alpha} and \eqref{large_alpha}. In Fig.~\ref{fig:fermionic_cond}, we plot the scaled diffusion coefficients $T\kappa_{11}/\eta$ (solid lines) and $T\kappa_{12}/\eta$ (dashed lines) as functions of $\alpha_\phi$ for the ferminoic case in the conformal limit, and for various values of $\alpha_\xi$. As observed in the bosonic case in Fig.~\ref{fig:bosonic_cond}, we see a similar scaling feature in the limit of small $\alpha_\phi$.

A comparison can be made with the results previously obtained in Ref.~\cite{Jain:2009pw} and Ref.~\cite{jaiswal:2015mxa} in the context of a strongly coupled and a weakly coupled fluid, respectively. Notably, for $N=1$, and at small as well as large chemical potential, our results, Eqs.~\eqref{small_alpha} and \eqref{large_alpha}, are exactly the same as obtained from kinetic theory description~\cite{jaiswal:2015mxa}. For $N=1$, the ADS/CFT result is given by~\cite{Jain:2009pw}
\begin{equation}\label{ADS/CFT}
    \frac{\kappa_{T}}{\eta}=C\pi^2\frac{T}{\mu^{2}} \,,
\end{equation}
which agrees with our results up to a constant of proportionality $C$. On the other hand, for $N>1$, the ratio $\frac{T\kappa_{T}}{\eta}$ for AdS/CFT result~\cite{Jain:2009pw}
\begin{equation} \label{ADS/CFT2}
    \frac{\kappa_{T}}{\eta}=C\pi^2\frac{T}{\sum_{a}\mu_{a}^{2}},
\end{equation}
agrees with our results for small chemical potential, again up to a constant of proportionality $C$. However, at large chemical potential, our results differs from the AdS/CFT result not only by a constant prefactor but also by the form of expression, which we obtain for the leading term in Eq.~\eqref{large_alpha}, as
\begin{equation}
    \frac{\kappa_{T}}{\eta}=\frac{5\pi^{2}T}{3}\frac{\sum_{a}\mu_{a}^{2}}{\sum_{a}\mu_{a}^{4}}.
\end{equation}
Note the difference between the form of the above equation and that given in Eq.~\eqref{ADS/CFT2}.

\section{Summary and outlook}
\label{sec:summary}
In this paper, we have presented the derivation of the elements of the diffusion matrix in terms of Kubo relations. For the same, we have used Zubarev's non-equilibrium statistical operator method to derive the Kubo formula for the diffusion coefficients for systems with multiple conserved charges, for e.g., baryon number, electric charge and strangeness in the context of heavy-ion collisions. The elements of the diffusion matrix are given by the zero-momentum spectral functions of the diffusive currents in the small frequency limit as in Eq.~\eqref{eq:jmua} and Eq.~\eqref{dimjj}. Such expressions are rather general and can be used for estimating the diffusion coefficients using non-perturbative methods such as lattice QCD or using any effective models of strong interactions such as the Nambu--Jona-Lasinio (NJL) model \cite{Buballa:2003qv}, quasi particle models~\cite{Peshier:2005pp,Berrehrah:2016vzw,Fotakis:2021diq}, or sigma models coupled to quarks \cite{Schaefer:2009ui,Rai:2023vnp}. This can also be written down in terms of the single-particle spectral functions as in Eq.~\eqref{kappaaab} along with other thermodynamic quantities like enthlapy density and number densities of the thermodynamic equilibrium systems and can be evaluated within any effective model as above.

As an explicit example, we have applied this correlator method to obtain the charge diffusion coefficients in a toy model of two  coupled charged scalar fields having two charges that are conserved. It may be mentioned here that, for scalar fields other transport coefficients like the viscosity coefficients and electrical/thermal conductivity has been derived using Kubo relations while diffusion has not been considered. We have also presented here, in some details, the calculations for the the lowest order thermal decay width that enters in the Kubo formula. Such a width that arises from the sunset diagram in the lowest order had earlier been considered \cite{Nishikawa:2003js} but with vanishing chemical potential. The expressions given here for the diffusion coefficient can be applicable to other models involving scalar fields with multiple conserved charges for hot and dense matter both in QGP as well as in nuclear astrophysics and cosmology~\cite{Bastero-Gil:2010dgy}. The present work is especially relevant in discussions concerning the physics of compressed baryonic matter at forthcoming FAIR and NICA facilities or for interpreting recent findings from the isobar run at RHIC. 

In the context of neutron stars, it will be interesting to investigate the effects of diffusion on the oscillation of neutron stars as a result of an internal instability or an external perturbation. It may be noted that whether an oscillation mode is excited or not depends upon the interplay of the excitation rate and how efficiently the dissipative mechanisms in the stellar matter counteract this excitation. Much attention has been given  on the viscous effects~\cite{Ofengeim:2019fjy,Huang:2009ue} while the effects of diffusion on neutron star oscillations has only been looked into recently for the neutron proton electron matter~\cite{Kraav:2021iit}. In this context, for neutron stars with a hyperon core~\cite{Jha:2010an} or a quark matter core~\cite{Kumar:2021hzo} the effect of diffusion of particles will be relevant and important to give insight into the structure of stellar matter. Moreover, a reaction of three distinct species of particles can be viewed as a bulk viscous fluid, which may have important consequences for neutron star physics~\cite{Gavassino:2023eoz}.

It will be interesting to apply this method for diffusion within a three flavor NJL model or different quasi-particle models to discuss the importance of the off-diagonal elements of the diffusion matrix. It will also be interesting to examine, through diagonalisation of the diffusion matrix, which combination of quark flavours will diffuse faster giving an insight to the diffusive process that will be relevant for low energy heavy-ion collisions. Further, in the context of off central heavy-ion collisions, where strong electromagnetic fields can be produced \cite{Panda:2024ccj}, the effect of strong electromagnetic field in the diffusion of different charges could also be interesting and relevant. Some of these problems will be investigated in future works.


\appendix
\section{Appendix-A}

\subsection{Retarded Green's function and spectral relations.}\label{AppenA}

Our interest here is to estimate the two-point correlation functions of elementary and composite operators to calculate the response function at linear order in the thermal force as in Eq.~\eqref{Charge kubo1} or Eqs.~\eqref{otherkubo1} and \eqref{otherkubo2}. This expression is written regarding KMB inner product as in Eq.~\eqref{eq:Kubo_corr_nota}. We shall show in the following subsection that KMB inner products are related to the retarded Green's functions and spectral function. Here we shall discuss some useful relations involving this Green's functions.

The retarded Green's function up to arbitrary bosonic operators $\{\hat{\mathcal{O}}_{a},\hat{\mathcal{O}}_{b}^{\dagger}\}$ in Minkowski space $\Ch\equiv(\xv,t)$ is defined as \cite{Laine:2016hma, Kovtun:2012rj}
\begin{align}
    G^{R}_{\hat{\mathcal{O}}_{a}\hat{\mathcal{O}}_{b}}(\Ch)=i\left\langle\theta(t)\bigg[\hat{\mathcal{O}}_{a}(\mathbf{x}, t), \hat{\mathcal{O}}^{\dagger}_{b}\left(\mathbf{0},0\right)\bigg]\,\,\right\rangle_{l} .\label{GRAB}
\end{align}
In addition, we define Green's function called the spectral function, which is essentially the Fourier transform of the commutator of these operators as 
\begin{align}
    \rho_{\hat{\mathcal{O}}_{a}\hat{\mathcal{O}}_{b}}(\Ka)=\frac{1}{2}\int_{\Ch}\,e^{i\Ka\cdot\Ch}\left\langle\Big[\hat{\mathcal{O}}_{a}(\Ch),\hat{\mathcal{O}}_{b}^{\dagger}(0)\Big]\right\rangle_{l}\label{SpdA},
\end{align}
where $\Ka=(k^{0},\km)$. Using the integral representation of the theta function, 
\begin{align}
    \theta(t)=i\int_{-\infty}^{\infty}\frac{d\omega}{2\pi}\frac{e^{-i\omega t}}{\omega+i0^{+}}\,,
\end{align}
one can show that these two Green's functions are related as 
\begin{align}
    G^{R}_{\hat{\mathcal{O}}_{a}\hat{\mathcal{O}}_{b}}(\Ka)&=-i\int_{0}^{\infty}dt\int d^{3}\xv\,e^{i\omega t-i\km\cdot\xv}\left\langle\theta(t)\bigg[\hat{\mathcal{O}}_{a}(\mathbf{x}, t), \hat{\mathcal{O}}^{\dagger}_{b}\left(\mathbf{0},0\right)\bigg]\,\,\right\rangle_{l}\,\label{FGR}\\
    &=\int_{-\infty}^{\infty}\frac{dp^{0}}{\pi}\frac{\rho_{\hat{\mathcal{O}}_{a}\hat{\mathcal{O}}_{b}}(\mathcal{P})}{p^{0}-k^{0}-i0^{+}} . \label{analyticconti}
\end{align}
Therefore, we see that the retarded Green's function $ G^{R}(\Ka)$ can be obtained from the spectral function $\rho(\mathcal{P})$ through analytic continuation in the upper half plane of $p^{0}$.

Furthermore, if the operators $\hat{\mathcal{O}}_{a}$ and $\hat{\mathcal{O}}_{b}$ are hermitian, one can show that $G^{R}_{\hat{\mathcal{O}}_{a}\hat{\mathcal{O}}_{b}}(\Ch)$ is real. Moreover, for the Hermitian operators, the retarded correlation function in the momentum space satisfies the relations 
\begin{align}
    \text{Re}\Big(G^{R}_{\hat{\mathcal{O}}_{a}\hat{\mathcal{O}}_{b}}(\Ka)\Big)=\text{Re}\Big(G^{R}_{\hat{\mathcal{O}}_{a}\hat{\mathcal{O}}_{b}}(-\Ka)\Big),\quad\text{Im}\Big(G^{R}_{\hat{\mathcal{O}}_{a}\hat{\mathcal{O}}_{b}}(\Ka)\Big)=-\text{Im}\Big(G^{R}_{\hat{\mathcal{O}}_{a}\hat{\mathcal{O}}_{b}}(-\Ka)\Big)\label{RIGR}.
\end{align}
Similarly, for the spectral function, one can show using the definition Eq.~\eqref{SpdA} that $\rho_{\hat{\mathcal{O}}_{a}\hat{\mathcal{O}}_{b}}(\Ka)$ is an odd function of $\Ka$
\begin{align}
    \rho_{\hat{\mathcal{O}}_{a}\hat{\mathcal{O}}_{b}}(\Ka)=-\rho_{\hat{\mathcal{O}}_{a}\hat{\mathcal{O}}_{b}}(-\Ka) . \label{Spdodd}
\end{align}
Further, if one assumes the spectral function to be real, one can identify the imaginary part of the retarded Green's function as\,\footnote{This identification can be made by making use of this mnemonic expression $\frac{1}{\Delta\pm i0^{+}}=P(\frac{1}{\Delta})\mp i\pi\delta(\Delta)$ and where $P$ denotes the principle value of the integral.}
\begin{align}
    \text{Im}\Big(G^{R}_{\hat{\mathcal{O}}_{a}\hat{\mathcal{O}}_{b}}(\Ka)\Big)=\rho_{\hat{\mathcal{O}}_{a}\hat{\mathcal{O}}_{b}}(\Ka)\,.\label{ImGs}
\end{align}
The above equation relates the imaginary part of the retarded Green's function to the spectral function. With the help of Eq.~\eqref{Spdodd}, one can show using L'Hôpital's rule that
\begin{align}
\lim_{\omega\rightarrow0}\frac{\rho_{\hat{\mathcal{O}}_{a}\hat{\mathcal{O}}_{b}}(\Ka)}{\omega}=\frac{\partial}{\partial \omega}\text{Im}\Big(G^{R}_{\hat{\mathcal{O}}_{a}\hat{\mathcal{O}}_{b}}(\Ka)\Big)\bigg|_{\omega\rightarrow 0}. \label{dGtorho}
\end{align}
The above relation helps us to express Eq.~\eqref{dimjj} in the form of Eq.~\eqref{dspmjj}.

Next, we discuss some other relations in the presence of finite chemical potentials in the context of Kubo-Martin-Schwinger (KMS) relations while considering the thermal averages of operators. The commutator of the operators corresponding to the total number operator $\hat{Q}_{a}$ (spatially integrated number density operator) is given by 
\begin{align}
    \Big[\hat{\mathcal{O}}_{a}(\Ch),\hat{Q}_{b}\Big]=\delta_{ab}\hat{\mathcal{O}}_{a}(\Ch)\,\label{ComQP}.
\end{align}
The KMS relation for such an operator can be written as \cite{Laine:2016hma}
\begin{align}
    \left\langle\hat{\mathcal{O}}_{a}(t-i\beta,\xv)\hat{\mathcal{O}}_{b}^{\dagger}(0,\textbf{0}) \right\rangle_{l}&=e^{-\beta\mu_{a}}\left\langle\hat{\mathcal{O}}_{b}^{\dagger}(0,\textbf{0})\hat{\mathcal{O}}_{a}(t,\xv) \right\rangle_{l} , \label{KMSOO}
\end{align}
where, Eq.~\eqref{ComQP} has been used. Such relations are valid for operators which are odd in powers of fields $\phi_{a}$. In particular, for equal time, Eq.~\eqref{KMSOO} reduce to 
\begin{align}
    \left\langle\hat{\mathcal{O}}_{a}(-i\beta,\xv)\hat{\mathcal{O}}_{b}(0,\textbf{0}) \right\rangle_{l}&=e^{-\beta\mu_{a}}\left\langle\hat{\mathcal{O}}_{a}(0,\xv)\hat{\mathcal{O}}_{b}(0,\textbf{0}) \right\rangle_{l}
\end{align}
The exponential factor $e^{-\beta\mu_{a}}$ thus spoils the periodicity of this correlation function. The corresponding Euclidean correlator is defined as 
\begin{align}
    \mathcal{G}^{E}_{\hat{\mathcal{O}}_{a}\hat{\mathcal{O}}_{b}}(K)=\int_{0}^{\beta}d\tau\int_{\xv}e^{(i k_{n}+\mu_{a})\tau-i\km\cdot\xv}\left\langle\hat{\mathcal{O}}_{a}(X)\hat{\mathcal{O}}_{b}(0) \right\rangle_{l}\,.\label{density GR}
\end{align}
The additional term in the Fourier transform with respect to $\tau$ cancels the multiplicative factor $e^{-\beta\mu_{a}}$ at $\tau=\beta$ so that the $\tau$ integrand remains periodic, leading to the bosonic Matsubara frequencies $k_{n}$.

Further, for the operators which commute with the total number operator, i.e., 
\begin{align}
    \Big[\hat{\mathcal{O}}_{a}(\Ch),\hat{Q}_{b}\Big]=0\,,\label{ComQPz}
\end{align}
the KMS relation is given by
\begin{align}
    \left\langle\hat{\mathcal{O}}_{a}(t-i\beta,\xv)\hat{\mathcal{O}}_{b}(0,\textbf{0}) \right\rangle_{l}&=\left\langle\hat{\mathcal{O}}_{b}(0,\textbf{0})\hat{\mathcal{O}}_{a}(t,\xv) \right\rangle_{l},\label{KMSOOz}
\end{align}
with no extra chemical potential dependent suppression factor, which may be noted. In particular, for equal time, Eq.~\eqref{KMSOOz} reduce to
\begin{align}
    \left\langle\hat{\mathcal{O}}_{a}(-i\beta,\xv)\hat{\mathcal{O}}_{b}(0,\textbf{0}) \right\rangle_{l}&=\left\langle\hat{\mathcal{O}}_{a}(0,\xv)\hat{\mathcal{O}}_{b}(0,\textbf{0}) \right\rangle_{l}\,,
\end{align}
which is a periodic function with respect to imaginary time. Therefore, in contrast to Eq.~\eqref{density GR}, the Fourier transformation of the Euclidean two-point function can be defined as
\begin{align}
    \mathcal{G}^{E}_{\hat{\mathcal{O}}_{a}\hat{\mathcal{O}}_{b}}(K)=\int_{X}e^{i K\cdot X}\left\langle\hat{\mathcal{O}}_{a}(X)\hat{\mathcal{O}}_{b}^{\dagger}(0)\right\rangle_{l}\,.
\end{align}
Here $K=(k_{n},\km)$ is the Euclidean momentum and this correlation function is time-ordered by definition $(0\leq\tau\leq\beta)$. Since Green's function $\mathcal{G}^{E}_{\hat{\mathcal{O}}_{a}\hat{\mathcal{O}}_{b}}(X)$ is periodic, we conclude that $k_{n}$ is a bosonic Matsubara frequency.

Similar to the retarded Green's function $G^{R}_{\hat{\mathcal{O}}_{a}\hat{\mathcal{O}}_{b}}(\Ka)$, the spectral function $\rho_{\hat{\mathcal{O}}_{a}\hat{\mathcal{O}}_{b}}(\Ka)$ also has different momentum space representations for the two cases. To evaluate the spectral function representation of the Euclidean Green's function, one needs to make a connection between the Euclidean Green's function $\langle\hat{\mathcal{O}}_{a}(X)\hat{\mathcal{O}}^{\dagger}_{b}(0)\rangle_{l}$ and the Wightman Green's function $\langle\hat{\mathcal{O}}_{a}(\Ch)\hat{\mathcal{O}}^{\dagger}_{b}(0)\rangle_{l}$ which are functions of Minkowski coordinates and Euclidean coordinates, respectively. This is possible by performing the analytic continuation $t\rightarrow -i\tau$, with $0\leq\tau\leq\beta$, on the Wightman function $\langle\hat{\mathcal{O}}_{a}(\Ch)\hat{\mathcal{O}}^{\dagger}_{b}(0)\rangle_{l}=\int_{\Ka}e^{i\Ka\cdot\Ch}G_{\hat{\mathcal{O}}_{a}\hat{\mathcal{O}}_{b}}(\Ka)$. Therefore, the Matsubara Fourier transformed Euclidean Green's function, as given in Eq.~\eqref{density GR}, can be expressed as
\begin{align}
    \mathcal{G}^{E}_{\hat{\mathcal{O}}_{a}\hat{\mathcal{O}}_{b}}(K)=\int_{0}^{\beta}d\tau\int_{\xv}e^{(i k_{n}+\mu_{a})\tau-i\km\cdot\xv}\Big[\int_{\mathcal{P}}e^{-\mathcal{P}\cdot\Ch}G_{\hat{\mathcal{O}}_{a}\hat{\mathcal{O}}_{b}}(\mathcal{P})\Big]_{it\rightarrow\tau}\,.\label{GEW}
\end{align}
Here, the exponential factor $e^{\mu_{a}\tau}$ appears for those non-commuting operators as given in Eq.~\eqref{ComQP}. To explore the relation between Euclidean Green's function and the spectral function, one needs to employ the spectral function representation of the Wightman function as \cite{Laine:2016hma}
\begin{align}
    G_{\hat{\mathcal{O}}_{a}\hat{\mathcal{O}}_{b}}(\Ka)=2\Big(1+f_{a}(k^{0}-\mu_{a})\Big)\rho_{\hat{\mathcal{O}}_{a}\hat{\mathcal{O}}_{b}}(\Ka)\,.\label{WtoSPD}
\end{align}
Using this equation into the Eq.~\eqref{GEW}, one can show that
\begin{align}
    \mathcal{G}^{E}_{\hat{\mathcal{O}}_{a}\hat{\mathcal{O}}_{b}}(K)&=\int_{0}^{\beta}d\tau\,e^{(ik_{n}+\mu)\tau}\int_{-\infty}^{\infty}\frac{dp^{0}}{2\pi}e^{-p^{0}\tau}2\Big(1+f_{a}(p^{0}-\mu_{a})\Big)\rho_{\hat{\mathcal{O}}_{a}\hat{\mathcal{O}}_{b}}(p^{0},\km)\nonumber\\
    &=\int_{-\infty}^{\infty}\frac{dp^{0}}{2\pi}2\Big(1+f_{a}(p^{0}-\mu_{a})\Big)\rho_{\hat{\mathcal{O}}_{a}\hat{\mathcal{O}}_{b}}(p^{0},\km)\int_{0}^{\beta}d\tau\,e^{(ik_{n}+\mu-p^{0})\tau}\nonumber\\
    &=\int_{-\infty}^{\infty}\frac{dp^{0}}{2\pi}2\Big(1+f_{a}(p^{0}-\mu_{a})\Big)\rho_{\hat{\mathcal{O}}_{a}\hat{\mathcal{O}}_{b}}(p^{0},\km)\Bigg[\frac{e^{(ik_{n}+\mu-p^{0})\tau}}{ik_{n}+\mu-p^{0}}\Bigg]_{0}^{\beta}\nonumber\\
    &=\int_{-\infty}^{\infty}\frac{dp^{0}}{2\pi}2\Big(1+f_{a}(p^{0}-\mu_{a})\Big)\rho_{\hat{\mathcal{O}}_{a}\hat{\mathcal{O}}_{b}}(p^{0},\km)\Bigg[\frac{e^{(\mu-p^{0})\beta}-1}{ik_{n}+\mu-p^{0}}\Bigg]\nonumber\\
    &=\int_{-\infty}^{\infty}\frac{dp^{0}}{\pi}\rho_{\hat{\mathcal{O}}_{a}\hat{\mathcal{O}}_{b}}(p^{0},\km)\Bigg[\frac{e^{(p^{0}-\mu)\beta}}{e^{(p^{0}-\mu)\beta}-1}\Bigg]\Bigg[\frac{e^{(\mu-p^{0})\beta}-1}{ik_{n}+\mu-p^{0}}\Bigg]\nonumber\\
    &=\int_{-\infty}^{\infty}\frac{dk^{0}}{\pi}\frac{\rho_{\hat{\mathcal{O}}_{a}\hat{\mathcal{O}}_{b}}(k^{0},\km)}{k^{0}-i[k_{n}-i\mu]}\,.\label{GESPDNC}
\end{align}
One can invert the above relation by using the Sokhotski-Plemelj theorem as 
\begin{align}
   \rho_{\hat{\mathcal{O}}_{a}\hat{\mathcal{O}}_{b}}(\Ka)=\frac{1}{2i}\text{Disc}\,\mathcal{G}^{E}_{\hat{\mathcal{O}}_{a}\hat{\mathcal{O}}_{b}}(k_{n}-i\mu\rightarrow-ik^{0},\km),
\end{align}
where the discontinuity is defined in Eq.~\eqref{SpdtoGE}.

Similar to Eq.~\eqref{GESPDNC}, one can also derive the relation between the Euclidean Green's function and spectral function for those bosonic operators, which satisfies the the commutation relation with the total charge operator as given in Eq.~\eqref{ComQPz}. To do so, one needs to employ a similar relation between the Wightman function and the spectral function as
\begin{align}
    G_{\mathcal{O}_{a}\mathcal{O}_{b}}(\Ka)=2\Big(1+f_{B}(k^{0})\Big)\rho_{\hat{\mathcal{O}}_{a}\hat{\mathcal{O}}_{b}}(\Ka)\,,
\end{align}
where $f_{B}(k^{0})=\frac{1}{\text{exp}(\beta k^{0})-1}$ is the bosonic distribution which does not carry the chemical potential, as opposed to Eq.~\eqref{WtoSPD}. Therefore, one can derive the relation between the Euclidean Green's function to the spectral function as follows 
\begin{align}
    \mathcal{G}^{E}_{\hat{\mathcal{O}}_{a}\hat{\mathcal{O}}_{b}}(K)=\int_{-\infty}^{\infty}\frac{dk^{0}}{\pi}\frac{\rho_{\hat{\mathcal{O}}_{a}\hat{\mathcal{O}}_{b}}(k^{0},\km)}{k^{0}-ik_{n}}\,.\label{gfao2}
\end{align}
Here, it is important to note that the absence of the chemical potential in the denominator gives a contrasting representation compared to Eq.~\eqref{GESPDNC}. Moreover, the above relation can be inverted to obtain
\begin{align}
   \rho_{\hat{\mathcal{O}}_{a}\hat{\mathcal{O}}_{b}}(\Ka)=\frac{1}{2i}\text{Disc}\,\mathcal{G}^{E}_{\hat{\mathcal{O}}_{a}\hat{\mathcal{O}}_{b}}(k_{n}\rightarrow-ik^{0},\km).
\end{align}
A comparison to the description in Section-(\ref{sec:transport}) can be made if one identifies $\phi$ and $\xi$ as those operators which satisfy Eq.~\eqref{ComQP}, that reflects in the spectral representation of the Euclidean Green's function Eq.~\eqref{GEtoSpd} with a resemblance to Eq.~\eqref{GESPDNC}. Similarly, recognising the operators like particle (charge) diffusion current, $\mathcal{J}^{\mu}_{a}~(\mathcal{J}^{\mu}_{A})$, as those operators which commute with the total charge operator, as given in Eq.~\eqref{ComQPz}, leads to the resemblance between Eq.~\eqref{gfao} and Eq.~\eqref{gfao2}.


\subsection{The KMB inner product and the Retarded Green's function. }\label{A2}

In this appendix, we show the relation between the first-order transport coefficients written in terms of KMB inner product, e.g. in Eq.~\eqref{Dab}, and retarded Green's function which is defined in Sec-(\ref{sec:neso}). A similar calculation can also be found in  Ref.~\cite{Hosoya:1983id, Huang:2011dc}. The KMB inner product is given by 
\begin{equation}
    \left(\hat{\mathcal{O}}_{a}(\mathbf{x}, t), \hat{\mathcal{O}}_{b}\left(\mathbf{x}_{1}, t_{1}\right)\right) =\frac{1}{\beta} \int_{0}^{\beta} d \tau\left\langle\hat{\mathcal{O}}_{a}(\mathbf{x}, t)\left[\hat{\mathcal{O}}_{b}(\textbf{x}_{1},t_{1}+i\tau) - \left\langle \hat{\mathcal{O}}_{b}(\textbf{x}_{1},t_{1}+i\tau)\right\rangle_{l}\right]\right\rangle_{l} .\label{KMBtau}
\end{equation}
Here, $\hat{\mathcal{O}}_{b}(\textbf{x}_{1},t_{1}+i\tau)=e^{-\tau \hat{H} } \hat{\mathcal{O}}_{b}(\textbf{x}_{1},t_{1}) e^{\tau \hat{H} }$, with $\hat{H}=u_{\mu}\hat{P}^{\mu}-\sum_{a}\mu_{a}\hat{n}_{a}$ being the Hamiltonian.

To extract the transport coefficients from the definition, for e.g. Eq.~\eqref{eq:EMT_in_Kubo_form}, the values of these transport coefficients are computed at the global equilibrium instead of the local equilibrium. Therefore, this global equilibrium distribution is defined with average thermodynamic parameters, like temperature $T=\beta^{-1}$ and the chemical potentials $\mu_{a}$ ($\mu_{A}$) in particle (charge) basis. In equilibrium, with proper definition of thermodynamic parameters, one can define the global grand-canonical statistical operator $\hat{\varrho}_{eq}$ in fluid rest frame with the time-independent Hamiltonian $\hat{H}=u_{\mu}\hat{P}^{\mu}-\sum_{a}\mu_{a}\hat{n}_{a}$ in particle basis or equivalently $\hat{H}=u_{\mu}\hat{P}^{\mu}-\sum_{A}\mu_{A}\hat{n}_{A}$ in charge basis, with $\hat{P}^{\mu}$ being the operator for 4-momentum. A comparison with Eq.~\eqref{eq:equi_part_Z} leads to the recognition that $\hat{\mathcal{A}}=\beta \hat{H}$. Therefore one can re-parameterize Eqs.~\eqref{TTofX} and \eqref{eq:Kubo_corr_nota} by $\tau=\beta\lambda$ to obtain
\begin{equation}
    \left(\hat{\mathcal{O}}_{a}(\mathbf{x}, t), \hat{\mathcal{O}}_{b}\left(\mathbf{x}_{1}, t_{1}\right)\right) =\frac{1}{\beta} \int_{0}^{\beta} d \tau\left\langle\hat{\mathcal{O}}_{a}(\mathbf{x}, t)\left[\hat{\mathcal{O}}_{b}(\textbf{x}_{1},t_{1}+i\tau) - \left\langle \hat{\mathcal{O}}_{b}(\textbf{x}_{1},t_{1}+i\tau)\right\rangle_{l}\right]\right\rangle_{l}~.\label{KMBtau}
\end{equation}
Here, $\hat{\mathcal{O}}_{b}(\textbf{x}_{1},t_{1}+i\tau)=e^{-\tau \hat{H} } \hat{\mathcal{O}}_{b}(\textbf{x}_{1},t_{1}) e^{\tau \hat{H} }$, which can be interpreted as the operator $\hat{\mathcal{O}}_{b}(\textbf{x},t)$ evolving in Heisenberg picture under a time displacement $\delta t=i\tau$. Further, by performing analytic continuation, one can arrive at the relations
\begin{align}
    \langle\hat{\mathcal{O}}_{b}(\textbf{x},t+i\tau)\rangle_{l}&=\langle\hat{\mathcal{O}}_{b}(\textbf{x},t)\rangle_{l}\,,\label{itau0}\\
    \langle\hat{\mathcal{O}}_{a}(\textbf{x},t)\hat{\mathcal{O}}_{b}(\textbf{x}^{\prime},t^{\prime}+i\beta)\rangle_{l}&=\langle\hat{\mathcal{O}}_{b}(\textbf{x}^{\prime},t^{\prime})\hat{\mathcal{O}}_{a}(\textbf{x},t)\rangle_{l}.\label{tauco}
\end{align}
The last relation is known as Kubo-Martin-Schwinger (KMS) relation.

If we consider that the correlation vanishes in the limit $t^{\prime}\rightarrow-\infty$, i.e., \cite{Hosoya:1983id,Huang:2011dc}
\begin{align}
    \lim_{t^{\prime}\rightarrow-\infty}\langle\hat{\mathcal{O}}_{a}(\textbf{x},t)\hat{\mathcal{O}}_{b}(\textbf{x}^{\prime},t^{\prime}+i\tau)\rangle_{l}=\lim_{t^{\prime}\rightarrow-\infty}\langle\hat{\mathcal{O}}_{a}(\textbf{x},t)\rangle_{l}\langle\hat{\mathcal{O}}_{b}(\textbf{x}^{\prime},t^{\prime}+i\tau)\rangle_{l}, \label{unco}
\end{align}
then we can rewrite the Eq.~\eqref{KMBtau} by using Eqs.~\eqref{itau0}, \eqref{tauco} and \eqref{unco} as
\begin{align}
\left(\hat{\mathcal{O}}_{a}(\mathbf{x}, t), \hat{\mathcal{O}}_{b}\left(\mathbf{x}_{1}, t_{1}\right)\right)&=\frac{1}{\beta}\int_{0}^{\beta}d\tau\Bigg[\left\langle\hat{\mathcal{O}}_{a}(\mathbf{x}, t) \hat{\mathcal{O}}_{b}\left(\mathbf{x}_{1}, t_{1}+i\tau\right)\right\rangle_{l} \nonumber\\
&\qquad\qquad\qquad -\left\langle\hat{\mathcal{O}}_{a}(\mathbf{x}, t)\right\rangle_{l}\left\langle \hat{\mathcal{O}}_{b}\left(\mathbf{x}_{1}, t_{1}+i\tau\right)\right\rangle_{l}\Bigg]\nonumber\\
&=\frac{1}{\beta}\int_{0}^{\beta}d\tau\int_{-\infty}^{t_{1}} dt^{\prime}\frac{d}{dt^{\prime}}\Bigg[\left\langle\hat{\mathcal{O}}_{a}(\mathbf{x}, t) \hat{\mathcal{O}}_{b}\left(\mathbf{x}_{1}, t_{1}+i\tau\right)\right\rangle_{l} \nonumber\\
&\qquad\qquad\qquad -\left\langle\hat{\mathcal{O}}_{a}(\mathbf{x}, t)\right\rangle_{l}\left\langle \hat{\mathcal{O}}_{b}\left(\mathbf{x}_{1}, t_{1}+i\tau\right)\right\rangle_{l}\Bigg].
\end{align}
Here, one needs to use the Heisenberg equation of motion of an operator $\hat{\mathcal{O}}(\textbf{x},t)$.

The Heisenberg equation of motion can be derived if it is evolved using $\hat{\mathcal{O}}_{b}(\textbf{x},t+i\tau)=e^{-\tau \hat{H} } \hat{\mathcal{O}}_{b}(\textbf{x},t) e^{\tau \hat{H} }$, leading to
\begin{align}
    \frac{d}{dt}\hat{\mathcal{O}}(\textbf{x},t+i\tau)=-i\frac{d}{d\tau}\hat{\mathcal{O}}(\textbf{x},t+i\tau).
\end{align}
Using the above relation, along with Eqs.~\eqref{itau0} and \eqref{tauco}, one can turn the KMB inner product of two operators, Eq.~\eqref{KMBtau}, into a commutator of two operators
\begin{align}
    \left(\hat{\mathcal{O}}_{a}(\mathbf{x}, t), \hat{\mathcal{O}}_{b}\left(\mathbf{x}_{1}, t_{1}\right)\right)&=\frac{i}{\beta}\int^{t_{1}}_{-\infty}\,dt^{\prime}\left\langle\bigg[\hat{\mathcal{O}}_{a}(\mathbf{x}, t), \hat{\mathcal{O}}_{b}\left(\mathbf{x}_{1}, t^{\prime}\right)\bigg]\right\rangle_{l}
\end{align}
In order to take into account the causality constraint, $t_{\prime}\leq t_{1}\leq t$, one needs to write the last equation in terms of retarded Green's function as,
\begin{align}
    \left(\hat{\mathcal{O}}_{a}(\mathbf{x}, t), \hat{\mathcal{O}}_{b}\left(\mathbf{x}_{1}, t_{1}\right)\right)&=-\frac{1}{\beta}\int^{t_{1}}_{-\infty}\,dt^{\prime}G^{R}_{\hat{\mathcal{O}}_{a}\hat{\mathcal{O}}_{b}}\big(\textbf{x}-\textbf{x}_{1},t-t^{\prime}\big)\,,\label{GRO}
\end{align}
where two-point Retarded Green's function for uniform medium can be written as
\begin{align}
    G^{R}_{\hat{\mathcal{O}}_{a}\hat{\mathcal{O}}_{b}}\big(\textbf{x}-\textbf{x}_{1},t-t^{\prime}\big)&=-i\theta(t-t^{\prime})\left\langle\bigg[\hat{\mathcal{O}}_{a}(\mathbf{x}, t), \hat{\mathcal{O}}_{b}\left(\mathbf{x}_{1}, t^{\prime}\right)\bigg]\right\rangle_{l}\,.
\end{align}

In order to calculate the transport coefficients, one needs to compute the integral such as those appearing in Eq.~\eqref{eq:EMT_in_Kubo_form}. In the following, similar integrals are computed with a set of operators $\{\hat{\mathcal{O}}_{a},\hat{\mathcal{O}}_{b}\}$
\begin{align}
    \mathcal{I}\big[\hat{\mathcal{O}}_{a}\hat{\mathcal{O}}_{b}\big]=\lim_{\epsilon\rightarrow0}\int d^{3}\textbf{x}_{1}\int_{-\infty}^{t}dt^{\prime}\,e^{-\epsilon(t-t^{\prime})}\left(\hat{\mathcal{O}}_{a}(\mathbf{x}, t), \hat{\mathcal{O}}_{b}\left(\mathbf{x}_{1}, t^{\prime}\right)\right)\,,
\end{align}
Now, using Eq.~\eqref{GRO}, one can compute the integral
\begin{align}
    \mathcal{I}\Big[\hat{\mathcal{O}}_{a}\hat{\mathcal{O}}_{b}\Big]&=-\lim_{\epsilon\rightarrow0}\int d^{3}\textbf{x}^{\prime}\int^{t}_{-\infty}dt^{\prime}e^{\epsilon(t^{\prime}-t)}\frac{1}{\beta}\int^{t^{\prime}}_{-\infty}dsG^{R}_{\hat{\mathcal{O}}_{a}\hat{\mathcal{O}}_{b}}\big(\textbf{x}-\textbf{x}^{\prime},t-s\big).
\end{align}
This can be conveniently done by taking the Fourier transformation $\tilde{G}^{R}(\omega,\km)$ of the retarded Green's function $G^{R}(\xv,t)$, leading to
\begin{align}
   \mathcal{I}\Big[\hat{\mathcal{O}}_{a}\hat{\mathcal{O}}_{b}\Big]&=-\lim_{\epsilon\rightarrow0}T\int\,\frac{d\omega}{2\pi}\frac{1}{i\omega(i\omega+\epsilon)}\lim_{\km\rightarrow0}G^{R}_{\hat{\mathcal{O}}_{a}\hat{\mathcal{O}}_{b}}(\km,\omega).
\end{align}
The right-hand side can be rewritten using partial fraction of the integrand and using the residue theorem. This leads to 
\begin{align}
     \mathcal{I}\Big[\hat{\mathcal{O}}_{a}\hat{\mathcal{O}}_{b}\Big]&=-\lim_{\epsilon\rightarrow0}\frac{T}{\epsilon}\lim_{\omega\rightarrow0}\lim_{\km\rightarrow0}\Big[G^{R}_{\hat{\mathcal{O}}_{a}\hat{\mathcal{O}}_{b}}(\km,\omega)-G^{R}_{\hat{\mathcal{O}}_{a}\hat{\mathcal{O}}_{b}}(\km,\omega+i\epsilon)\Big].
\end{align}
Eventually, one needs to take the limit $\epsilon\rightarrow0$ of the integral $\mathcal{I}\Big[\hat{\mathcal{O}}_{a}\hat{\mathcal{O}}_{b}\Big]$, as mentioned in the definition of relevant statistical operator, Eq.~\eqref{eq:Z_function_def}, leading to
\begin{align}
\mathcal{I}\Big[\hat{\mathcal{O}}_{a}\hat{\mathcal{O}}_{b}\Big]=iT\lim_{\omega\rightarrow0}\partial_{\omega}G^{R}_{\hat{\mathcal{O}}_{a}\hat{\mathcal{O}}_{b}}(\textbf{0},\omega).\label{DefI}
\end{align}
Let us consider $G^{R}_{\hat{\mathcal{O}}_{a}\hat{\mathcal{O}}_{b}}(\omega,\km)$ as given in Eq.~\eqref{FGR}. We have,
\begin{align}
   G^{R}_{\hat{\mathcal{O}}_{a}\hat{\mathcal{O}}_{b}}(\omega,\km)=-i\int_{0}^{\infty}dt\int d^{3}\xv\,e^{i\omega t-i\km\cdot\xv}\left\langle\theta(t)\bigg[\hat{\mathcal{O}}_{a}(\mathbf{x}, t), \hat{\mathcal{O}}_{b}\left(\mathbf{0},0\right)\bigg]\,\,\right\rangle_{l}\,.
\end{align}
It is clear from the above equation that
\begin{align}
    \Big( G^{R}_{\hat{\mathcal{O}}_{a}\hat{\mathcal{O}}_{b}}(\textbf{0},\omega)\Big)^{*}=G^{R}_{\hat{\mathcal{O}}_{a}\hat{\mathcal{O}}_{b}}(\textbf{0},-\omega)\,.
\end{align}
Using this, one can write for the diffusive current-current correlation as
\begin{align}
    \mathcal{I}\Big[\hat{\mathcal{J}}_{a}\hat{\mathcal{J}}_{b}\Big]=-T\lim_{\omega\rightarrow0}\partial_{\omega}\text{Im}\Bigg(G^{R}_{\hat{\mathcal{J}}_{a}\hat{\mathcal{J}}_{b}}(\textbf{0},\omega)\Bigg).\label{A39}
\end{align}
Using Eq.~\eqref{A39} in Eq.~\eqref{Dab}, one obtains Eq.~\eqref{dimjj}.


\section{Self Energy}\label{SETad}
\subsection{Real part of the self-energy to the lowest order: tadpole diagrams}\label{SETad}

In this Appendix, we compute the self-energies $\Sigma_{a}$ of the fields $\phi$ and $\xi$ with the Lagrangian given in Eq.~\eqref{chargedlagrangiaan}. For this purpose, we note that the Euclidean two point function for these fields can be written in terms of the single particle RSF $\rho_{a}(k_{0},\km)$, as given in Eq.~\eqref{GEtoSpd}. When the imaginary part of the self-energy vanishes, the spectral functions are given by 
\begin{equation}
    \rho_{a}(k_{0},\km)=\frac{\pi}{2\omega_{a}(\km)}\Big[\delta(k_{0}-\omega_{a}(\km))-\delta(k_{0}+\omega_{a}(\km))\Big], \label{FSPD}
\end{equation}
where, $\omega_{a}(\km)$ are the quasiparticle energies, given by 
\begin{equation}
    \omega_{a}^{2}(\km)=\km^{2}+m_{a}^{2}+\text{Re}(\Sigma_{a}(\omega_{a}(\km),\km))\equiv\km^{2}+M_{a}^{2}.
\end{equation}
On the other hand, when there is non-vanishing imaginary part of the self energy, the spectral functions take the form 
\begin{equation}
    \rho_{a}(k_{0},\km)=\frac{\text{Im}(\Sigma_{a}(k_{0},\km))}{\omega_{a}^{2}-\km^{2}-M_{a}^{2}+|\text{Im}(\Sigma_{a}(k_{0},\km))|^{2}}.
\end{equation}
%


\begin{figure}[h]
      \begin{tikzpicture}[baseline=(current bounding box.center)] 
                \begin{feynman}
                    \vertex (x);
                    \vertex[right=1.5cm of x] (a);
                    \vertex[right=4.5cm of x] (b);
                    \vertex[right=3cm of x] (c);
                    \diagram*{
                    (x) --[fermion] (a),
                    (a) --[fermion] (c),
                    a --[fermion, out=135, in=45, min distance=3.5cm] a,
                    };
                    \vertex [left=0.15em of x] {\(\left(p_n-i\mu_{\phi},\textbf{p}\right)\)};
                    \vertex [right=0.15em of c] {\(\left(p_n-i\mu_{\phi},\textbf{p}\right)\)};
                    \vertex [above=2.55em of a] {\(\left(r_n,\textbf{r}\right)\)};
                    \vertex [below=0.5em of a] {\(\lambda_1\)};
                \end{feynman}
            \end{tikzpicture}
            \hspace{0.25em}
            \begin{tikzpicture}[baseline=(current bounding box.center)] 
                \begin{feynman}
                    \vertex (x);
                    \vertex[right=1.5cm of x] (a);
                    \vertex[right=4.5cm of x] (b);
                    \vertex[right=3cm of x] (c);
                    \diagram*{
                    (x) --[fermion] (a)[dot],
                    (a) --[fermion] (c),
                    a --[charged scalar, out=135, in=45, min distance=3.5cm] a,
                    };
                    \vertex [left=0.15em of x] {\(\left(p_n-i\mu_{\phi},\textbf{p}\right)\)};
                    \vertex [right=0.15em of c] {\(\left(p_n-i\mu_{\phi},\textbf{p}\right)\)};
                    \vertex [above=2.55em of a] {\(\left(r_n,\textbf{r}\right)\)};
                    \vertex [below=0.5em of a] {\(g\)};
                \end{feynman}
            \end{tikzpicture}
                \caption{Tadpole diagram for $\phi$ field with self/mutual-coupling depicts in left/right side.}
            \label{fig:FD1}
        \end{figure}
\begin{figure}[h]
      \begin{tikzpicture}[baseline=(current bounding box.center)] 
                \begin{feynman}
                    \vertex (x);
                    \vertex[right=1.5cm of x] (a);
                    \vertex[right=4.5cm of x] (b);
                    \vertex[right=3cm of x] (c);
                    \diagram*{
                    (x) --[charged scalar] (a)[dot],
                    (a) --[charged scalar] (c),
                    a --[fermion, out=135, in=45, min distance=3.5cm] a,
                    };
                    \vertex [left=0.15em of x] {\(\left(p_n-i\mu_{\xi},\textbf{p}\right)\)};
                    \vertex [right=0.15em of c] {\(\left(p_n-i\mu_{\xi},\textbf{p}\right)\)};
                    \vertex [above=2.55em of a] {\(\left(r_n,\textbf{r}\right)\)};
                    \vertex [below=0.5em of a] {\(g\)};
                \end{feynman}
            \end{tikzpicture}
            \hspace{0.25em}
            \begin{tikzpicture}[baseline=(current bounding box.center)] 
                \begin{feynman}
                    \vertex (x);
                    \vertex[right=1.5cm of x] (a);
                    \vertex[right=4.5cm of x] (b);
                    \vertex[right=3cm of x] (c);
                    \diagram*{
                    (x) --[charged scalar] (a)[dot],
                    (a) --[charged scalar] (c),
                    a --[charged scalar, out=135, in=45, min distance=3.5cm] a,
                    };
                    \vertex [left=0.15em of x] {\(\left(p_n-i\mu_{\xi},\textbf{p}\right)\)};
                    \vertex [right=0.15em of c] {\(\left(p_n-i\mu_{\xi},\textbf{p}\right)\)};
                    \vertex [above=2.55em of a] {\(\left(r_n,\textbf{r}\right)\)};
                    \vertex [below=0.5em of a] {\(\lambda_{2}\)};
                \end{feynman}
            \end{tikzpicture}
                \caption{Tadpole diagram for $\phi$ field with self/mutual-coupling depicts in left/right side.}
            \label{fig:FD2}
        \end{figure}


The lowest order contribution to the self-energy comes from one-loop tadpole diagrams as shown in Fig.~\ref{fig:FD1} and Fig.~\ref{fig:FD2}. These tadpole diagrams do not lead to imaginary part of the self energy and these contributions do not depend upon energy or momentum. From the tadpole diagrams in Fig.~\ref{fig:FD1}, contribution to $\text{Re}(\Sigma_{\phi})$ is given  by,
\begin{equation}
    \Sigma_{\phi}=\frac{\lambda_{1}}{4}\sumintb_{P}\mathcal{G}^{E}_{\phi}(P)+\frac{g}{2}\sumintb_{P}\mathcal{G}^{E}_{\xi}(P).
\end{equation}
Using the single particle RSF in Eq.~\eqref{deltaanti} and summing over the Matsubara modes, we have \cite{Kapusta:2023eix, Roder:2005vt} 
\begin{align}
    \Sigma_{\phi}
    &=4\pi\left(\frac{\lambda_{1}}{4}\right)\int_{0}^{\infty} \frac{dp_{0}}{\pi}\int_{0}^{\infty} \frac{dp}{(2\pi)^{3}} p^{2}\Big[1+f_{\phi}(p_{0})+\bar{f}_{\phi}(p_{0})\Big]\rho_{\phi}(p_{0},p)\nonumber\\
    &+4\pi\left(\frac{g}{2}\right)\int_{0}^{\infty} \frac{dp_{0}}{\pi}\int_{0}^{\infty} \frac{dp}{(2\pi)^{3}} p^{2}\Big[1+f_{\xi}(p_{0})+\bar{f}_{\xi}(p_{0})\Big]\rho_{\xi}(p^{0},p),\label{Tadphisig}
\end{align}
where, $f_{a}(p_{0})=\big(e^{\beta(p_{0}-\mu_{a})}+1\big)^{-1}$ is the equilibrium bosonic particle distribution function with chemical potential $\mu_{a}$ and $\bar{f}_{a}(p_{0})=\big(e^{\beta(p_{0}+\mu_{a})}+1\big)^{-1}$ is the corresponding anti-particle distribution function. 

Similarly for the $\xi$ fields the contribution to the $\Sigma_{\xi}$ is calculated for tadpole diagrams in Fig.~\ref{fig:FD2}, and is given by 
\begin{align}
    \Sigma_{\xi}&=\frac{\lambda_{2}}{4}\sumintb_{P}\mathcal{G}^{E}_{\xi}(P)+\frac{g}{2}\sumintb_{P}\mathcal{G}^{E}_{\phi}(P)\nonumber\\
    &=4\pi\left(\frac{\lambda_{2}}{4}\right)\int_{0}^{\infty} \frac{dp_{0}}{\pi}\int_{0}^{\infty} \frac{dp}{(2\pi)^{3}} p^{2}\Big[1+f_{\xi}(p_{0})+\bar{f}_{\xi}(p_{0})\Big]\rho_{\xi}(p_{0},p)\nonumber\\
    &\quad+4\pi\left(\frac{g}{2}\right)\int_{0}^{\infty} \frac{dp_{0}}{\pi}\int_{0}^{\infty} \frac{dp}{(2\pi)^{3}} p^{2}\Big[1+f_{\phi}(p_{0})+\bar{f}_{\phi}(p_{0})\Big]\rho_{\phi}(p_{0},p).\label{Tadxisig}
\end{align}
Since the real part does not depend on energy and momentum, they only modify the masses of $\phi$ and $\xi$ fields. The thermal parts of the self-energies in Eqs.~\eqref{Tadphisig} and \eqref{Tadxisig} $\sim f_{a}(p_{0})$ and are ultraviolet finite. On the other hand, vacuum parts $(\sim 1)$ and are divergent which need to be renormalized \cite{Laine:2016hma, Kapusta:2023eix}. Note that these are coupled equation for the self-energies $\Sigma_{a}$ which need to be solved self-consistently. For $O(N)$ sigma model these has been investigated in \cite{Roder:2005vt}.


\begin{figure}[t]
            \centering
            \begin{tikzpicture}[baseline=(current bounding box.center)] 
                \begin{feynman}
                    \vertex (x);
                    \vertex[right=7cm of x] (y);
                    \vertex[right=1.5cm of x] (a);
                    \vertex[right=5.5cm of x] (b);
                    \vertex[right=3.5cm of x] (c);
                    \diagram*{
                    (x) --[fermion] (a),
                    (b) --[fermion] (y),
                    (b) --[fermion] (a),
                    a --[fermion, out=45, in=135, min distance=2cm] b,
                    a --[fermion, out=-45, in=-135, min distance=2cm] b,
                    };
                    \vertex [left=0.15em of x] {\(\left(p_n-i\mu_{\phi},\textbf{p}\right)\)};
                    \vertex [right=0.15em of y] {\(\left(p_n-i\mu_{\phi},\textbf{p}\right)\)};
                    \vertex [above=0.09em of c] {\(\left(s_n-i\mu_{\xi},\textbf{s}\right)\)};
                    \vertex [above=3em of c] {\(\left(p_n+l_{n}-i\mu_{\phi},\textbf{p}+\textbf{l}\right)\)};
                    \vertex [below=3em of c] {\(\left( s_n-l_{n}-i\mu_{\xi}, \textbf{s} - \textbf{l} \right)\)};
                    \vertex [below=0.5em of a] {\(\lambda_{1}\)};
                    \vertex [below=0.5em of b] {\(\lambda_{1}\)};
                \end{feynman}
            \end{tikzpicture}
            \caption{Sunset diagram for the $\phi$ field with self coupling.}
            \label{fig:FD4b}
        \end{figure}
\begin{figure}[t]
            \centering
            \begin{tikzpicture}[baseline=(current bounding box.center)] 
                \begin{feynman}
                    \vertex (x);
                    \vertex[right=7cm of x] (y);
                    \vertex[right=1.5cm of x] (a);
                    \vertex[right=5.5cm of x] (b);
                    \vertex[right=3.5cm of x] (c);
                    \diagram*{
                    (x) --[fermion] (a),
                    (b) --[fermion] (y),
                    (b) --[fermion] (a),
                    a --[charged scalar, out=45, in=135, min distance=2cm] b,
                    a --[charged scalar, out=-45, in=-135, min distance=2cm] b,
                    };
                    \vertex [left=0.15em of x] {\(\left(p_n-i\mu_{\phi},\textbf{p}\right)\)};
                    \vertex [right=0.15em of y] {\(\left(p_n-i\mu_{\phi},\textbf{p}\right)\)};
                    \vertex [above=0.09em of c] {\(\left(s_n-i\mu_{\xi},\textbf{s}\right)\)};
                    \vertex [above=3em of c] {\(\left(p_n+l_{n}-i\mu_{\phi},\textbf{p}+\textbf{l}\right)\)};
                    \vertex [below=3em of c] {\(\left( s_n-l_{n}-i\mu_{\xi}, \textbf{s} - \textbf{l} \right)\)};
                    \vertex [below=0.5em of a] {\(g\)};
                    \vertex [below=0.5em of b] {\(g\)};
                \end{feynman}
            \end{tikzpicture}
            \caption{Sunset diagram for the $\phi$ field with mutual coupling.}
            \label{fig:FD4a}
        \end{figure}
\begin{figure}[t]
            \centering
           \begin{tikzpicture}[baseline=(current bounding box.center)] 
                \begin{feynman}
                    \vertex (x);
                    \vertex[right=5.5cm of x] (y);
                    \vertex[right=1cm of x] (a);
                    \vertex[right=4.5cm of x] (b);
                    \vertex[right=2.75cm of x] (c);
                    \diagram*{
                    (x) --[charged scalar] (a),
                    (b) --[charged scalar] (y),
                    (b) --[charged scalar] (a),
                    a --[charged scalar, out=45, in=135, min distance=2cm] b,
                    a --[charged scalar, out=-45, in=-135, min distance=2cm] b,
                    };
                    \vertex [left=0.15em of x] {\(\left(p_n-i\mu_{\xi},\textbf{p}\right)\)};
                    \vertex [right=0.15em of y] {\(\left(p_n-i\mu_{\xi},\textbf{p}\right)\)};
                    \vertex [above=0.15em of c] {\(\left(s_{n}-i\mu_{\xi},\textbf{s}\right)\)};
                    \vertex [above=3em of c] {\(\left(p_n+l_{n}-i\mu_{\xi},\textbf{p}+\textbf{l}\right)\)};
                    \vertex [below=2.5em of c] {\(\left( s_n-l_{n}-i\mu_{\xi},\textbf{s} - \textbf{l}\right)\)};
                    \vertex [below=0.5em of a] {\(\lambda_{2}\)};
                    \vertex [below=0.5em of b] {\(\lambda_{2}\)};
                \end{feynman}
            \end{tikzpicture}
          \caption{Sunset diagram for the $\xi$ field with self coupling.}
            \label{fig:FD5}
            \vspace{1em}
            \begin{tikzpicture}[baseline=(current bounding box.center)] 
                \begin{feynman}
                    \vertex (x);
                    \vertex[right=5.5cm of x] (y);
                    \vertex[right=1cm of x] (a);
                    \vertex[right=4.5cm of x] (b);
                    \vertex[right=2.75cm of x] (c);
                    \diagram*{
                    (x) --[charged scalar] (a),
                    (b) --[charged scalar] (y),
                    (b) --[charged scalar] (a),
                    a --[fermion, out=45, in=135, min distance=2cm] b,
                    a --[fermion, out=-45, in=-135, min distance=2cm] b,
                    };
                    \vertex [left=0.15em of x] {\(\left(p_n-i\mu_{\xi},\textbf{p}\right)\)};
                    \vertex [right=0.15em of y] {\(\left(p_n-i\mu_{\xi},\textbf{p}\right)\)};
                    \vertex [above=0.15em of c] {\(\left(s_{n}-i\mu_{\xi},\textbf{s}\right)\)};
                    \vertex [above=3em of c] {\(\left(p_n+l_{n}-i\mu_{\phi},\textbf{p}+\textbf{l}\right)\)};
                    \vertex [below=2.5em of c] {\(\left( s_n-l_{n}-i\mu_{\phi},\textbf{s} - \textbf{l}\right)\)};
                    \vertex [below=0.5em of a] {\(g\)};
                    \vertex [below=0.5em of b] {\(g\)};
                \end{feynman}
            \end{tikzpicture}
            \caption{Sunset diagram for the $\xi$ field with mutual coupling.}
            \label{fig:FD6}
        \end{figure}


\subsection{Imaginary part of the self-energy to the lowest order; sunset diagram}\label{SEsun}

In the following, we compute the lowest order contribution to the imaginary part of the self-energy that appears at two-loop from the sunset diagrams, given in Figs.~\ref{fig:FD4a} and \ref{fig:FD4b}. For the $\phi$ fields this contribution is given by
\begin{equation}
    \Sigma^{\text{sunset}}_{\phi}(K)=\Sigma^{\text{sunset}}_{(a)\phi}(K)+\Sigma^{\text{sunset}}_{(b)\phi}(K)
\end{equation}
where, $\Sigma_{(a)\phi}^{\text{sunset}}\sim{\lambda_{1}^{2}}$ arises from Fig.~\ref{fig:FD4a}
\begin{align}
    \Sigma_{(a)\phi}^{\text{sunset}}(P)=-\frac{\lamo^{2}}{2!\times(4)^{2}}\sumintb_{L S}\Big(8\times\mathcal{G}^{E}_{\phi}(P+L)\mathcal{G}^{E}_{\phi}(S)\mathcal{G}^{E}_{\phi}(S-L)\Big),
\end{align}
and $\Sigma_{(b)\phi}^{\text{sunset}}\sim{g^{2}}$ arises from the Fig.~\ref{fig:FD4b}
\begin{align}
    \Sigma_{(b)\phi}^{\text{sunset}}(P)=-\frac{g^{2}}{2!\times(2)^{2}}\sumintb_{LS}\Big(\mathcal{G}^{E}_{\xi}(S)\mathcal{G}^{E}_{\xi}(S-L)\mathcal{G}^{E}_{\phi}(P+L)\Big).
\end{align}
Representing these sunset contribution of the self-energy in terms of the single particle RSF can be done using Eq.~\eqref{GEtoSpd}, i.e.,
\begin{align}
    \Sigma^{\text{sunset}}_{\phi}(P)&=-\frac{g^{2}}{2^{2}\times2!}\Bigg[\sumtintb_{LS}\int\prod_{i=1}^{3}\frac{d\omega_{i}}{\pi}\Bigg(\frac{\rho_{\phi}(\omega_{1},\pmo+\lm)}{\omega_{1}-i(p_{n}+l_{n}-i\mu_{\phi})}\Bigg)\Bigg(\frac{\rho_{\xi}(\omega_{2},\sm)}{\omega_{2}+i(s_{n}+i\mu_{\xi})}\Bigg)\nonumber\\
    &\Bigg(\frac{\rho_{\xi}(\omega_{3},\sm-\lm)}{\omega_{1}+i(s_{n}-l_{n}+i\mu_{\xi})}\Bigg)\Bigg]-8\times\!\frac{\lambda_{1}^{2}}{4^{2}\times2!}\Bigg[\sumtintb_{LS}\int\prod_{i=1}^{3}\frac{d\omega_{i}}{\pi}\Bigg(\frac{\rho_{\phi}(\omega_{1},\pmo+\lm)}{\omega_{1}-i(p_{n}+l_{n}-i\mu_{\phi})}\Bigg)\nonumber\\
    &\Bigg(\frac{\rho_{\phi}(\omega_{2},\sm)}{\omega_{2}+i(s_{n}+i\mu_{\phi})}\Bigg)\Bigg(\frac{\rho_{\phi}(\omega_{3},\sm-\lm)}{\omega_{1}+i(s_{n}-l_{n}+i\mu_{\phi})}\Bigg)\Bigg]\,.\label{Imph}
\end{align}
In what follows, without loss of generality, we assume that the self-coupling $\lambda_{1}$ to be much smaller than the cross-coupling $g$, and consider only the first term of Eq.~\eqref{Imph}.

Before proceeding further, we shall show a general identity that will be useful to perform the Matsubara frequency sums in Eq.\eqref{Imph}. A double Matsubara sum,
\begin{align}
    \mathcal{S}(\omega_{1}&,\omega_{2},\omega_{3},p_{n}-i\mu_{a})\nonumber\\
    &=T^{2}\sum_{s_{n},l_{n}}\frac{1}{\Big(\omega_{1}-i(p_{n}+l_{n}-i\mu_{a})\Big)\Big(\omega_{2}+i(s_{n}+i\mu_{b})\Big)\Big(\omega_{3}+i(s_{n}-l_{n}+i\mu_{b})\Big)} \nonumber\\
    &=T\sum_{l_{n}}\frac{1}{\Big(\omega_{1}-i(p_{n}+l_{n}-i\mu_{a})\Big)}\times\mathcal{I}_{n}(\omega_{2},\omega_{3},l_{n},\mu_{b})\label{ICal},
\end{align}
can be evaluated by performing two successive summation with respect $l_{n}$ and $s_{n}$ respectively. The summation over $s_{n}$ is performed as
\begin{align}
\mathcal{I}_{n}(\omega_{2},\omega_{3},l_{n},\mu_{b})&=T\sum_{s_{n}}\frac{1}{\Big(\omega_{2}+i(s_{n}+i\mu_{b})\Big)\Big(\omega_{3}+i(s_{n}-l_{n}+i\mu_{b})\Big)}\label{Ident1}\,.
\end{align}
This can be calculated by using $\beta\delta_{j_{n},l_{n}}=\int_{0}^{\beta}d\tau\,e^{i\tau(j_{n}-l_{n})}$ and rewriting the above equation as 
\begin{align}
\mathcal{I}_{n}(\omega_{2},\omega_{3},l_{n},\mu_{b})&=\int_{0}^{\beta}d\tau e^{il_{n}\tau}\Bigg(T\sum_{s_{n}}\frac{e^{-is_{n}\tau}}{\omega_{2}+i(s_{n}+i\mu_{b})}\Bigg)\Bigg(T\sum_{s'_{n}}\frac{e^{is'_{n}\tau}}{\omega_{3}+i(s'_{n}+i\mu_{b})}\Bigg)\nonumber\\
&=\Bigg(f_{b}(\omega_{2})-f_{b}(\omega_{3})\Bigg)\times\Bigg(\frac{1}{\omega_{3}-\omega_{2}-il_{n}}\Bigg)\label{Ident2}.
\end{align}
A similar technique can be repeated while summing over $l_{n}$ of Eq.~\eqref{ICal}. Doing this, we obtain
\begin{align}
\mathcal{S}(\omega_{1}&,\omega_{2},\omega_{3},p_{n}-i\mu_{a})\nonumber\\
    &=\frac{\Big(1+f_{a}(\omega_{1})\Big)\Big(1+f_{b}(\omega_{3})\Big)f_{b}(\omega_{2})}{\omega_{1}-\omega_{2}+\omega_{3}-i(p_{n}-i\mu_{a})}\Big[1-\text{exp}\Big(-\beta(\omega_{1}-\mu_{a}+\omega_{3}-\omega_{2})\Big)\Big]\label{Ident}.
\end{align}

Further, by applying the identities in Eq.~\eqref{Ident}, performing the analytic continuation $i(p_{n}-i\mu_{\phi})\rightarrow\omega+i0^{+}$ in Eq.~\eqref{Imph}, and taking the imaginary part, we obtain
\begin{align}
    \text{Im}\Big(\Sigma_{\phi}(\omega,\pmo)\Big)=&\,\frac{g^{2}}{2^{2}\times 2!}\Big(1-e^{-\beta(\omega-\mu_{\phi})}\Big)\!\int\! d\lm \int\! d\sm \int \frac{d\omega_{2}}{\pi}\int\frac{d\omega_{3}}{\pi}\Bigg[\Big(1+f_{\phi}(\omega-\omega_{2}-\omega_{3})\Big)\nonumber\\
    &\times\Big[1+\bar{f}_{\xi}(\omega_{2})\Big]\Big[1+f_{\xi}(\omega_{3})\Big]\rho_{\phi}(\omega-\omega_{2}-\omega_{3},\pmo+\lm)\,\rho_{\xi}(\omega_{2},\sm)\,\rho_{\xi}(\omega_{3},\sm-\lm)\Bigg].\label{wCf}
\end{align}
Moreover, using the fact that the spectral functions do not depend on the direction of the three momentum, angular integration can be carried out by using the identity
\begin{align}
    \int\,d\textbf{p}\,\mathcal{F}(|\textbf{k}-\textbf{p}|)\mathcal{G}(|\textbf{p}|)&=\frac{1}{(2\pi)^{2}|\textbf{k}|}\int_{0}^{\infty}dl_{1}dl_{2}\,l_{1}l_{2}\mathcal{F}(l_{1})\mathcal{G}(l_{2})\nonumber\\
    &\times\Theta(|l_{1}-l_{2}|\leq|\textbf{k}|\leq l_{1}+l_{2}),
\end{align}
resulting in the expression for Eq.~\eqref{wCf} as
\begin{align}
    \text{Im}\Big[\Sigma_{\phi}(\omega,\pmo)\Big] &= \frac{g^{2}\Big(1-e^{-\beta(\omega-\mu_{\phi})} \Big)}{2^{2}\times 2!\times (2\pi)^5} \int \frac{d\omega_{2}}{\pi} \frac{d\omega_{3}}{\pi} \int_{0}^{\infty} dl_{1}\, dl_{2}\, dl_{3}\, \Theta(|l_{2}-l_{3}|\leq l_{1}\leq l_{2}+l_{3})\nonumber\\
    &\quad\times l_1\,l_2\,l_3\, \rho_{\phi}(\omega-\omega_{2}-\omega_{3},|\pmo+\lm_{1}|)\, \rho_{\xi}(\omega_{2},l_{2})\, \rho_{\xi}(\omega_{3},l_{3}) \nonumber\\
    &\quad\times \Big[1+f_{\phi}(\omega-\omega_{2}-\omega_{3})\Big] \Big[1+\bar{f}_{\xi}(\omega_{2})\Big] \Big[1+f_{\xi}(\omega_{3})\Big]. \label{Gammaphi}
\end{align}
Here, one can think that $l_{1}$,$l_{2}$ and $l_{3}$ are the magnitude of the three momentums $\textbf{l}_{1}$,$\textbf{l}_{2}$ and $\textbf{l}_{3}$ respectively. The above form reveals the domain of momentum integration through the step function in this expression. A similar calculation can be done for $\xi$ fields (See Fig.~\ref{fig:FD5} and \ref{fig:FD6}), which leads to the following result
\begin{align}
    \text{Im}\Big(\Sigma_{\xi}(\omega,\pmo)\Big)=\text{Im}\Big(\Sigma_{\phi}(\omega,\pmo)\Big)\Big|_{\phi\leftrightarrow\xi}\,.\label{Gammaxi}
\end{align}
The expression as given in Eq.~\eqref{Gammaphi} and Eq.~\eqref{Gammaxi} can be written as a sum of different physical processes contributing to imaginary part of the self energy, which we discuss next. For this purpose, we rewrite Eq.~\eqref{wCf} as with the four vectors $\Tilde{l}_{i}\equiv(\omega_{i},\textbf{l}_{i})$ and $\Tilde{p}\equiv(\omega,\textbf{p})$
\begin{align}
     \text{Im}\Big(\Sigma_{\phi}(\tilde{p})\Big)=&\,\frac{g^{2}}{2!}\Big(1-e^{-\beta(\omega-\mu_{\phi})}\Big)(2\pi)^{4}\!\int\! \frac{d\tilde{l}_{1}}{(2\pi)^{4}}\frac{d\tilde{l}_{2}}{(2\pi)^{4}}\frac{d\tilde{l}_{3}}{(2\pi)^{4}}\Bigg[\Big(1+f_{\phi}(\omega_{1})\Big)\nonumber\\
    &\times\Big[1+\bar{f}_{\xi}(\omega_{2})\Big]\Big[1+f_{\xi}(\omega_{3})\Big]\rho_{\phi}(\omega_{\phi},\lm_{1})\,\rho_{\xi}(\omega_{2},\lm_{2})\,\rho_{\xi}(\omega_{3},\lm_{3})\delta^{4}(\tilde{l}_{1}+\tilde{l}_{2}+\tilde{l}_{3}-\tilde{p})\Bigg].\label{tildeSig}
\end{align}
We next note the following relation for the bosonic distribution functions with finite chemical potential
\begin{align}
    &\Big(1-e^{-\beta(\omega_{1}+\omega_{2}+\omega_{3}-\mu_{\phi})}\Big)\Big(1+f_{\phi}(\omega_{1})\Big)\Big(1+\bar{f}_{\xi}(\omega_{2})\Big)\Big(1+f_{\xi}(\omega_{3})\Big)\nonumber\\
    &=\Bigg[\Big(1+f_{\phi}(\omega_{1})\Big)\Big(1+\bar{f}_{\xi}(\omega_{2})\Big)\Big(1+f_{\xi}(\omega_{3})\Big)-f_{\phi}(\omega_{1})\bar{f}_{\xi}(\omega_{2})f_{\xi}(\omega_{3})\Bigg]\label{disid}
\end{align}
We then substitute the expression for the spectral function as given in Eq.~\eqref{FSPD} and use the identity Eq.~\eqref{disid} in Eq.~\eqref{tildeSig}. The two delta functions in each of the three spectral functions occurring in Eq.~\eqref{tildeSig} leads to eight different processes that contribute to the imaginary part of the self energy. Therefore we have
\begin{align}
    \text{Im}\Big(\Sigma_{\phi}(\omega,\pmo)\Big)&=\sum_{a=1}^{8}\text{Im}\Big(\Sigma_{\phi}(\omega,\pmo)\Big)_{a},\\
    \text{Im}\Big(\Sigma_{\xi}(\omega,\pmo)\Big)&=\sum_{a=1}^{8}\text{Im}\Big(\Sigma_{\xi}(\omega,\pmo)\Big)_{a}.
\end{align}
For instance, one can write the first process as
\begin{align}
    \text{Im}\Big(\Sigma_{\phi}(\omega,\pmo)\Big)_{1} =&\,\frac{g^{2}\pi^{4}}{8}\int d\textbf{l}_{1}\, d\textbf{l}_{2}\, d\textbf{l}_{3}\,\delta^{3}(\textbf{l}_{1}+\textbf{l}_{2}+\textbf{l}_{3}-\textbf{p})\times\frac{1}{E_{1}E_{2}E_{3}}\Bigg[\Big(1+f_{\phi}(E_{1})\Big)\nonumber\\
    &\times\Big(1+\bar{f}_{\xi}(E_{2})\Big)\Big(1+f_{\xi}(E_{3})\Big)-f_{\phi}(E_{1})\bar{f}_{\xi}(E_{2})f_{\xi}(E_{3})\Bigg]\nonumber\\
    &\times\delta(\omega-E_{2}-E_{3}-E_{1}),\label{fIGph}
\end{align}
where, $E_{1}=\sqrt{\lm_{1}^{2}+m_{\phi}^{2}}$ and $E_{2}, E_{3}$ are $\sqrt{\textbf{l}_{2}^{2}+m_{\xi}^{2}},\sqrt{\textbf{l}_{3}^{2}+m_{\xi}^{2}}$ respectively. This term can be interpreted as the probability for the ($\phi\rightarrow\phi\xi\bar{\xi})$ with a corresponding bose enhancement factor $(1+f_{\phi})(1+f_{\xi})(1+\bar{f}_{\xi})$ minus the probability for ($\phi\bar{\xi}\xi\rightarrow\phi$) with the weight $f_{\phi}\bar{f}_{\xi}f_{\xi}$. 

Second process can be interpreted as the anti-particle counterpart of the first process and given as,
\begin{align}
    \text{Im}\Big(\Sigma_{\phi}(\omega,\pmo)\Big)_{2} =&\, \frac{g^{2}\pi^{4}}{8}\int d\textbf{l}_{1}\, d\textbf{l}_{2}\, d\textbf{l}_{3}\,\delta^{3}(\textbf{l}_{1}+\textbf{l}_{2}+\textbf{l}_{3}-\textbf{p})\times\frac{1}{E_{1}E_{2}E_{3}}\Bigg[\bar{f}_{\phi}(E_{1})f_{\xi}(E_{2})\bar{f}_{\xi}(E_{3})\nonumber\\
    &-\Big(1+\bar{f}_{\phi}(E_{1})\Big)\Big(1+f_{\xi}(E_{2})\Big)\Big(1+\bar{f}_{\xi}(E_{3})\Big)\Bigg]\times\delta(E_{1}+E_{2}+E_{3}+\omega).
\end{align}
Both these processes contribute in the zero temperature limit $(T\rightarrow0$ and $f_{a}\rightarrow0$) due to the presence of statistical weight $(1+f_{\phi})(1+f_{\xi})(1+\bar{f}_{\xi})$. Further, in the third process, the contribution arises from the probability of annihilation of a particle-antiparticle pair of $\phi$ fields and creation of a particle-antiparticle pair of $\xi$ fields ($\phi\bar{\phi}\rightarrow\xi\bar{\xi}$) minus the probability for the reverse process ($\xi\bar{\xi}\rightarrow\phi\bar{\phi}$)
\begin{align}
    \text{Im}\Big(\Sigma_{\phi}(\omega,\pmo)\Big)_{3} =&\, \frac{g^{2}\pi^{4}}{8}\int d\textbf{l}_{1}\, d\textbf{l}_{2}\, d\textbf{l}_{3}\,\delta^{3}(\textbf{l}_{1}+\textbf{l}_{2}+\textbf{l}_{3}-\textbf{p})\times \frac{1}{E_{1}E_{2}E_{3}}\Bigg[\bar{f}_{\phi}(E_{1})\Big(1+\bar{f}_{\xi}(E_{2})\Big)\nonumber\\
    &\times\Big(1+f_{\xi}(E_{3})\Big)-\Big(1+\bar{f}_{\phi}(E_{1})\Big)\bar{f}_{\xi}(E_{2})f_{\xi}(E_{3})\Bigg]\times\delta(E_{3}+E_{2}-E_{1}-\omega).
\end{align}

Fourth term is due to the scattering process between $\phi$ and $\bar{\xi}$ particles from different energy states ($\phi\bar{\xi}\rightarrow\phi\bar{\xi}$). The fifth process is different from the fourth one, by changing $\bar{\xi}\rightarrow\xi$ in the same scattering channels ($\phi\xi\rightarrow\phi\xi$). We have
\begin{align}
    \text{Im}\Big(\Sigma_{\phi}(\omega,\pmo)\Big)_{4} =&\, \frac{g^{2}\pi^{4}}{8}\int d\textbf{l}_{1}\, d\textbf{l}_{2}\, d\textbf{l}_{3}\,\delta^{3}(\textbf{l}_{1}+\textbf{l}_{2}+\textbf{l}_{3}-\textbf{p})\times\frac{1}{E_{1}E_{2}E_{3}}\Bigg[\Big(1+f_{\phi}(E_{1})\Big)\nonumber\\
    &\Big(1+\bar{f}_{\xi}(E_{2})\Big)\bar{f}_{\xi}(E_{3})-f_{\phi}(E_{1})\bar{f}_{\xi}(E_{2})\Big(1+\bar{f}_{\xi}(E_{3})\Big)\Bigg]\nonumber\\
   &\times\delta(E_{3}-E_{3}+E_{1}-\omega),
\end{align}
and
\begin{align}
    \text{Im}\Big(\Sigma_{\phi}(\omega,\pmo)\Big)_{5} =&\, \frac{g^{2}\pi^{4}}{8}\int d\textbf{l}_{1}\, d\textbf{l}_{2}\, d\textbf{l}_{3}\,\delta^{3}(\textbf{l}_{1}+\textbf{l}_{2}+\textbf{l}_{3}-\textbf{p})\times\frac{1}{E_{1}E_{2}E_{3}}\Bigg[\Big(1+f_{\phi}(E_{1})\Big)\nonumber\\
    &f_{\xi}(E_{2})\Big(1+f_{\xi}(E_{3})\Big)-f_{\phi}(E_{1})\Big(1+f_{\xi}(E_{2})\Big)f_{\xi}(E_{3})\Bigg]\nonumber\\
    &\times\delta(E_{3}-E_{2}-E_{1}-\omega),
\end{align}

Similarly, the sixth term arises from the probability of three body inelastic collision of $\phi\bar{\phi}\bar{\xi}\rightarrow\bar{\xi}$ minus the probability $\bar{\xi}\rightarrow\bar{\phi}\bar{\xi}\phi$ process, and the seventh one take place with a replacement of particle to anti-particle of $\xi$ field of the same inelastic process. 
\begin{align}
    \text{Im}\Big(\Sigma_{\phi}(\omega,\pmo)\Big)_{6} =&\, \frac{g^{2}\pi^{4}}{8}\int d\textbf{l}_{1}\,d\textbf{l}_{2}\, d\textbf{l}_{3}\,\delta^{3}(\textbf{l}_{1}+\textbf{l}_{2}+\textbf{l}_{3}-\textbf{p})\times\frac{1}{E_{1}E_{2}E_{3}}\Bigg[\bar{f}_{\phi}(E_{1})\Big(1+\bar{f}_{\xi}(E_{2})\Big)\nonumber\\
    &\times\bar{f}_{\xi}(E_{3})-\Big(1+\bar{f}_{\phi}(E_{1})\Big)\bar{f}_{\xi}(E_{2})\Big(1+\bar{f}_{\xi}(E_{3})\Big)\Bigg]\nonumber\\
    &\times\delta(E_{2}-E_{3}-E_{1}-\omega).
\end{align}
and
\begin{align}
    \text{Im}\Big(\Sigma_{\phi}(\omega,\pmo)\Big)_{7} =&\, \frac{g^{2}\pi^{4}}{8}\int\,d\textbf{l}_{1}\, d\textbf{l}_{2}\, d\textbf{l}_{3}\,\delta^{3}(\textbf{l}_{1}+\textbf{l}_{2}+\textbf{l}_{3}-\textbf{p})\times\frac{1}{E_{1}E_{2}E_{3}}\Bigg[\bar{f}_{\phi}(E_{1})f_{\xi}(E_{2})\nonumber\\
    &\Big(1+f_{\xi}(E_{3})\Big)-\Big(1+\bar{f}_{\phi}(E_{1})\Big)\Big(1+f_{\xi}(E_{2})\Big)f_{\xi}(E_{3})\Bigg]\nonumber\\
    &\times\delta(E_{3}-E_{2}-E_{1}-\omega).
\end{align}
The eighth term comes from the difference of the probability of $(\phi\xi\bar{\xi}\rightarrow\phi)$ and it's reverse processes $(\phi\rightarrow\xi\bar{\xi}\phi)$ 
\begin{align}
    \text{Im}\Big(\Sigma_{\phi}(\omega,\pmo)\Big)_{8} =&\, \frac{g^{2}\pi^{4}}{8}\int d\textbf{l}_{1}\, d\textbf{l}_{2}\, d\textbf{l}_{3}\,\delta^{3}(\textbf{l}_{1}+\textbf{l}_{2}+\textbf{l}_{3}-\textbf{p})\times\frac{1}{E_{1}E_{2}E_{3}}\Bigg[\Big(1+f_{\phi}(E_{1})\Big)f_{\xi}(E_{2})\nonumber\\
    &\bar{f}_{\xi}(E_{3})-f_{\phi}(E_{1})\Big(1+f_{\xi}(E_{2})\Big)\Big(1+\bar{f}_{\xi}(E_{3})\Big)\Bigg]\nonumber\\
    &\times\delta(E_{1}-E_{2}-E_{3}-\omega).\label{lImGph}
\end{align}
But this process is different than the first term due to the difference of processes that takes place in internal lines of sunset diagram Fig.~\ref{fig:FD4b}, which leads to a distinction in statistical weight of internal lines.

Next, we determine the region of $k^{2}=\omega^{2}-\pmo^{2}$ where all the eight physical processes contained in Eq.~\eqref{Gammaphi} are contingent according to the kinematic constraints, i.e., the integral over $l_{1}$ in Eq.~\eqref{Gammaphi} survives. Kinematically first and second term survives if $k^{2}>(m_{\phi}+M_{\xi\phi})^{2}$, where $M_{\xi\phi}$ is the invariant mass of the $\phi$ and $\xi$. Therefore, the processes are possible when $k^{2}>(2m_{\xi}+m_{\phi})^{2}$ since $M_{\xi\phi}>m_{\xi}+m_{\phi}$. The third and eighth terms are allowed when it satisfies the condition $k^{2}>(M_{\xi\xi}-m_{\phi})^{2}>(2m_{\xi}-m_{\phi})^{2}$ since $M_{\xi\xi}>2m_{\xi}$. The similarities of kinematical conditions between third and eighth processes can be understood by changing external incoming/ougoing $\phi$-line to outgoing/incoming $\bar{\phi}$-line in any one of the process. Similarly, fourth, fifth, sixth and seventh term needs to satisfy the same kinematic condition, which is $k^{2}<(M_{\xi\phi}-m_{\xi})^{2}$. The similitude of fourth and fifth processes happens because these processes involve particles and anti-particle of $\xi$ fields, respectively. Furthermore, an equivalence with sixth and seventh processes can be made by changing external incoming/outgoing $\phi$ by its outgoing/incoming $\bar{\phi}$. Therefore,these four scattering processes take place at arbitrary $k^{2}$. A similar calculation can be performed for the $\xi$ field by changing each $\phi$'s propagator to $\xi$'s propagator, and resulting in similar processes can be occurred as described in Eq.~\eqref{fIGph}-\eqref{lImGph} with an appropriate changes in physical parameters under $\phi\leftrightarrow\xi$.

Thus, the thermal width of the quasi-particle profile of the spectral function can be estimated, which is $\Gamma_{a}(E_{a\,\pmo},\pmo)=\frac{\text{Im}(\Sigma_{a}(E_{p},\pmo))}{2E_{\pmo}}$ and at the weak coupling limit its contribution start from $g^{2}$ order.


\section{Euclidean Green's function for the arbitrary bilinear Hermitian bosonic operator.}\label{AppenB}

In this section, we provide some details of the computation of the Euclidean Green's function for two arbitrary bilinear Hermitian bosonic operators $\hat{\mathcal{O}}_{1}$ and $\hat{\mathcal{O}}_{2}$. We consider two such operators,
\begin{align}
    \mathcal{O}_{1}(X)&=\sum_{a}\phi_{a}^{\dagger}\overleftrightarrow{\mathcal{D}}_{1a}(X)\phi_{a},\label{Od1}\\
    \mathcal{O}_{2}(X)&=\sum_{b}\phi_{b}^{\dagger}\overleftrightarrow{\mathcal{D}}_{2b}(X)\phi_{b},\label{Od2}
\end{align}
where, each operator can carry Lorentz and/or particle indices, and that is taken care of by the indices of this $ \overleftrightarrow {\mathcal{D}}_{1}$ and $\overleftrightarrow{\mathcal{D}}_{2}$. In general, $\overleftrightarrow{\mathcal{D}}_{a}$'s can be expressed as polynomials of left and right partial derivative, $\overleftarrow{\partial}_{X}$ and $\overrightarrow{\partial}_{X}$, which acts on $\phi_{a}(X)$ and $(\phi_{a}^{\dagger}(X)$, respectively. For example, if we consider an operator $\hat{\mathcal{N}}^{i}_{a}$ and the operator's structure as given in Eq.~\eqref{Nph}, we obtain $\overleftrightarrow{\mathcal{D}}_{a}=i\sum_{b}\delta_{ba}(\overrightarrow{\partial}^{i}-\overleftarrow{\partial}^{i})$. Similar construction can be done with the diffusion current $\hat{\mathcal{J}}_{a}^{i}$ and the heat flux $\hat{\mathcal{Q}}^{i}$ as given in Eq.~\eqref{dmj} and Eq.~\eqref{Hph}, respectively.

Next, we calculate the Equilibrium Green's function for $\hat{\mathcal{O}}_{1}=\hat{\mathcal{J}}_{a}^{i}$ and $\hat{\mathcal{O}}_{2}=\hat{\mathcal{J}}_{b}^{j}$, as required in Eq.~\eqref{Jabdef}, i.e.,
\begin{align}
    \Delta^{ij}\Big\langle\hat{\mathcal{J}}_{a\,i}(X)\hat{\mathcal{J}}_{b\,j}(X^{\prime})\Big\rangle_{l}&=\Delta^{ij}\Bigg[\Big\langle\hat{\mathcal{N}}_{a\,i}(X)\hat{\mathcal{N}}_{b\,j}(X^{\prime})\Big\rangle_{l}-\frac{n_{b}}{h}\Big\langle\hat{\mathcal{N}}_{a\,i}(X)\hat{\mathcal{Q}}_{j}(X^{\prime})\Big\rangle_{l}\num\\
    &\qquad\quad -\frac{n_{a}}{h}\Big\langle\hat{\mathcal{Q}}_{i}(X)\hat{\mathcal{N}}_{b\,j}(X^{\prime})\Big\rangle_{l} +\frac{n_{a}n_{b}}{h^{2}}\Big\langle\hat{\mathcal{Q}}_{i}(X)\hat{\mathcal{Q}}_{j}(X^{\prime})\Big\rangle_{l}\Bigg],\label{AJab}
\end{align}
where, $a,~b$ are the indices for particle types. Nevertheless, we present a calculation with a generic set of composite bosonic and bilinear operators $\hat{\mathcal{O}}_{1}$ and $\hat{\mathcal{O}}_{2}$. If these operators contain the differential operator as given in Eqs.~\eqref{Od1} and \eqref{Od2}, then in the Euclidean momentum space it will be a function of Matsubara frequency and the 3-momentum,\footnote{By applying the Euclidean Fourier decomposition of fields as Eq.~\eqref{FdPh} one can show that the differential operator becomes a function of Euclidean momentum, i.e. $\mathcal{O}_{1}(X)=\sum_{a}\phi_{a}^{\dagger}\overleftrightarrow{\mathcal{D}}_{1a}\phi_{a}=\sum_{a}e^{i(Q-P)\cdot X}\mathcal{D}_{1a}(P,Q)\Iphi_{a}^{\dagger}(P)\Iphi_{a}(Q)$. If we again take same example $\mathcal{O}_{1}(X)=\mathcal{N}^{i}_{a}(X)$, one can show that $\mathcal{D}_{1a}(P,Q)=(\pmo^{i}+\qm^{i})$.}. Therefore, the two-point Euclidean Green's function of these generic operators can be written as
\begin{align}
    \left\langle\hat{\mathcal{O}}_{1}(X)\hat{\mathcal{O}}_{2}(X^{\prime})\right\rangle&=\sum_{kl}\overleftrightarrow{\mathcal{D}}_{1k}(X)\overleftrightarrow{\mathcal{D}}_{2l}(X^{\prime})\left\langle\phi_{k}^{\dagger}(X)\phi_{k}(X)\phi_{l}^{\dagger}(X^{\prime})\phi_{l}(X^{\prime})\right\rangle\nonumber\\
    &=\sum_{kl}\overleftrightarrow{\mathcal{D}}_{1k}(X)\overleftrightarrow{\mathcal{D}}_{2l}(X^{\prime})\mathcal{G}^{E}_{lk}(X-X^{\prime})\mathcal{G}^{E}_{kl}(X^{\prime}-X).\label{OGE}
\end{align}
Here, one needs to define the action of $\overleftrightarrow{\mathcal{D}}_{1}(X)$ and $\overleftrightarrow{\mathcal{D}}_{2}(X')$ on the Euclidean correlation function $\langle\phi_{a}(X)\phi_{b}^{\dagger}(X^{\prime})\rangle=\mathcal{G}_{ab}^{E}(X^{\prime},X)=\mathcal{G}_{ab}^{E}(X-X^{\prime})$. This can be obtained from the distinctive action of the differential operator $\overleftarrow{\partial}_{X}$ and $\overrightarrow{\partial}_{X}$ on $\phi^{\dagger}_{a}$ and $\phi_{a}$. Hence, the left and right partial derivative $\overleftarrow{\partial}$ and $\overrightarrow{\partial}$ acts on the left and right variable of the arguments of two-point particle Euclidean correlation function $\mathcal{G}_{ab}^{E}(X^{\prime},X)$, respectively.

Functionally, we can show these properties with an example. For instance, a differential operator which contains a linear combination of left and right derivatives, $\overleftrightarrow{\mathcal{D}}(X)=\overrightarrow{\partial}_{X}-\overleftarrow{\partial}_{X}$, operates like
\begin{align}
    \overleftrightarrow{\mathcal{D}}(X)\mathcal{G}^{E}(X^{\prime},X)&=\partial_{X}\mathcal{G}^{E}(X^{\prime},X)\,,\nonumber\\
    \overleftrightarrow{\mathcal{D}}(X)\mathcal{G}^{E}(X,X^{\prime})&=-\partial_{X}\mathcal{G}^{E}(X,X^{\prime})\,,\nonumber\\
    \overleftrightarrow{\mathcal{D}}(X^{\prime})\mathcal{G}^{E}(X^{\prime},X)&=-\partial_{X'}\mathcal{G}^{E}(X^{\prime},X)\,,\nonumber\\
    \overleftrightarrow{\mathcal{D}}(X')\mathcal{G}^{E}(X,X^{\prime})&=\partial_{X'}\mathcal{G}^{E}(X,X^{\prime})\,.
\end{align}
Here, $\partial_{X}=(\partial_{\tau},\partial_{x^{i}})=(\partial_{\tau},\partial_{\textbf{x}})$ and $\partial_{X^{\prime}}=(\partial_{\tau^{\prime}},\partial_{x^{\prime\,i}})=(\partial_{\tau^{\prime}},\partial_{\textbf{x}^{\prime}})$. Using these properties of left/right towards derivative, one can simplify the following correlation functions.

\bigskip

\noindent\textbf{(a)} To evaluate the Euclidean $\hat{\mathcal{N}}_{a}$-$\hat{\mathcal{N}}_{b}$ correlation function, which is first term of Eq.~\eqref{AJab}, we can express Eq.~\eqref{OGE} by using Eqs.~\eqref{Od1}, \eqref{Od2}. This leads to the identification
\begin{align}
    \overleftrightarrow{\mathcal{D}}_{1\,a}(X)&=i\sum_{k}\delta_{k\,a}\Big(\overrightarrow{\partial}_{x^{i}}-\overleftarrow{\partial}_{x^{i}}\Big)\,, \\
    \overleftrightarrow{\mathcal{D}}_{2\,b}(X^{\prime})&=i\sum_{l}\delta_{lb}\Big(\overrightarrow{\partial}_{x^{\prime i}}-\overleftarrow{\partial}_{x^{\prime i}}\Big)\,.
\end{align}
Using these differential operators, one can write
\begin{align}
    &\Delta^{ij}\Big\langle \hat{\mathcal{N}}_{a\,i}(X)\hat{\mathcal{N}}_{b\,j}(X^{\prime})\Big\rangle_{E}\nonumber\\
    &=-\delta_{ab}\Delta^{ij}\Big(\overrightarrow{\partial}_{x^{i}}-\overleftarrow{\partial}_{x^{i}}\Big)\Big(\overrightarrow{\partial}_{x^{\prime j}}-\overleftarrow{\partial}_{x^{\prime j}}\Big)\mathcal{G}_{aa}^{E}(X^{\prime},X)\mathcal{G}_{aa}^{E}(X,X^{\prime})\nonumber\\
    &=-\delta_{ab}\Delta^{ij}\Big(\partial_{x^{i}}\mathcal{G}_{aa}^{E}(X^{\prime},X)\partial_{x^{\prime j}}\mathcal{G}_{aa}^{E}(X,X^{\prime})+\partial_{x^{\prime j}}\mathcal{G}_{aa}^{E}(X^{\prime},X)\partial_{x^{i}}\mathcal{G}_{aa}^{E}(X,X^{\prime})\nonumber\\
&\quad-\partial_{x^{\prime j}}\partial_{x^{i}}\mathcal{G}_{aa}^{E}(X^{\prime},X)\mathcal{G}_{aa}^{E}(X,X^{\prime})-\mathcal{G}_{aa}^{E}(X^{\prime},X)\partial_{x^{\prime j}}\partial_{x^{i}}\mathcal{G}_{aa}^{E}(X,X^{\prime})\Big).
\end{align}
Moreover, by introducing the Fourier decomposition of the two-point Euclidean correlation function, Eqs.~\eqref{FGE} and \eqref{GEtoSpd}, one can further simplify it as
\begin{align}
\Delta^{ij}\Big\langle \hat{\mathcal{N}}_{a\,i}(X)\hat{\mathcal{N}}_{b\,j}(X^{\prime})\Big\rangle_{E} = \delta_{ab}\sumintb_{P L }\,e^{iL\cdot (X-X^{\prime}) }&\, \Delta_{ij}(2\pmo^{j}-\lm^{j}) (2\pmo^{i}-\lm^{i}) \int \frac{d\omega_{1}}{\pi} \frac{d\omega_{2}}{\pi} \nonumber\\
& \times \left[\frac{\rho_{a}(\omega_{1},\pmo)\rho_{a}(\omega_{1},\pmo-\lm)}{\Big(\omega_{1}+i\tilde{p}_{n}\Big)\Big(\omega_{2}+i(\tilde{p}_{n}-l_{n})\Big)}\right], 
\end{align}
where the chemical potential dependence is hidden in the variable $\tilde{p}_{n}$ defined as $\tilde{p}_{n}=p_{n}+i\mu_{a}$. We now sum over the Matsubara frequency $p_{n}$ by using the identities in Eqs.~\eqref{Ident1} and\eqref{Ident2}, leading to the Euclidean correlation function $\mathcal{G}^{E}_{\mathcal{N}_{a}\mathcal{N}_{a}}$ of Eq.~\eqref{GJJEL} as
\begin{align}
     \mathcal{G}^{E}_{\mathcal{N}_{a}\mathcal{N}_{a}}(L)
      =&-\int d\pmo(2\pmo-\lm)^{2}\int\frac{d\omega_{1}}{\pi} \frac{d\omega_{2}}{\pi}\rho_{a}(\omega_{1},\pmo)\rho_{a}(\omega_{2},\pmo-\lm)\frac{f_{a}(\omega_{1})-f_{a}(\omega_{2})}{\omega_{2}-\omega_{1}+il_{n}}.\label{GENN}
\end{align}
Here, $f_{a}(\omega)=\big(\text{exp}\beta(\omega-\mu_{a})-1\big)^{-1}$ is the thermal distribution function of `a' type particle.

\bigskip

\noindent\textbf{(b)}  Similar calculation can be done for the second mixed correlation function of particle and heat current correlation function $\mathcal{N}_{a}$-$\mathcal{Q}$ of the Eq.~\eqref{AJab}. With the help of the Eqs.~\eqref{Od1} and \eqref{Od2}, it is easy to identify
\begin{align}
    \overleftrightarrow{\mathcal{D}}_{1}(X)&=i\Big(\overrightarrow{\partial}_{x^{i}}-\overleftarrow{\partial}_{x^{i}}\Big)\,, \\
    \overleftrightarrow{\mathcal{D}}_{2}(X^{\prime})&=i\Big(\overleftarrow{\partial_{\tau^{\prime}}}\,\overrightarrow{\partial}_{x^{\prime i}}+\overleftarrow{\partial}_{x^{\prime i}}\,\overrightarrow{\partial_{\tau^{\prime}}}\Big)\,.
\end{align}
With these differential operators, one can compute Euclidean Green's function in a similar manner as
\begin{align}
\mathcal{G}^{E}_{\mathcal{N}_{a}\mathcal{Q}}(L) =&\sum_{k}\delta_{ak}\Delta^{ij}\int d\pmo\,\int \frac{d\omega_{1}}{\pi}\frac{d\omega_{2}}{\pi}\Big[\omega_{1}(\pmo_{i}-\lm_{i})+\omega_{2}\pmo_{i}\Big] \Big(2\pmo_{j}-\lm_{j}\Big) \nonumber\\
&\quad\quad\quad\quad\quad\quad\quad\quad
\times\bigg[\rho_{k}(\omega_{2},\pmo-\lm)\,\rho_{k}(\omega_{1},\pmo)\,\frac{f_{k}(\omega_{1})-f_{k}(\omega_{2})}{\omega_{2}-\omega_{1}+il_{n}}\bigg]\, .\label{GENQ}
\end{align}
Likewise, the third mixed correlation function of Eq.~\eqref{AJab} can be written as
\begin{align}
\mathcal{G}^{E}_{\mathcal{Q}\mathcal{N}_{a}}(L) =&\sum_{k}\delta_{ak}\Delta^{ij}\int d\pmo\,\int \frac{d\omega_{1}}{\pi}\frac{d\omega_{2}}{\pi}\Big[\omega_{1}(\pmo_{j}-\lm_{j})+\omega_{2}\pmo_{j}\Big] \Big(2\pmo_{i}-\lm_{i}\Big)\nonumber\\
&\quad\quad\quad\quad\quad\quad\quad\quad\quad
\times \bigg[ \rho_{k}(\omega_{2},\pmo-\lm)\, \rho_{k}(\omega_{1},\pmo)\, \frac{f_{k}(\omega_{1})-f_{k}(\omega_{2})}{\omega_{2}-\omega_{1}+il_{n}}\bigg]\, .\label{GEQN}
\end{align}
Owing to the symmetry properties of the $\Delta^{ij}$, the second and third $\mathcal{N}_{a}$-$\mathcal{Q}$ mixed correlation functions are  equal.

\bigskip

\noindent\textbf{(c)} The last $\mathcal{Q}$-$\mathcal{Q}$ correlation function can be written in a similar manner
\begin{align}
\mathcal{G}^{E}_{\mathcal{Q}\mathcal{Q}}(L)&=\sum_{k}\Delta^{ij}\int\,d\pmo\int \frac{d\omega_{1}}{\pi}\frac{d\omega_{2}}{\pi}\Big[\omega_{1}(\pmo_{i}-\lm_{i})+\omega_{2}\pmo_{i}\Big]\Big[\omega_{1}(\pmo_{j}-\lm_{j})+\omega_{2}\pmo_{j}\Big]\nonumber\\
&\quad\quad\quad\quad\quad\quad\quad\quad\quad\quad\quad\quad
\times \bigg[\rho_{k}(\omega_{2},\pmo-\lm)\, \rho_{k}(\omega_{1},\pmo)\, \frac{f_{k}(\omega_{1})-f_{k}(\omega_{2})}{\omega_{2}-\omega_{1}+il_{n}}\bigg]\, .\label{GEQQ}
\end{align}
With the help of Eq.~\eqref{spO}, we evaluate the spectral function of each of the above-mentioned Euclidean Green's functions and the spectral function of $\mathcal{J}_{a}$-$\mathcal{J}_{b}$. As mentioned in Eq.~\eqref{spO}, one needs to employ the $\text{Disc}$ function on all of these Green's functions in Eqs.~\eqref{GENN}, \eqref{GENQ}, \eqref{GEQN} and \eqref{GEQQ}. The operation of the $\text{Disc}$ function will act on the part which contains a discontinuity of these Green's functions while taking the limit $il_{n}\rightarrow\omega\pm i0^{+}$. Such part in these Green's function is $\Big(\frac{1}{\omega_{2}-\omega_{1}+il_{n}}\Big)$, which makes them discontinuous across the real $\omega$ axis
\begin{align}
    \text{Disc}\Big(\frac{1}{\omega_{2}-\omega_{1}+il_{n}}\Big)=&\Bigg(\frac{1}{\omega_{2}-\omega_{1}+\omega+i0^{+}}-\frac{1}{\omega_{2}-\omega_{1}+\omega-i0^{+}}\Bigg)\nonumber\\
    =&-2i\pi\delta(\omega_{2}-\omega_{1}+\omega)\,.
\end{align}
Applying this identity, we derive the spectral functions for all of these Green's functions, which are given by 
\begin{align}
    \rho_{\mathcal{N}_{a}\mathcal{N}_{b}}(\omega,\lm)
      =&-\delta_{ab}\Delta_{ij}\int d\pmo(2\pmo^{j}-\lm^{j}) (2\pmo^{i}-\lm^{i})\int\frac{d\omega^{\prime}}{\pi}\,\rho_{a}(\omega^{\prime}+\omega,\pmo)\,\rho_{a}(\omega^{\prime},\pmo-\lm)\nonumber\\ &\quad\quad\quad\quad\quad\quad\quad\quad\quad\quad\quad\quad\quad\quad\quad\quad\times\Big[f_{a}(\omega^{\prime}+\omega)-f_{a}(\omega^{\prime})\Big],\\
      \rho_{\mathcal{N}_{a}\mathcal{Q}}(\omega,\lm)
      =&-\sum_{k}\delta_{ak}\Delta^{ij}\int d\pmo\,\int \frac{d\omega^{\prime}}{\pi}\Big[(\omega^{\prime}+\omega)(\pmo_{i}-\lm_{i})+\omega^{\prime}\pmo_{i}\Big] \Big(2\pmo_{j}-\lm_{j}\Big)\nonumber\\
      &\quad\quad\quad\quad\quad \times\Bigg[\rho_{k}(\omega^{\prime},\pmo-\lm)\,\rho_{k}(\omega^{\prime}+\omega,\pmo)\Big(f_{k}(\omega^{\prime}+\omega)-f_{k}(\omega^{\prime})\Big)\Bigg]\,,\\
      \rho_{\mathcal{Q}\mathcal{Q}}(\omega,\lm)=&-\sum_{k}\Delta^{ij}\int\,d\pmo\int \frac{d\omega^{\prime}}{\pi}\big((\omega^{\prime}+\omega)(\pmo_{i}-\lm_{i})+\omega^{\prime}\pmo_{i}\big)\big((\omega^{\prime}+\omega)(\pmo_{j}-\lm_{j})+\omega^{\prime}\pmo_{j}\big)\nonumber\\
      &\quad\quad\quad\quad\quad \times \Bigg[ \rho_{k}(\omega^{\prime},\pmo-\lm)\,\rho_{k}(\omega^{\prime}+\omega,\pmo)\,\Big(f_{k}(\omega^{\prime}+\omega)-f_{k}(\omega^{\prime})\Big)\Bigg]\, .
\end{align}
Adding these spectral functions according to the Eq.~\eqref{AJab}, we obtain the $\mathcal{J}_{a}$-$\mathcal{J}_{b}$ spectral function as
\begin{align}
    \rho_{\mathcal{J}_{a}\mathcal{J}_{b}}(\omega,\lm)=&\delta_{ab}\rho_{\mathcal{N}_{a}\mathcal{N}_{a}}(\omega,\lm)-\frac{n_{b}}{h}\rho_{\mathcal{N}_{a}\mathcal{Q}}(\omega,\lm)-\frac{n_{a}}{h}\rho_{\mathcal{N}_{b}\mathcal{Q}}(\omega,\lm)+\frac{n_{a}n_{b}}{h^{2}}\rho_{\mathcal{Q}\mathcal{Q}}(\omega,\lm)\,.\label{defSAJJ}
\end{align}
These results are used in Sec-\ref{sec:transport} to obtain the diffusion coefficient in Eq.~\eqref{dspmjj}.
Similar calculation can be done for the fermion fields, a similar method of calculation can be done with only difference,
\begin{equation}
    \mathcal{G}^{E}(X,X^{\prime})=-\mathcal{G}^{E}(X^{\prime},X)
\end{equation}
which is necessary to taken care of anti-commuting nature of fermion fields at equal time.\\
In a similar footing, one can compute the $\mathcal{J}_{a}$-$\mathcal{J}_{b}$ spectral function in terms of fermion particle spectral functions through generalization of the Eq.\eqref{defSAJJ}, which are 
\begin{align}
    \rho_{\mathcal{N}_{a}\mathcal{N}_{a}}(\omega,\lm)&=-\Delta^{ij}\pi\int d\pmo\int\frac{dp^{0}}{\pi}\frac{dk^{0}}{\pi}\delta(k_{0}-p_{0}-\omega)\Bigg[\text{Tr}\Big(\gamma_{i}\rho^{F}_{a}(p_{0},\textbf{p})\gamma_{j}\rho^{F}_{a}(k_{0},\textbf{p-l})\Big)\nonumber\\
    &\Big(f_{a}(k_{0})-f_{a}(p_{0})\Big)\Bigg]\,,
\end{align}
\begin{align}
    \rho_{\mathcal{N}_{a}\mathcal{Q}}(\omega,\lm)&=-\Delta^{ij}\frac{\pi}{4}\sum_{k}\delta_{ak}\int d\pmo\int\frac{dp^{0}}{\pi}\frac{dk^{0}}{\pi}\delta(k_{0}-p_{0}-\omega)\Bigg[-\pmo_{j}\text{Tr}\Big(\gamma_{0}\rho^{F}_{k}(p_{0},\textbf{p})\gamma_{j}\rho^{F}_{k}(k_{0},\textbf{p-l})\Big)\nonumber\\
    &+p_{0}\text{Tr}\Big(\gamma_{i}\rho^{F}_{k}(p_{0},\textbf{p})\gamma_{j}\rho^{F}_{k}(k_{0},\textbf{p-l})\Big)\Bigg]\Big(f_{k}(k_{0})-f_{k}(p_{0})\Big)\,,
\end{align}
and,
\begin{align}
    \rho_{\mathcal{Q}\mathcal{Q}}(\omega,\lm)&=-\Delta^{ij}\frac{\pi}{4^2}\sum_{k}\int d\pmo\int\frac{dp^{0}}{\pi}\frac{dk^{0}}{\pi}\delta(k_{0}-p_{0}-\omega)\Bigg[\big(2\pmo_{i}-\lm_{i}\big) \big(2\pmo_{j}-\lm_{j}\big)\text{Tr}\Big(\gamma_{0}\rho^{F}_{k}(p_{0},\textbf{p})\gamma_{0}\rho^{F}_{k}(k_{0},\textbf{p-l})\Big)\nonumber\\
    &-\big(p_{0}+k_{0}\big) \Bigg(\big(2\pmo_{i}-\lm_{i}\big)\text{Tr}\Big(\gamma_{0}\rho^{F}_{k}(p_{0},\textbf{p})\gamma_{j}\rho^{F}_{k}(k_{0},\textbf{p-l})\Big)+ \big(2\pmo_{j}-\lm_{j}\big)\text{Tr}\Big(\gamma_{i}\rho^{F}_{k}(p_{0},\textbf{p})\gamma_{0}\rho^{F}_{k}(k_{0},\textbf{p-l})\Big)\Bigg)\nonumber\\
    &+\big(p_{0}+k_{0}\big)^{2}\text{Tr}\Big(\gamma_{i}\rho^{F}_{k}(p_{0},\textbf{p})\gamma_{j}\rho^{F}_{k}(k_{0},\textbf{p-l})\Big)\Bigg]\Big(f_{k}(k_{0})-f_{k}(p_{0})\Big)\,.
\end{align}
Here, $f_{a}(k_{0})=\big(\text{exp}\beta(k_{0}-\mu_{a})+1\big)^{-1}$ is the thermal distribution function for the `a'-type fermion particles. To get the final expression one needs to employ this identity,
\begin{align}
   \text{Tr}\Big(\gamma_{\mu}\rho^{F}_{k}(p_{0},\textbf{p})\gamma_{\nu}\rho^{F}_{l}(k_{0},\textbf{k})\Big)&=\delta_{kl}\text{Tr}\Big(\gamma_{\mu}\left(\slashed{\mathcal{P}} + m\right)\gamma_{\alpha}\left(\slashed{\mathcal{K}} + m\right)\Big)\rho^{0}_{k}(p_{0},\textbf{p})\rho^{0}_{k}(k_{0},\textbf{k})\nonumber\\
    &=\delta_{kl}\Big[\Big(\mathcal{P}_{\mu}\mathcal{K}_{\alpha}+\mathcal{P}_{\alpha}\mathcal{K}_{\mu}\Big)-g_{\mu\alpha}\Big(\mathcal{P}\cdot\mathcal{K}-m^{2}\Big)\Big]\rho^{0}_{k}(p_{0},\textbf{p})\rho^{0}_{k}(k_{0},\textbf{k})\,,
\end{align}
where, $\mathcal{P}_{\mu}=(p_{0},\pmo)$ is the four vector. These results are helpful for the evaluation of the diffusion coefficients for fermionic system, leading to the Eq.\eqref{KABf} by using the Eq.~\eqref{dspmjj}. A similar calculation can be done for $\mathcal{T}^{\mu\nu}$-$\mathcal{T}_{\mu\nu}$ correlation function as explained in Eq.\eqref{kuboeta} to extract the shear and bulk viscosity for bosonic and fermionic system.
\section{Some useful integrals:}\label{Integrals}
In order to extract the transport coefficients from expressions like Eq.\eqref{Kphiphi}-\eqref{KAB}, one encounters following three kind of integration, such are
\begin{align}
    \mathcal{G}^{k}_{n}(z,\alpha)=\frac{1}{\Gamma(n)}\int_{0}^{\infty}dx\,x^{n-1}\Big(f^{k}(x,z,\alpha)\big(1+k f^{k}(x,z,\alpha)\big)+\bar{f}^{k}(x,z,\alpha)\big(1+k \bar{f}^{k}(x,z,\alpha)\big)\Big)\,,
\end{align}
\begin{align}
    \mathcal{H}^{k}_{n}(z,\alpha)=\frac{1}{\Gamma(n)}\int_{0}^{\infty}dx\,\frac{x^{n-1}}{\sqrt{x^2+z^2}}\Big(f^{k}(x,z,\alpha)\big(1+k f^{k}(x,z,\alpha)\big)-\bar{f}^{k}(x,z,\alpha)\big(1+k \bar{f}^{k}(x,z,\alpha)\big)\Big)
\end{align}
and,
\begin{align}
    \mathcal{I}^{k}_{n}(z,\alpha)=\int_{0}^{\infty}dx\,\frac{x^{n}}{x^2+z^2}\Big(f^{k}(x,z,\alpha)\big(1+k f^{k}(x,z,\alpha)\big)+\bar{f}^{k}(x,z,\alpha)\big(1+k \bar{f}^{k}(x,z,\alpha)\big)\Big)\,.
\end{align}
Where, $f^{k}(x,z,\alpha)=\frac{1}{\text{exp}\big(\sqrt{x^2+z^2}-\alpha\big)+k}$ and $\bar{f}^{k}(x,z,\alpha)=\frac{1}{\text{exp}\big(\sqrt{x^2+z^2}+\alpha\big)+k}$ are the thermal equilibrium distribution for boson and fermion for $k=-1$ and $k=+1$ respectively, also these integrals are function of these dimensionless parameters $\alpha=\frac{\mu}{T}$ and $z=\frac{m}{T}$.
Further using the following identities,
\begin{align}
    \partial_{\alpha}f^{k}(x,z,\alpha)&=f^{k}(x,z,\alpha)\Big(1+k f^{k}(x,z,\alpha)\Big)\\
    \partial_{\alpha}\bar{f}^{k}(x,z,\alpha)&=\bar{f}^{k}(x,z,\alpha)\Big(1+k \bar{f}^{k}(x,z,\alpha)\Big)\, ,
\end{align}
one can express these integrals as a derivative of the new set of integrals, which are,
\begin{align}
    \mathcal{G}^{k}_{n}(z,\alpha)&=\partial_{\alpha}G^{k}_{n}(z,\alpha)\,,\\
    \mathcal{H}^{k}_{n}(z,\alpha)&=\partial_{\alpha}H^{k}_{n}(z,\alpha)\,,\\
    \mathcal{I}^{k}_{n}(z,\alpha)&=\partial_{\alpha}I^{k}_{n}(z,\alpha)\,,\\
\end{align}
where,
\begin{align}
    G^{k}_{n}(z,\alpha)&=\frac{1}{\Gamma(n)}\int_{0}^{\infty}dx\,x^{n-1}\Big(f^{k}(x,z,\alpha)-\bar{f}^{k}(x,z,\alpha)\Big)\,,\label{Gn}\\
    H^{k}_{n}(z,\alpha)&=\frac{1}{\Gamma(n)}\int_{0}^{\infty}dx\,\frac{x^{n-1}}{\sqrt{x^2+z^2}}\Big(f^{k}(x,z,\alpha)+\bar{f}^{k}(x,z,\alpha)\Big)\,,\label{Hn}\\
    I^{k}_{n}(z,\alpha)&=\int_{0}^{\infty}dx\,\frac{x^{n-1}}{x^2+z^2}\Big(f^{k}(x,z,\alpha)-\bar{f}^{k}(x,z,\alpha)\Big)\,.\label{In}
\end{align}
To analyse the integral in the domain $|\alpha|\leq z$ as per our interest in the context of physical situations, particularly for bosonic systems; one can introduce the variable $r=\frac{\alpha}{z}$ $(|r|\leq1)$.
Three recursion relations among these three types of integral can be derived, which are
\begin{align}
    \partial_{z}G^{k}_{n+1}(z,r)&= n r H^{k}_{n+1}(z,r)-\frac{z}{n}G^{k}_{n-1}(z,r)+\frac{z^2 r}{l} H_{n-1}^{k}(z,r)\,,\label{recG}\\
    \partial_{z}H^{k}_{n+1}(z,r)&=\frac{r}{n}G^{k}_{n-1}(z,r)-\frac{z}{n}H_{n-1}(z,r)\,,\label{recH}\\
    I^{k}_{n+2}(z,r)&=\Gamma(n+1)G_{n+1}^{k}(z,r)-z^{2}I^{k}_{n}(z,r)\,.\label{recI}
\end{align}
Consequently, knowledge of the lower moments of $G^{k}_{n}(z,r)$ and $H^{k}_{n}(z,r)$ yields the higher moments $G^{k}_{n+1}(z,r)$ and $H^{k}_{n+1}(z,r)$ through the differential equation with respect to $z$ with the boundary condition,
\begin{align}
    G^{k}_{n}(0,0)&=0\,\,(n>0)\,,\\
    H^{k}_{n}(0,0)&=\frac{2 \text{Li}_{n-1}(-k)}{k(1-n)}\,\,(n>2)\,.\label{Hboun}
\end{align}
Hence, $G_{1}^{k}(z,r)$ and $H^{k}_{1}(z,r)$ can generate all the integrals $G^{k}_{n}(z,r)$ and $H^{k}_{n}(z,r)$ for any positive odd values of $n$\footnote{For even-numbered dimensional spacetime, $H^{\pm}_{n}$ and $G^{\pm}_{n}$ integrals are required for positive and odd values of $n$. On the other hand, for even positive $n$, these two integrals $H_{n}^{\pm}$ and $G^{\pm}_{n}$ can be computed from recursion relations for given $H_{0}^{\pm}$ and $G^{\pm}_{0}$ respectively, with the boundary conditions as given in Eq.\eqref{Hboun}.}.
In a similar manner, higher moments of integral  $I^{k}_{n+1}(z,r)$ are connected to the lower moments of $I^{k}_{n}(z,r)$ and $G^{k}_{n}(z,r)$ through the polynomial expression, and $I^{-}_{0}(z,r)$ is the lowest moments of the integral.
\subsection{Bosonic case $(k=-1)$}
In the context of massive bosonic systems with finite chemical potential has been well described in the Ref.\cite{Haber:1981fg}, and condensation-related subtleties were taken care of by introducing a dimensionless variable $\alpha=r z$, which helps to define the physical region $|r|\leq 1$ and $z>0$ of these integrals. In the following subsection, we adopted the method of integration as given in Ref.\cite{Haber:1981ts} to tackle the branch cuts that appear in these integrals Eq.\eqref{Gn}-\eqref{In} at high-temperature expansion.
\subsubsection{$G^{-}_{1}(z,r)$ and $H^{-}_{1}(z,r)$ integral at $z\ll 1$ limit :}
The high-temperature expansion can be done by expansion of $z\ll 1$ ( for any $|r|\leq 1$) of integrands in Eqs.\eqref{Gn}-\eqref{In}. A straightforward Taylor's series  expansion gives the diverging coefficients, in the resolution we employ the identity,
\begin{align}
    \frac{1}{e^{y}-1}=\frac{1}{y}-\frac{1}{2}+2\sum_{n=1}^{\infty}\frac{y}{y^2+(2\pi n)^2}\,.\label{ideney}
\end{align}
Using this expansion, one can list the expressions for the particle and antiparticle, respectively, which are,
\begin{align}
    f^{-}(x,z,r)=\frac{1}{\sqrt{x^2+z^2}-rz}-\frac{1}{2}+2\sum_{n=1}^{\infty}\frac{\sqrt{x^2+z^2}-rz}{(\sqrt{x^2+z^2}-rz)^2+(2\pi n)^2}\, \label{particleexpan},
\end{align}
\begin{align}
    \bar{f}^{-}(x,z,r)=\frac{1}{\sqrt{x^2+z^2}+rz}-\frac{1}{2}+2\sum_{n=1}^{\infty}\frac{\sqrt{x^2+z^2}+rz}{(\sqrt{x^2+z^2}+rz)^2+(2\pi n)^2}\,.\label{antiparticleexpan}
\end{align}
Applying these expansions in to the integrand of Eq.\eqref{Gn} after replacing the variable $\alpha=r z$, which can be expressed as 
\begin{align}
    G^{-}_{1}(z,r)=G^{-}_{(0)1}(z,r)+2\sum_{n=1}^{\infty}G^{-}_{(n)1}(z,r)\,,
\end{align}
where, 
\begin{align}
    G^{-}_{(0)1}(z,r)&=2rz\int_{0}^{\infty}dx\,\frac{x^{-\epsilon}}{x^2+z^2(1-r^2)}=\pi  r \left(1-r^2\right)^{-\frac{\epsilon }{2}-\frac{1}{2}} z^{-\epsilon } \sec \left(\frac{\pi  \epsilon }{2}\right)\\
    G^{-}_{(n)1}(z,r)&=2rz\int_{0}^{\infty}dx\,x^{-\epsilon}\frac{x^{2}+z^{2}(1-r^{2})-(2\pi n)^2}{(x^{2}+z^{2}(1-r^{2})+(2\pi n)^2)^{2}+(4\pi n r z)^2}\,\\
    &=-r z 4^{-\frac{\epsilon }{2}-\frac{1}{2}} \pi ^{-\epsilon} \epsilon  \left(n^2\right)^{-\frac{\epsilon }{2}-\frac{1}{2}} \sec \left(\frac{\pi  \epsilon }{2}\right)+\frac{1}{3} r z^3 2^{-(\epsilon +4)} \pi ^{-(\epsilon+2)} \nonumber\\
    &\times(\epsilon +1) (\epsilon +2) \left(n^2\right)^{-\frac{\epsilon }{2}-\frac{3}{2}} \left(r^2 \epsilon +3\right) \sec \left(\frac{\pi  \epsilon }{2}\right)+O\left(z^5\right)
\end{align}
Here, introducing converging term $x^{-\epsilon} (0<\epsilon<1)$ in the integrand helps to ensure the finiteness of the integration over $x$, and eventually one can safely take the limit $\epsilon\rightarrow0$. Moreover, summing over $n$ on $G^{-}_{(n)1}(z,r)$ yields the result up to the order $\mathcal{O}(z^3)$, listed as,
\begin{align}
    \sum_{n=1}^{\infty}G^{-}_{(n)1}(z,r)&=r \left(-2^{-(\epsilon +1)}\right) \pi ^{-\epsilon } \epsilon  \zeta (\epsilon +1) \sec \left(\frac{\pi  \epsilon }{2}\right) z+\frac{1}{3} r 2^{-(\epsilon +4)} \pi ^{-(\epsilon+2)} \left(\epsilon ^2+3 \epsilon +2\right) \\\nonumber
    &\times\left(r^2 \epsilon +3\right) \zeta (\epsilon +3) \sec \left(\frac{\pi  \epsilon }{2}\right) z^3+O\left(z^5\right)
\end{align}
Hence, after taking the limit $\epsilon\rightarrow0$ one can evaluate the the integral $G^{-}_{1}(z,r)$, given as
\begin{align}
    G^{-}_{1}(z,r)&=\frac{\pi  r}{\sqrt{1-r^2}}-r z+\sum_{k=1}^{\infty}(-1)^{k+1}a_{k}\zeta(2k+1)z^{2k+1}\,,
\end{align}
where, the coefficients of the term up to $\mathcal{O}(z^7)$ are listed as given below, 
\begin{align}
    a_{1}(r)&=\frac{r \zeta (3)}{4 \pi ^2}\,,\nonumber\\
    a_{2}(r)&=-\frac{r \left(4 r^2+3\right)  \zeta (5)}{32 \pi ^4}\,,\nonumber\\
    a_{3}(r)&=\frac{3 r \left(8 r^4+20 r^2+5\right) \zeta (7)}{512 \pi ^6}\,.
\end{align}
One must notice that the first term captures the branch cut related to confining our result in the domain of $|r|\leq1$.\footnote{Here one must notice that $G^{-}_{1}(z,r)$ is not continuous at $z=0$, but has definite limit $z\rightarrow0^{+}$ from positive side. }

Turning our interest to compute the integral $H^{-}_{1}(z,r)$, applying the Eqs.\eqref{Hn} and \eqref{ideney} helps to derive the expression as given in,
\begin{align}
    H^{-}_{1}(z,r)=H^{\prime-}_{1}(z,r)+H^{\prime\prime-}_{1}(z,r)+2\sum_{n=0}^{\infty}H^{-}_{(n)1}(z,r)\,,
\end{align}
where,
\begin{align}
    H^{\prime-}_{1}(z,r)&=2\int_{0}^{\infty}dx\,\frac{x^{-\epsilon}}{x^2+z^2(1-r^2)}=\pi  \left(1-r^2\right)^{-\frac{\epsilon }{2}-\frac{1}{2}} z^{-\epsilon -1} \sec \left(\frac{\pi  \epsilon }{2}\right)\,,\\
    H^{\prime\prime-}_{1}(z,r)&=-\int_{0}^{\infty}dx\,\frac{x^{-\epsilon}}{\sqrt{x^2+z^2}}=-\frac{z^{-\epsilon } \Gamma \left(\frac{1}{2}-\frac{\epsilon }{2}\right) \Gamma \left(\frac{\epsilon }{2}\right)}{2 \sqrt{\pi }}\,,\\
    H^{-}_{(n)1}(z,r)&=2\int^{\infty}_{0}dx\,\frac{x^{-\epsilon } \left(4 \pi ^2 n^2+\left(1-r^2\right) z^2+x^2\right)}{\left(4 \pi ^2 n^2+\left(1-r^2\right) z^2+x^2\right)^2+16 \pi ^2 n^2 r^2 z^2}\nonumber\,,\\
    &=\frac{1}{(2\pi)^{(\epsilon+1)}}n^{-(\epsilon+1)}  \pi \sec \left(\frac{\pi  \epsilon }{2}\right)-z^2 2^{-(\epsilon +4)} \pi ^{-(\epsilon +2)} (\epsilon +1)n^{-(\epsilon +3)}\\
    &\times  \left(r^2 (\epsilon +2)+1\right) \sec \left(\frac{\pi  \epsilon }{2}\right)+z^4 2^{-(\epsilon +8)}\pi ^{-(\epsilon+4)}\frac{(\epsilon +1) (\epsilon +3)  }{3 n^{\epsilon+5}}\nonumber\\
    &\times\big(r^2 (\epsilon +4) \left(r^2 (\epsilon +2)+6\right)+3\big) \sec \left(\frac{\pi  \epsilon }{2}\right)+\mathcal{O}(z^6)\,.
\end{align}
Furthermore, summing over $n$ of $H^{-}_{(n)1}(z,r)$ gives,
\begin{align}
    \sum_{n=0}^{\infty}H^{-}_{(n)1}(z,r)&=\frac{1}{2^{\epsilon +1} \pi ^{\epsilon }} \zeta (\epsilon +1) \sec \left(\frac{\pi  \epsilon }{2}\right)-\frac{z^2}{ 2^{2 \epsilon +4} \pi ^{2 \epsilon +2} }(\epsilon +1) \Big((r^2 \epsilon+1)  (2 \pi )^{\epsilon }+r^2 2^{\epsilon +1} \pi ^{\epsilon }\Big)\nonumber\\
    &\zeta (\epsilon +3) \sec \left(\frac{\pi  \epsilon }{2}\right)+\frac{z^4}{2^{\epsilon+8} \pi ^{\epsilon+4}} \frac{(\epsilon +1) (\epsilon +3)}{3}\Big(r^2 (\epsilon +4) \left(r^2 (\epsilon +2)+6\right)+3\Big)\nonumber\\
    &\times\zeta (\epsilon +5) \sec \left(\frac{\pi  \epsilon }{2}\right)+\mathcal{O}(z^6)\,.
\end{align}
One can see that the leading order term in  $ H^{-}_{(n)1}(z,r)$, which is the coefficient of the $z^0$ term along with $H^{\prime\prime-}_{1}(z,r)$, contains a diverging term at $\epsilon\rightarrow0$ due to the presence of a simple pole $\frac{1}{\epsilon}$, as shown below,
\begin{align}
   H^{\prime\prime-}_{1}(z,r) &=-\frac{1}{\epsilon }-(\log (2)-\log (z))+O\left(\epsilon ^1\right)\,,\nonumber\\
   2\sum_{n=0}^{\infty}H^{-}_{(n)1}(z,r)|_{z\rightarrow0}&=\frac{1}{\epsilon }+(\gamma -\log (2 \pi ))+O\left(\epsilon ^1\right)\,,
\end{align}
here, $\gamma$ is the Euler constant.
Hence, $H^{-}_{1}(z,r)$ does not diverge because two simple pole terms with opposite signs cancel each other, and safely taking limit $\epsilon\rightarrow0$ can be performed eventually, one can arrive at,
\begin{align}
    H_{1}^{-}(r,z)&=\frac{\pi }{z\sqrt{1-r^2} }+\log \left(\frac{z}{4 \pi }\right)+\gamma-\frac{\left(2 r^2+1\right) \zeta (3)}{8 \pi ^2}z^2+\frac{\left(8 \left(r^2+3\right) r^2+3\right) \zeta (5)}{128 \pi ^4}z^4\nonumber\\
    &-\frac{\left(16 r^6+120 r^4+90 r^2+5\right) \zeta (7)}{1024 \pi ^6}z^6+\mathcal{O}(z^8)\,.
\end{align}
A noteworthy point is that the first and second terms in the last expression contain the branch cut to confine our result within the domain $|r|\leq 1$ and $z>0$, respectively.
\subsubsection{$I^{-}_{0}(z,r)$ integral at $z\ll 1$ limit :}
Focusing our analysis on the computation of the integral $I^{-}_{0}(z,r)$ which is defined in Eq.\eqref{In} (here $\alpha=r z$); by using the expansions given in Eq.\eqref{particleexpan} and Eq.\eqref{antiparticleexpan}, one may derive it in the following form,
\begin{align}
    I^{-}_{0}(z,r)=I^{-}_{(0)0}(z,r)+2\sum_{n=1}^{\infty}I^{-}_{(n)0}(z,r)\,,
\end{align}
where,
\begin{align}
    I^{-}_{(0)0}(z,r)&=2rz\int^{\infty}_{0}\frac{x^{-\epsilon}}{\big(x^2+z^2\big)\big(x^2+z^2(1-r^2)\big)}\nonumber\\
    &=-\frac{\pi  \left(\left(1-r^2\right)^{\frac{1}{2}-\frac{\epsilon }{2}}+r^2-1\right) z^{-(\epsilon+2)} \sec \left(\frac{\pi  \epsilon }{2}\right)}{r \left(r^2-1\right)} \\
    I^{-}_{(n)0}(z,r)&=2rz\int_{0}^{\infty}dx\,\frac{x^{-\epsilon}}{x^2+z^2}\frac{x^{2}+z^{2}(1-r^{2})-(2\pi n)^2}{(x^{2}+z^{2}(1-r^{2})+(2\pi n)^2)^{2}+(4\pi n r z)^2}\,\\
    &=z^{-\epsilon } \left(-\frac{r \sec \left(\frac{\pi  \epsilon }{2}\right)}{4 \left(n^2 \pi \right)}+\frac{r^3 \sec \left(\frac{\pi  \epsilon }{2}\right) z^2}{16 n^4 \pi ^3}+O\left(z^4\right)\right)+ \Bigg(\frac{r z}{2^{\epsilon+3}\pi ^{\epsilon+2}}  \frac{(\epsilon +2)}{n^{3+\epsilon}} \sec \left(\frac{\pi  \epsilon }{2}\right)\nonumber\\
    &-z^3\frac{r}{ 2^{\epsilon +6} \pi^{\epsilon +4}}\frac{\epsilon +4}{3n^{5+\epsilon}}\big(r^2 (\epsilon +2) (\epsilon +3)+3 (\epsilon +1)\big) \sec \left(\frac{\pi  \epsilon }{2}\right)+O\left(z^4\right)\Bigg)
\end{align}
Likewise, applying the eventual limit $\epsilon\rightarrow0$ on $I^{-}_{(0)0}$ and $I^{-}_{(n)0}$; after performing sum over $n$ of $I^{-}_{(n)0}$ gives the result up to $O(z^5)$,
\begin{align}
    I^{-}_{0}(z,r)&=\frac{\pi}{z^2}\left(\frac{1}{\sqrt{1-r^2}}-1\right)-\frac{\pi  r}{4}+\frac{7 r z \zeta (3)}{2 \pi ^2}+\frac{1}{48} \pi  r^3 z^2-\frac{31 z^3 \left(2 r^3+r\right) \zeta (5)}{8 \pi ^4}\nonumber\\
    &-\frac{1}{480} \left(\pi  r^5\right) z^4+O\left(z^5\right) \,\,\,\,(\text{for}\, z>0 \,\text{and}\,|r|\leq 1)\,.
\end{align}
Facilitated by the recursion relations as given in Eqs.\eqref{recG}-\eqref{recI}, one can derive higher moments of these integrals. However, in our calculation, we need to know the integral $H^{-}_{n}(z,r)$ and $G^{-}_{n}(z,r)$ up to the value $n=5$, and the integrals $I^{-}_{n}(z,r)$ up to the value $n=6$.
These are listed here up to the $O(z^5)$,
\begin{align}
    G^{-}_{3}(z,r)&=\frac{1}{3} \pi ^2 r z-\frac{1}{2} \pi  r z^2\sqrt{1-r^2}+\left(\frac{r}{4}-\frac{r^3}{6}\right) z^3-\frac{r z^5 \zeta (3)}{32 \pi ^2}+O(z^6)\,,\\
    G^{-}_{5}(z,r)&=\frac{1}{45} \pi ^4 r z+\frac{1}{72} \pi ^2 r \left(4 r^2-3\right) z^3+\frac{1}{24} \pi  r \left(1-r^2\right)^{3/2} z^4\nonumber\\
    &+\frac{1}{960} \left(-8 r^5+20 r^3-15 r\right) z^5+O\left(z^6\right)\,,\\
    H^{-}_{3}(z,r)&=\frac{\pi ^2}{6}-\frac{1}{2} \left(\pi  \sqrt{1-r^2}\right) z+\frac{1}{8} z^2 \left(-2 r^2-2 \log (z)-2 \gamma +1+2 \log (4 \pi )\right)\nonumber\\
    &+\frac{\left(4 r^2+1\right) z^4 \zeta (3)}{64 \pi ^2}+O\left(z^6\right)\,,\\
    H^{-}_{5}(z,r)&=\frac{\pi ^4}{180}+\frac{1}{48} \pi ^2 \left(2 r^2-1\right) z^2++\frac{1}{24} \pi  \left(1-r^2\right)^{3/2} z^3+\frac{1}{768} z^4 \Big(-8 r^4+24 r^2 \nonumber\\
    &+12 \log(z)+12 \gamma -9-12 \log(4 \pi )\Big)+O(z^6)\,,\\
    I^{-}_{4}(z,r)&=\frac{2}{3} \pi ^2 r z+\frac{\pi  \left(r^3-\sqrt{1-r^2}-2 r+1\right) z^2}{\sqrt{1-r^2}}+\frac{1}{6} r \left(9-2 r^2\right) z^3\nonumber\\
    &-\frac{1}{4} (\pi  r) z^4+\frac{51 r z^5 \zeta (3)}{16 \pi ^2}+O\left(z^6\right)\,,\\  
    I^{-}_{6}(z,r)&=\frac{8}{15} \pi ^4 r z+\frac{1}{3} \pi ^2 r \left(4 r^2-5\right) z^3+\frac{\pi  \left(r^5-3 r^3+\sqrt{1-r^2}+3 r-1\right) z^4}{\sqrt{1-r^2}}\nonumber\\
    &+\frac{1}{120} \left(-24 r^5+100 r^3-225 r\right) z^5+O\left(z^6\right)\,.
\end{align}
\subsection{ Fermionic case $(k=1)$ :}
In the context of the massive fermionic case, with the same condition $\alpha\leq z$ at the high-temperature limit, a similar calculation can be performed as previously explained for the bosonic case.
\subsubsection{$G^{+}_{1}(z,r)$, $H^{+}_{1}(z,r)$ and $I^{+}_{0}(z,r)$ integral at $z\ll 1$ limit :}
At the $z\ll1$ limit ( for any $|r|\leq1$ ), and by juxtaposing the bosonic identity given in Eq.\eqref{ideney}, a systematic calculation can be performed by employing this identity.
\begin{align}
    \frac{1}{e^y+1}=\frac{1}{2}-2\sum_{n=0}^{\infty}\frac{y}{y^2+(2n+1)^2\pi^2}\,.
\end{align}
Substituting this identity into the integrands of $G^{+}_{n}(z,r)$, $H^{+}_{n}(z,r)$, as given in Eq.\eqref{Gn},\eqref{Hn} ,respectively, and evaluate the lowest moment of the integral for $n=1$ yields:
\begin{align}
    G^{+}_{1}(z,r)&=r z-\frac{7 z^3 r \zeta (3)}{4 \pi ^2}+\frac{31 r \left(4 r^2+3\right) z^5 \zeta (5)}{32 \pi ^4}+O\left(z^6\right),,\\
    H^{+}_{1}(z,r)&=-\Big(\log (\frac{z}{\pi})+\gamma \Big)+\frac{7 \left(2 r^2+1\right) z^2 \zeta (3)}{8 \pi ^2}-\frac{31 z^4 \left(8 r^4+24 r^2+3\right) \zeta (5)}{128 \pi ^4}+O\left(z^6\right)\,.
\end{align}
Moreover, from the definition Eq.\eqref{In} one can compute the integral $I^{+}_{1}(z,r)$, one can  write it up to $O(z^5)$ as follows,
\begin{align}
    I^{+}_{0}(z,r)&=\frac{\pi  r}{4}-\frac{7 z (r \zeta (3))}{2 \pi ^2}-\frac{1}{48} \left(\pi  r^3\right) z^2+\frac{31 \left(2 r^3+r\right) z^3 \zeta (5)}{8 \pi ^4}+\frac{1}{480} \pi  r^5 z^4+O\left(z^5\right)\,.
\end{align}
Besides, higher moments can be derived from the recursion relations as given in  Eqs.\eqref{recG} and \eqref{recH}. These can be listed up to $O(z^6)$ as follows, 
\begin{align}
    G^{+}_{3}(z,r)&=\frac{1}{6} \pi ^2 r z+\frac{1}{12} r \left(2 r^2-3\right) z^3+\frac{7 r z^5 \zeta (3)}{32 \pi ^2}+O\left(z^6\right)\,,\\
    G^{+}_{5}(z,r)&=\frac{7}{360} \pi ^4 r z+\frac{1}{144} \pi ^2 r \left(4 r^2-3\right) z^3+\frac{1}{960} r \left(8 r^4-20 r^2+15\right) z^5+O\left(z^6\right)\,,\\
    H^{+}_{3}(z,r)&=\frac{\pi ^2}{12}+\frac{1}{8} z^2 \left(2 r^2+2 \log (z)+2 \gamma -1-2 \log (\pi )\right)\nonumber\\
    &-\frac{7 z^4 \left(4 r^2+1\right) \zeta (3)}{64 \pi ^2}+O\left(z^6\right)\,,\\
    H^{+}_{5}(z,r)&=\frac{7 \pi ^4}{1440}+\frac{ \pi^2}{96} \left(2 r^2-1\right) z^2+\frac{1}{768} z^4 \Big(8 r^4-24 r^2-12 \log \Big(\frac{z}{\pi}\Big)\nonumber\\
    &-12 \gamma +9\Big)+O\left(z^6\right)\,,\\
    I^{+}_{4}(z,r)&=\frac{1}{3} \pi ^2 r z+\frac{1}{6} r \left(2 r^2-9\right) z^3+\frac{1}{4} \pi  r z^4-\frac{21 z^5 (r \zeta (3))}{16 \pi ^2}+O\left(z^6\right)\,,\\
    I^{+}_{6}(z,r)&=\frac{7}{15} \pi ^4 r z+\frac{1}{6} \pi ^2 r \left(4 r^2-5\right) z^3+\left(\frac{r^5}{5}-\frac{5 r^3}{6}+\frac{15 r}{8}\right) z^5+O\left(z^6\right)\,.
\end{align}
\subsubsection{Integrals related to the conformal fermionic system at finite chemical potentials :}
In the case of massless fermions with finite chemical potential, to compute the transport coefficients, one may need to know the following integrals,
\begin{align}
    \mathcal{J}_{n}(\alpha)&=\int_{0}^{\infty}dx\,x^{n-1}\Big(f(x,\alpha)\big(1-f(x,\alpha)\big)-\bar{f}(x,\alpha)\big(1-\bar{f}(x,\alpha)\big)\Big)\nonumber\\
    &=\int^{\infty}_{0}dx\,x^{n-1}\partial_{\lambda}\Big(f(x,\lambda)+\bar{f}(x,\lambda)\Big)\Big|_{\lambda\rightarrow\alpha}\nonumber\\
    &=\partial_{\lambda}J_{n}(\alpha)\Big|_{\lambda\rightarrow\alpha}\,,\\
    \mathcal{L}_{n}(\alpha)&=\int_{0}^{\infty}dx\,x^{n-1}\Big(f(x,\alpha)\big(1-f(x,\alpha)\big)+\bar{f}(x,\alpha)\big(1-\bar{f}(x,\alpha)\big)\Big)\\
    &=\int^{\infty}_{0}dx\,x^{n-1}\partial_{\lambda}\Big(f(x,\lambda)-\bar{f}(x,\lambda)\Big)\Big|_{\lambda\rightarrow\alpha}\nonumber\\
    &=\partial_{\lambda}L_{n}(\alpha)\Big|_{\lambda\rightarrow\alpha}\,,\\
\end{align}
where, $f(x,\alpha)=\frac{1}{e^{x-\alpha}+1}$ and  $\bar{f}(x,\alpha)=\frac{1}{e^{x+\alpha}+1}$ are the distribution function of fermion particles and anti-particles, respectively at finite chemical potential and in the conformal limit. A straightforward calculation can show that,
\begin{align}
    J_{n}(\alpha)=-\Gamma (n) \left(\text{Li}_n\left(-e^{-\alpha }\right)+\text{Li}_n\left(-e^{\alpha }\right)\right)\,,\\
    L_{n}(\alpha)=\Gamma (n) \left(\text{Li}_n\left(-e^{\alpha }\right)-\text{Li}_n\left(-e^{-\alpha }\right)\right)\,.
\end{align}
These functions help to understand the scaling properties of dimensionless quantities, which are constructed from the ratios of transport coefficients, e.g.,$\frac{T\kappa_{T}}{\eta}$ and $\frac{T\kappa_{AB}}{\eta}$ (See Fig.~\ref{fig:thermal cond} and Fig~\ref{fig:fermionic_cond}).
\section{Pinching Pole Approximation}\label{Pinchingpole}

To extract the transport coefficients, Eq.~\eqref{Kappat}, from Eq.~\eqref{rspdtoD} with the help of representation of the particle RSF as given in Eq.~\eqref{rspdtoD}, we consider the finite thermal width. Physically that corresponds to a transport coefficients arising due to a finite mean free path (or lifetime) in the presence of interactions. On the other hand, transport coefficients are ill-defined in the free theory due to the vanishing lifetime of the quasi-particle excitation. In the weakly coupling limit with non-zero but small (by ignoring the terms of $\mathcal{O}(\Gamma_{a\,\pmo}/E_{a\,\pmo})$) thermal width, one can approximate the spectral function as Eq.~\eqref{rspdtoD} which is written as a difference of the propagators $\Delta_{a\,1}(k_{0},\km)$ and $\Delta_{a\,2}(k_{0},\km)$. Therefore, the dependency on the single-particle spectral function in a quadratic form needs to be evaluated, and finding the dominant contribution is termed as pinching pole approximation. The quadratic term of the spectral function consists of two kind of terms, $\Delta_{a\,1}^{2}(k_{0},\km)$ and $\Delta_{a\,2}^{2}(k_{0},\km)$ having poles of order two on the same side of the frequency $k_{0}$-axis. These lead to contribution smaller than the contribution coming from mixing term $\Delta_{a\,1}(k_{0},\km)\Delta_{a\,2}(k_{0},\km)$ which has symmetrically distributed simple poles across the real $k_{0}$ axis. In the computation of the transport coefficients in Eq.~\eqref{kappaaab}, the opposite lying poles with respect to the $k_{0}$-axis gives rise to the pinching of the contour and has dominant contribution 
\begin{align}
    &\int\frac{dk_{0}}{2\pi}\Delta_{a\,1}(k_{0},\km)\Delta_{a\,2}(k_{0},\km) F(k_{0},\km) = \label{residue}\\
    &\frac{1}{8E_{a\,\km}\Gamma_{a\,\km}}\Bigg[\frac{F(E_{a\,\km}-i\Gamma_{a\,\km},\km)}{E_{a\,\km}-i\Gamma_{a\,\km}}+\frac{F(-E_{a\,\km}-i\Gamma_{a\,\km},\km)}{E_{a\,\km}+i\Gamma_{a\,\km}}\Bigg]
    -2\pi i\sum_{n}\text{Res}(F(k_{0},\km),k_{0}=\alpha_{n}). \nonumber
\end{align}
Here, $\alpha_{n}$ is the position of the pole of the function $F(k_{0},\km)$ on the complex plane of $k_{0}$. Hence, we can expand the last result around small values of thermal width for weakly coupled system and are able to write Eq.~\eqref{residue} as
\begin{align}
    \int\frac{dk_{0}}{2\pi}\Delta_{a\,1}(k_{0},\km)\Delta_{a\,2}(k_{0},\km) F(k_{0},\km)=&\, \frac{1}{8E^{2}_{a\,\km}\Gamma_{a\,\km}}\Big[F(E_{a\,\km},\km)+F(-E_{a\,\km},\km)\Big] \nonumber\\
    &\, +\sum_{n=0}\mathcal{C}_{n}(\Gamma_{a\,\km}/E_{a\,\km})^{n}.\label{PinchingpoleD2}
\end{align}
For weakly coupled theory one can ignore terms of the form $\mathcal{O}(\Gamma_{a\,\pmo}/E_{a\,\pmo})^{n}$ for $n\geq0$ compared to the first term due to the presence of pinching pole contribution $\mathcal{O}(1/\Gamma_{a\,\km})$. Similar calculation can be performed for Eq.~\eqref{kappaaab} and one can show the result as given in Eq.~\eqref{Kappat}. 


\acknowledgments
S.D. thanks Arpan Das, Arghya Mukherjee and Ritesh Ghosh for useful discussions regarding thermal field theory. S.D. thanks Samapan Bhadury for assisting with the Feynman diagrams. S.D. acknowledges the kind hospitality of Jagiellonian University, where part of this work was carried out. S.D. was supported in part through the academic co-operation agreement between Jagiellonian University, Krakow and National Institute of Science Education and Research, Jatni.



\bibliographystyle{JHEP}
\bibliography{ref}

\end{document}

%% file: shortcut.tex
\def\sumintb{\sum\!\!\!\!\!\!\!\!\int\limits}

\def\sumtintb{\sum\!\!\!\!\!\!\!\!\!\int\limits}

\def\deltacut{\tilde{\delta}}

\def\Iphi{\tilde{\phi}}

\def\heat{\hat{\mathcal{Q}}}
\def\del{\partial}
\def\0E{\bar{0}}
\def\Ch{\mathcal{X}}

\def\Ka{\mathcal{K}}
\def\pmo{ {\bf p} }
\def\qm{ {\bf q} }

\def\sm{ {\bf s} }
\def\lm{ {\bf l} }
\def\km{ {\bf k} }

\def\xv{ {\bf x} }

\def\Oc{\mathcal{O}}

\def\lamo{\lambda_{1}}

\def\phd{\phi^{\dagger}}

\def\num{\nonumber}